\documentclass[rmp,twocolumn,twoside]{revtex4}

\usepackage{graphicx}
\usepackage{rotating}
\usepackage{amsmath}
\usepackage{amssymb}

\bibpunct{[}{]}{,}{n}{}{,}
\renewcommand{\sectionmark}[1]{\markboth{The structure and function of
complex networks}{\thesection\quad #1}}

\newcommand{\defn}{\textit}
\newcommand{\half}{\mbox{$\frac12$}}

\newcommand{\ord}{\mathrm{O}}
\renewcommand{\d}{\mathrm{d}}

\newcommand{\e}{\mathrm{e}}
\newcommand{\set}[1]{\lbrace#1\rbrace}
\newcommand{\av}[1]{\langle#1\rangle}
\newcommand{\eref}[1]{(\ref{#1})}
\newcommand{\citen}{\onlinecite}
\newcommand{\etal}{{\it{}et~al.}}
\newcommand{\vA}{\mathbf{A}}
\newcommand{\vE}{\mathbf{E}}
\newcommand{\vI}{\mathbf{I}}
\newcommand{\ve}{\mathbf{e}}
\newcommand{\vm}{\mathbf{m}}
\newcommand{\vx}{\mathbf{x}}
\newcommand{\vy}{\mathbf{y}}
\newcommand{\cG}{\mathcal{G}}

\newcommand{\norm}[1]{\|\,#1\,\|}
\newcommand{\Tr}{\mathop{\mathrm{Tr}}}
\newcommand{\Li}{\mathop{\mathrm{Li}}}
\newcommand{\Beta}{\mathrm{B}}

\begin{document}

\title{The structure and function of complex networks}
\author{M. E. J. Newman}
\affiliation{Department of Physics, University of Michigan, Ann Arbor, MI
48109, U.S.A.}
\affiliation{Santa Fe Institute, 1399 Hyde Park Road, Santa Fe, NM 87501,
U.S.A.}

\begin{abstract}
Inspired by empirical studies of networked systems such as the Internet,
social networks, and biological networks, researchers have in recent years
developed a variety of techniques and models to help us understand or
predict the behavior of these systems.  Here we review developments in this
field, including such concepts as the small-world effect, degree
distributions, clustering, network correlations, random graph models,
models of network growth and preferential attachment, and dynamical
processes taking place on networks.
\end{abstract}
\maketitle

\tableofcontents
\onecolumngrid

\begin{quote}
\begin{acknowledgments}
\footnotesize
For useful feedback on early versions of this article, the author would
particularly like to thank Lada Adamic, Michelle Girvan, Petter Holme,
Randy LeVeque, Sidney Redner, Ricard Sol\'e, Steve Strogatz, Alexei
V\'azquez, and an anonymous referee.  For other helpful conversations and
comments about networks thanks go to Lada Adamic, L\'aszl\'o Barab\'asi,
Stefan Bornholdt, Duncan Callaway, Peter Dodds, Jennifer Dunne, Rick
Durrett, Stephanie Forrest, Michelle Girvan, Jon Kleinberg, James Moody,
Cris Moore, Martina Morris, Juyong Park, Richard Rothenberg, Larry Ruzzo,
Matthew Salganik, Len Sander, Steve Strogatz, Alessandro Vespignani, Chris
Warren, Duncan Watts, and Barry Wellman.  For providing data used in
calculations and figures, thanks go to Lada Adamic, L\'aszl\'o Barab\'asi,
Jerry Davis, Jennifer Dunne, Ram\'on Ferrer i Cancho, Paul Ginsparg, Jerry
Grossman, Oleg Khovayko, Hawoong Jeong, David Lipman, Neo Martinez, Stephen
Muth, Richard Rothenberg, Ricard Sol\'e, Grigoriy Starchenko, Duncan Watts,
Geoffrey West, and Janet Wiener.  Figure~\ref{examples}a was kindly
provided by Neo Martinez and Richard Williams and Fig.~\ref{school} by
James Moody.  This work was supported in part by the US National Science
Foundation under grants DMS--0109086 and DMS--0234188 and by the James
S. McDonnell Foundation and the Santa Fe Institute.
\end{acknowledgments}
\end{quote}

\clearpage
\twocolumngrid

\section{Introduction}
\begin{figure}[b]
\resizebox{5cm}{!}{\includegraphics{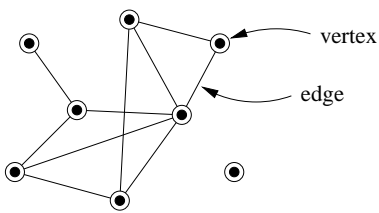}}
\caption{A small example network with eight vertices and ten edges.}
\label{network}
\end{figure}

\begin{figure*}
\resizebox{15cm}{!}{\includegraphics{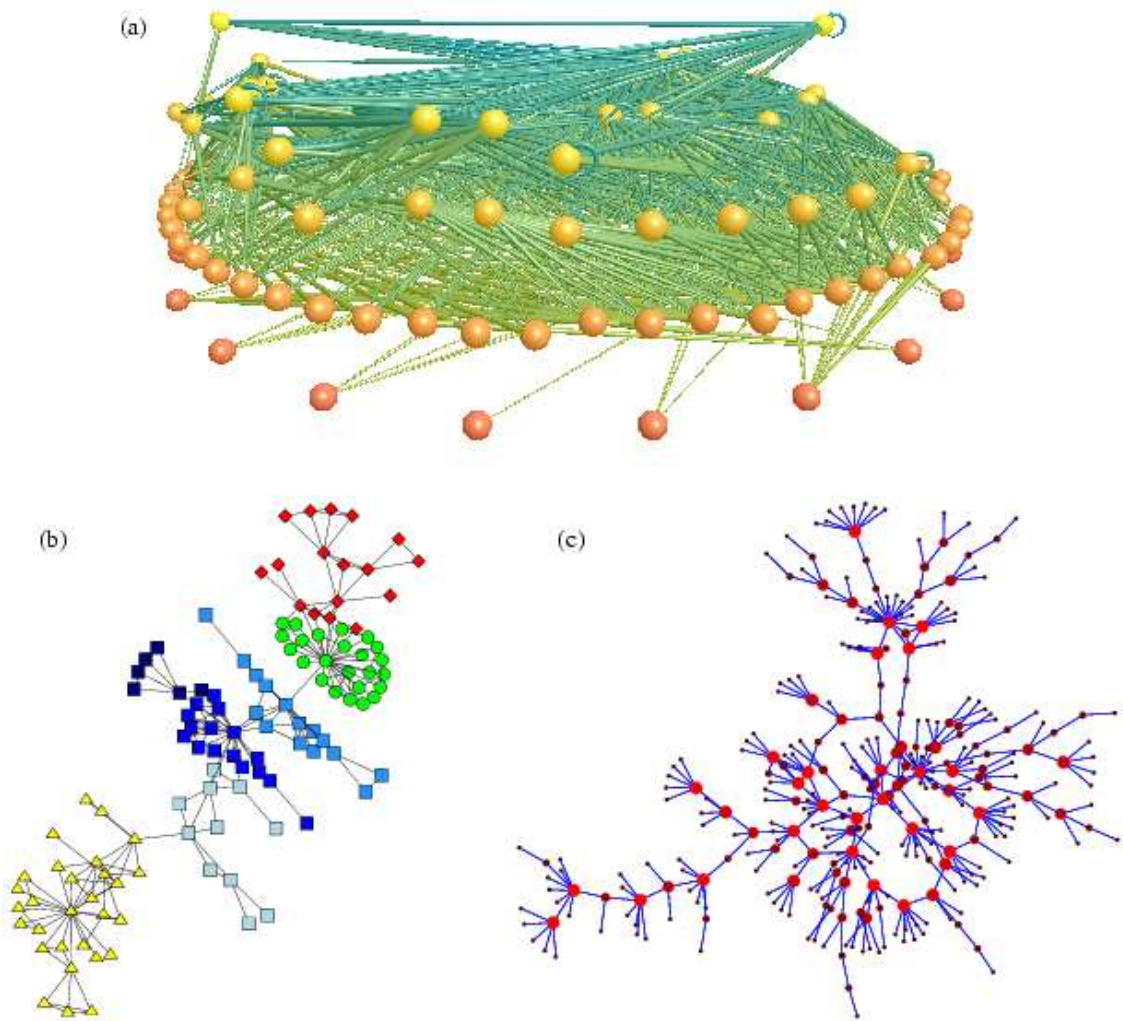}}
\caption{Three examples of the kinds of networks that are the topic of
this review.  (a)~A food web of predator-prey interactions between species
in a freshwater lake~\cite{Martinez91}.  Picture courtesy of Neo Martinez
and Richard Williams.  (b)~The network of collaborations between scientists
at a private research institution~\cite{GN02}.  (c)~A network of sexual
contacts between individuals in the study by
Potterat~\etal~\cite{Potterat02}.}
\label{examples}
\end{figure*}

A \defn{network} is a set of items, which we will call \defn{vertices} or
sometimes nodes, with connections between them, called
\defn{edges} (Fig.~\ref{network}).  Systems taking the form of networks
(also called ``graphs'' in much of the mathematical literature) abound in
the world.  Examples include the Internet, the World Wide Web, social
networks of acquaintance or other connections between individuals,
organizational networks and networks of business relations between
companies, neural networks, metabolic networks, food webs, distribution
networks such as blood vessels or postal delivery routes, networks of
citations between papers, and many others (Fig.~\ref{examples}).  This
paper reviews recent (and some not-so-recent) work on the structure and
function of networked systems such as these.

The study of networks, in the form of mathematical graph theory, is one of
the fundamental pillars of discrete mathematics.  Euler's celebrated 1735
solution of the K\"onigsberg bridge problem is often cited as the first
true proof in the theory of networks, and during the twentieth century
graph theory has developed into a substantial body of knowledge.

Networks have also been studied extensively in the social sciences.
Typical network studies in sociology involve the circulation of
questionnaires, asking respondents to detail their interactions with
others.  One can then use the responses to reconstruct a network in which
vertices represent individuals and edges the interactions between them.
Typical social network studies address issues of centrality (which
individuals are best connected to others or have most influence) and
connectivity (whether and how individuals are connected to one another
through the network).

Recent years however have witnessed a substantial new movement in network
research, with the focus shifting away from the analysis of single small
graphs and the properties of individual vertices or edges within such
graphs to consideration of large-scale statistical properties of graphs.
This new approach has been driven largely by the availability of computers
and communication networks that allow us to gather and analyze data on a
scale far larger than previously possible.  Where studies used to look at
networks of maybe tens or in extreme cases hundreds of vertices, it is not
uncommon now to see networks with millions or even billions of vertices.
This change of scale forces upon us a corresponding change in our analytic
approach.  Many of the questions that might previously have been asked in
studies of small networks are simply not useful in much larger networks.  A
social network analyst might have asked, ``Which vertex in this network
would prove most crucial to the network's connectivity if it were
removed?''  But such a question has little meaning in most networks of a
million vertices---no single vertex in such a network will have much effect
at all when removed.  On the other hand, one could reasonably ask a
question like, ``What percentage of vertices need to be removed to
substantially affect network connectivity in some given way?''\ and this
type of statistical question has real meaning even in a very large network.

However, there is another reason why our approach to the study of networks
has changed in recent years, a reason whose importance should not be
underestimated, although it often is.  For networks of tens or hundreds of
vertices, it is a relatively straightforward matter to draw a picture of
the network with actual points and lines (Fig.~\ref{examples}) and to
answer specific questions about network structure by examining this
picture.  This has been one of the primary methods of network analysts
since the field began.  The human eye is an analytic tool of remarkable
power, and eyeballing pictures of networks is an excellent way to gain an
understanding of their structure.  With a network of a million or a billion
vertices however, this approach is useless.  One simply cannot draw a
meaningful picture of a million vertices, even with modern 3D computer
rendering tools, and therefore direct analysis by eye is hopeless.  The
recent development of statistical methods for quantifying large networks is
to a large extent an attempt to find something to play the part played by
the eye in the network analysis of the twentieth century.  Statistical
methods answer the question, ``How can I tell what this network looks like,
when I can't actually look at it?''

The body of theory that is the primary focus of this review aims to do
three things.  First, it aims to find statistical properties, such as path
lengths and degree distributions, that characterize the structure and
behavior of networked systems, and to suggest appropriate ways to measure
these properties.  Second, it aims to create models of networks that can
help us to understand the meaning of these properties---how they came to be
as they are, and how they interact with one another.  Third, it aims to
predict what the behavior of networked systems will be on the basis of
measured structural properties and the local rules governing individual
vertices.  How for example will network structure affect traffic on the
Internet, or the performance of a Web search engine, or the dynamics of
social or biological systems?  As we will see, the scientific community
has, by drawing on ideas from a broad variety of disciplines, made an
excellent start on the first two of these aims, the characterization and
modeling of network structure.  Studies of the effects of structure on
system behavior on the other hand are still in their infancy.  It remains
to be seen what the crucial theoretical developments will be in this area.

\subsection{Types of networks}
\label{types}
\begin{figure}[t]
\resizebox{6.5cm}{!}{\includegraphics{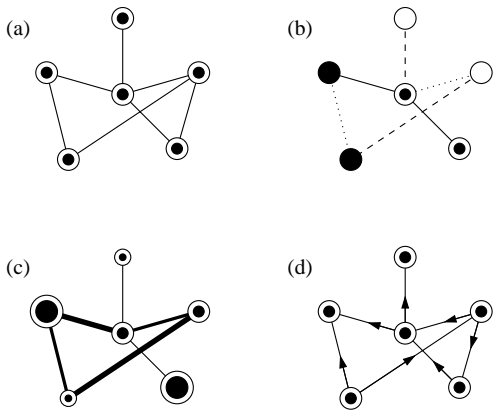}}
\caption{Examples of various types of networks: (a)~an undirected network
with only a single type of vertex and a single type of edge; (b)~a network
with a number of discrete vertex and edge types; (c)~a network with varying
vertex and edge weights; (d)~a directed network in which each edge has a
direction.}
\label{figtypes}
\end{figure}

A set of vertices joined by edges is only the simplest type of network;
there are many ways in which networks may be more complex than this
(Fig.~\ref{figtypes}).  For instance, there may be more than one different
type of vertex in a network, or more than one different type of edge.  And
vertices or edges may have a variety of properties, numerical or otherwise,
associated with them.  Taking the example of a social network of people,
the vertices may represent men or women, people of different nationalities,
locations, ages, incomes, or many other things.  Edges may represent
friendship, but they could also represent animosity, or professional
acquaintance, or geographical proximity.  They can carry weights,
representing, say, how well two people know each other.  They can also be
directed, pointing in only one direction.  Graphs composed of directed
edges are themselves called directed graphs or sometimes \defn{digraphs},
for short.  A graph representing telephone calls or email messages between
individuals would be directed, since each message goes in only one
direction.  Directed graphs can be either cyclic, meaning they contain
closed loops of edges, or acyclic meaning they do not.  Some networks, such
as food webs, are approximately but not perfectly acyclic.

One can also have \defn{hyperedges}---edges that join more than two
vertices together.  Graphs containing such edges are called
\defn{hypergraphs}.  Hyperedges could be used to indicate family ties in a
social network for example---$n$~individuals connected to each other by
virtue of belonging to the same immediate family could be represented by an
$n$-edge joining them.  Graphs may also be naturally partitioned in various
ways.  We will see a number of examples in this review of \defn{bipartite
graphs}: graphs that contain vertices of two distinct types, with edges
running only between unlike types.  So-called \defn{affiliation networks}
in which people are joined together by common membership of groups take
this form, the two types of vertices representing the people and the
groups.  Graphs may also evolve over time, with vertices or edges appearing
or disappearing, or values defined on those vertices and edges changing.
And there are many other levels of sophistication one can add.  The study
of networks is by no means a complete science yet, and many of the
possibilities have yet to be explored in depth, but we will see examples of
at least some of the variations described here in the work reviewed in this
paper.

The jargon of the study of networks is unfortunately confused by differing
usages among investigators from different fields.  To avoid (or at least
reduce) confusion, we give in Table~\ref{glossary} a short glossary of terms as they are
used in this paper.

\newsavebox{\gloss}
\sbox{\gloss}{\fbox{%
\begin{minipage}{12cm}
\setlength{\parskip}{4pt}
\setlength{\parindent}{2em}
\small

\textit{Vertex (pl.~vertices):} The fundamental unit of a network,
also called a site (physics), a node (computer science), or an actor
(sociology).

\textit{Edge:} The line connecting two vertices.  Also
called a bond (physics), a link (computer science), or a tie (sociology).

\textit{Directed/undirected:} An edge is directed if it runs in only
one direction (such as a one-way road between two points), and undirected
if it runs in both directions.  Directed edges, which are sometimes called
\defn{arcs}, can be thought of as sporting arrows indicating their
orientation.  A graph is directed if all of its edges are directed.  An
undirected graph can be represented by a directed one having two edges
between each pair of connected vertices, one in each direction.

\textit{Degree:} The number of edges connected to a vertex.  Note
that the degree is not necessarily equal to the number of vertices adjacent
to a vertex, since there may be more than one edge between any two
vertices.  In a few recent articles, the degree is referred to as the
``connectivity'' of a vertex, but we avoid this usage because the word
connectivity already has another meaning in graph theory.  A directed graph
has both an in-degree and an out-degree for each vertex, which are the
numbers of in-coming and out-going edges respectively.

\textit{Component:} The component to which a vertex belongs is that
set of vertices that can be reached from it by paths running along edges of
the graph.  In a directed graph a vertex has both an in-component and an
out-component, which are the sets of vertices from which the vertex can be
reached and which can be reached from it.

\textit{Geodesic path:} A geodesic path is the shortest path through
the network from one vertex to another.  Note that there may be and often
is more than one geodesic path between two vertices.

\textit{Diameter:} The diameter of a network is the length (in number of
edges) of the longest geodesic path between any two vertices.  A few
authors have also used this term to mean the \emph{average} geodesic
distance in a graph, although strictly the two quantities are quite
distinct.

\end{minipage}
}}
\begin{table*}
\usebox{\gloss}
\caption{A short glossary of terms.}
\label{glossary}
\end{table*}

\subsection{Other resources}
A number of other reviews of this area have appeared recently, which the
reader may wish to consult.  Albert and Barab\'asi~\cite{AB02} and
Dorogovtsev and Mendes~\cite{DM02} have given extensive pedagogical reviews
focusing on the physics literature.  Both devote the larger part of their
attention to the models of growing graphs that we describe in
Sec.~\ref{growing}.  Shorter reviews taking other viewpoints have been
given by Newman~\cite{Newman00b} and Hayes~\cite{Hayes00a,Hayes00b}, who
both concentrate on the so-called ``small-world'' models (see
Sec.~\ref{sw}), and by Strogatz~\cite{Strogatz01}, who includes an
interesting discussion of the behavior of dynamical systems on networks.

A number of books also make worthwhile reading.  Dorogovtsev and
Mendes~\cite{DM03b} have expanded their above-mentioned review into a book,
which again focuses on models of growing graphs.  The edited volumes by
Bornholdt and Schuster~\cite{BS03} and by Pastor-Satorras and
Rubi~\cite{PR03} both contain contributed essays on various topics by
leading researchers.  Detailed treatments of many of the topics covered in
the present work can be found there.  The book by Newman~\etal~\cite{NBW03}
is a collection of previously published papers, and also contains some
review material by the editors.

Three popular books on the subject of networks merit a mention.
Albert-L\'aszl\'o Barab\'asi's \textit{Linked}~\cite{Barabasi02a} gives a
personal account of recent developments in the study of networks, focusing
particularly on Barab\'asi's work on scale-free networks.  Duncan Watts's
\textit{Six Degrees}~\cite{Watts03} gives a sociologist's view, partly
historical, of discoveries old and new.  Mark Buchanan's
\textit{Nexus}~\cite{Buchanan02} gives an entertaining portrait of the
field from the point of view of a science journalist.

Farther afield, there are a variety of books on the study of networks in
particular fields.  Within graph theory the books by Harary~\cite{Harary95}
and by Bollob\'as~\cite{Bollobas98} are widely cited and among social
network theorists the books by Wasserman and Faust~\cite{WF94} and by
Scott~\cite{Scott00}.  The book by Ahuja~\etal~\cite{AMO93} is a useful
source for information on network algorithms.

\subsection{Outline of the review}
The outline of this paper is as follows.  In Sec.~\ref{data} we describe
empirical studies of the structure of networks, including social networks,
information networks, technological networks and biological networks.  In
Sec.~\ref{props} we describe some of the common properties that are
observed in many of these networks, how they are measured, and why they are
believed to be important for the functioning of networked systems.
Sections~\ref{rg} to~\ref{growing} form the heart of the review.  They
describe work on the mathematical modeling of networks, including random
graph models and their generalizations, exponential random graphs,
p$^*$~models and Markov graphs, the small-world model and its variations,
and models of growing graphs including preferential attachment models and
their many variations.  In Sec.~\ref{onnets} we discuss the progress, such
as it is, that has been made on the study of processes taking place on
networks, including epidemic processes, network failure, models displaying
phase transitions, and dynamical systems like random Boolean networks and
cellular automata.  In Sec.~\ref{concs} we give our conclusions and point
to directions for future research.

\section{Networks in the real world}
\label{data}
In this section we look at what is known about the structure of networks of
different types.  Recent work on the mathematics of networks has been
driven largely by observations of the properties of actual networks and
attempts to model them, so network data are the obvious starting point for
a review such as this.  It also makes sense to examine simultaneously data
from different kinds of networks.  One of the principal thrusts of recent
work in this area, inspired particularly by a groundbreaking 1998 paper by
Watts and Strogatz~\cite{WS98}, has been the comparative study of networks
from different branches of science, with emphasis on properties that are
common to many of them and the mathematical developments that mirror those
properties.  We here divide our summary into four loose categories of
networks: social networks, information networks, technological networks and
biological networks.

\subsection{Social networks}
\label{socnet}
A social network is a set of people or groups of people with some pattern
of contacts or interactions between them~\cite{WF94,Scott00}.  The patterns
of friendships between individuals~\cite{Moreno34,RH61}, business
relationships between companies~\cite{Mariolis75,Mizruchi82}, and
intermarriages between families~\cite{PA93} are all examples of networks
that have been studied in the past.\footnote{Occasionally social networks
of animals have been investigated also, such as dolphins~\cite{CHB99}, not
to mention networks of fictional characters, such as the protagonists of
Tolstoy's \textit{Anna Karenina}~\cite{Knuth93} or Marvel Comics
superheroes~\cite{AMR02}.}  Of the academic disciplines the social sciences
have the longest history of the substantial quantitative study of
real-world networks~\cite{Scott00,Freeman96}.  Of particular note among the
early works on the subject are: Jacob Moreno's work in the 1920s and 30s on
friendship patterns within small groups~\cite{Moreno34}; the so-called
``southern women study'' of Davis~\etal~\cite{DGG41}, which focused on the
social circles of women in an unnamed city in the American south in 1936;
the study by Elton Mayo and colleagues of social networks of factory
workers in the late 1930s in Chicago~\cite{RD39}; the mathematical models
of Anatol Rapoport~\cite{Rapoport57}, who was one of the first theorists,
perhaps \emph{the} first, to stress the importance of the degree
distribution in networks of all kinds, not just social networks; and the
studies of friendship networks of school children by Rapoport and
others~\cite{RH61,FS64}.  In more recent years, studies of business
communities~\cite{Mariolis75,GM78,Galaskiewicz85} and of patterns of sexual
contacts~\cite{Klovdahl94,Morris97,Liljeros01,Potterat02,BMS02,JH03} have
attracted particular attention.

Another important set of experiments are the famous ``small-world''
experiments of Milgram~\cite{Milgram67,TM69}.  No actual networks were
reconstructed in these experiments, but nonetheless they tell us about
network structure.  The experiments probed the distribution of path lengths
in an acquaintance network by asking participants to pass a
letter\footnote{Actually a folder containing several documents.} to one of
their first-name acquaintances in an attempt to get it to an assigned
target individual.  Most of the letters in the experiment were lost, but
about a quarter reached the target and passed on average through the hands
of only about six people in doing so.  This experiment was the origin of
the popular concept of the ``six degrees of separation,'' although that
phrase did not appear in Milgram's writing, being coined some decades later
by Guare~\cite{Guare90}.  A brief but useful early review of Milgram's work
and work stemming from it was given by Garfield~\cite{Garfield79}.

Traditional social network studies often suffer from problems of
inaccuracy, subjectivity, and small sample size.  With the exception of a
few ingenious indirect studies such as Milgram's, data collection is
usually carried out by querying participants directly using questionnaires
or interviews.  Such methods are labor-intensive and therefore limit the
size of the network that can be observed.  Survey data are, moreover,
influenced by subjective biases on the part of respondents; how one
respondent defines a friend for example could be quite different from how
another does.  Although much effort is put into eliminating possible
sources of inconsistency, it is generally accepted that there are large and
essentially uncontrolled errors in most of these studies.  A review of the
issues has been given by Marsden~\cite{Marsden90}.

Because of these problems many researchers have turned to other methods for
probing social networks.  One source of copious and relatively reliable
data is collaboration networks.  These are typically affiliation networks
in which participants collaborate in groups of one kind or another, and
links between pairs of individuals are established by common group
membership.  A classic, though rather frivolous, example of such a network
is the collaboration network of film actors, which is thoroughly documented
in the online Internet Movie
Database.\footnote{\texttt{http://www.imdb.com/}} In this network actors
collaborate in films and two actors are considered connected if they have
appeared in a film together.  Statistical properties of this network have
been analyzed by a number of authors~\cite{WS98,ASBS00,AH00,NSW01}.  Other
examples of networks of this type are networks of company directors, in
which two directors are linked if they belong to the same board of
directors~\cite{Mariolis75,DG97,DYB01}, networks of coauthorship among
academics, in which individuals are linked if they have coauthored one or
more
papers~\cite{GI95,MP96,DG99,BG00,BM00,Newman01a,Newman01b,Newman01c,Barabasi02b,Moody03},
and coappearance networks in which individuals are linked by mention in the
same context, particularly on Web pages~\cite{KSS97,AA01} or in newspaper
articles~\cite{CKMD02} (see Fig.~\ref{examples}b).

Another source of reliable data about personal connections between people
is communication records of certain kinds.  For example, one could
construct a network in which each (directed) edge between two people
represented a letter or package sent by mail from one to the other.  No
study of such a network has been published as far as we are aware, but some
similar things have.  Aiello~\etal~\cite{ACL00,ACL02} have analyzed a
network of telephone calls made over the AT\&T long-distance network on a
single day.  The vertices of this network represent telephone numbers and
the directed edges calls from one number to another.  Even for just a
single day this graph is enormous, having about 50~million vertices, one of
the largest graphs yet studied after the graph of the World Wide Web.
Ebel~\etal~\cite{EMB02} have reconstructed the pattern of email
communications between five thousand students at Kiel University from logs
maintained by email servers.  In this network the vertices represent email
addresses and directed edges represent a message passing from one address
to another.  Email networks have also been studied by
Newman~\etal~\cite{NFB02} and by Guimer\`a~\etal~\cite{Guimera03}, and
similar networks have been constructed for an ``instant messaging'' system
by Smith~\cite{Smith02}, and for an Internet community Web site by
Holme~\etal~\cite{HEL02}.  Dodds~\etal~\cite{DMW03} have carried out an
email version of Milgram's small-world experiment in which participants
were asked to forward an email message to one of their friends in an effort
to get the message ultimately to some chosen target individual.  Response
rates for the experiment were quite low, but a few hundred completed chains
of messages were recorded, enough to allow various statistical analyses.

\subsection{Information networks}
\label{infonets}
Our second network category is what we will call \defn{information
networks} (also sometimes called ``knowledge networks'').  The classic
example of an information network is the network of citations between
academic papers~\cite{ER90}.  Most learned articles cite previous work by
others on related topics.  These citations form a network in which the
vertices are articles and a directed edge from article~A to article~B
indicates that A cites~B.  The structure of the citation network then
reflects the structure of the information stored at its vertices, hence the
term ``information network,'' although certainly there are social aspects
to the citation patterns of papers too~\cite{WWN03}.

\begin{figure}
\resizebox{8cm}{!}{\includegraphics{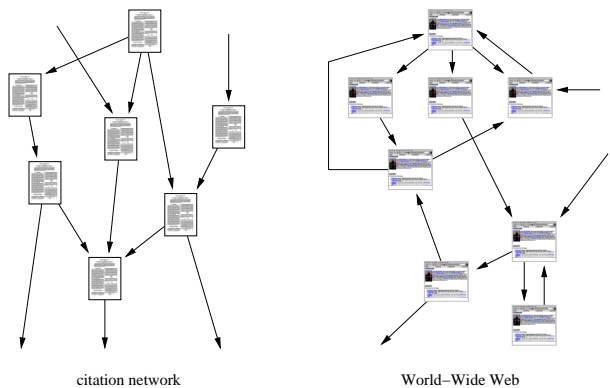}}
\caption{The two best studied information networks.  Left: the citation
network of academic papers in which the vertices are papers and the
directed edges are citations of one paper by another.  Since papers can
only cite those that came before them (lower down in the figure) the graph
is acyclic---it has no closed loops.  Right: the World Wide Web, a network
of text pages accessible over the Internet, in which the vertices are pages
and the directed edges are hyperlinks.  There are no constraints on the Web
that forbid cycles and hence it is in general cyclic.}
\label{citeweb}
\end{figure}

Citation networks are acyclic (see Sec.~\ref{types}) because papers can
only cite other papers that have already been written, not those that have
yet to be written.  Thus all edges in the network point backwards in time,
making closed loops impossible, or at least extremely rare (see
Fig.~\ref{citeweb}).

As an object of scientific study, citation networks have a great advantage
in the copious and accurate data available for them.  Quantitative study of
publication patterns stretches back at least as far as Alfred Lotka's
groundbreaking 1926 discovery of the so-called Law of Scientific
Productivity, which states that the distribution of the numbers of papers
written by individual scientists follows a power law.  That is, the number
of scientists who have written $k$ papers falls off as $k^{-\alpha}$ for
some constant~$\alpha$.  (In fact, this result extends to the arts and
humanities as well.)  The first serious work on citation patterns was
conducted in the 1960s as large citation databases became available through
the work of Eugene Garfield and other pioneers in the field of
bibliometrics.  The network formed by citations was discussed in an early
paper by Price~\cite{Price65}, in which among other things, the author
points out for the first time that both the in- and out-degree
distributions of the network follow power laws, a far-reaching discovery
which we discuss further in Sec.~\ref{dd}.  Many other studies of citation
networks have been performed since then, using the ever better resources
available in citation databases.  Of particular note are the studies by
Seglen~\cite{Seglen92} and Redner~\cite{Redner98}.\footnote{An interesting
development in the study of citation patterns has been the arrival of
automatic citation ``crawlers'' that construct citation networks from
online papers.  Examples include Citeseer
(\texttt{http://citeseer.nj.nec.com/}), SPIRES
(\texttt{http://www.slac.stanford.edu/spires/hep/}) and Citebase
(\texttt{http://citebase.eprints.org/}).}

Another very important example of an information network is the World Wide
Web, which is a network of Web pages containing information, linked
together by hyperlinks from one page to another~\cite{Huberman01}.  The Web
should not be confused with the Internet, which is a physical network of
computers linked together by optical fibre and other data
connections.\footnote{While the Web is primarily an information network,
it, like citation networks, has social aspects to its structure
also~\cite{AA01}.}  Unlike a citation network, the World Wide Web is
cyclic; there is no natural ordering of sites and no constraints that
prevent the appearance of closed loops (Fig.~\ref{citeweb}).  The Web has
been very heavily studied since its first appearance in the early 1990s,
with the studies by Albert~\etal~\cite{AJB99,BAJ00},
Kleinberg~\etal~\cite{Kleinberg99b}, and Broder~\etal~\cite{Broder00} being
particularly influential.  The Web also appears to have power-law in- and
out-degree distributions (Sec.~\ref{dd}), as well as a variety of other
interesting
properties~\cite{AJB99,Kleinberg99b,Adamic99,Kumar00,Broder00,FLGC02}.

One important point to notice about the Web is that our data about it come
from ``crawls'' of the network, in which Web pages are found by following
hyperlinks from other pages~\cite{Broder00}.  Our picture of the network
structure of the World Wide Web is therefore necessarily biased.  A page
will only be found if another page points to it,\footnote{This is not
always strictly true.  Some Web search engines allow the submission of
pages by members of the public for inclusion in databases, and such pages
need not be the target of links from any other pages.  However, such pages
also form a very small fraction of all Web pages, and certainly the biases
discussed here remain very much present.} and in a crawl that covers only a
part of the Web (as all crawls do at present) pages are more likely to be
found the more other pages point to them~\cite{LG99}.  This suggests for
instance that our measurements of the fraction of pages with low in-degree
might be an underestimate.\footnote{The degree distribution for the Web
shown in Fig.~\ref{degree} falls off slightly at low values of the
in-degree, which may perhaps reflect this bias.}  This behavior contrasts
with that of a citation network.  A paper can appear in the citation
indices even if it has never been cited (and in fact a plurality of papers
in the indices are never cited).

A few other examples of information networks have been studied to a lesser
extent.  Jaffe and Trajtenberg~\cite{JT02}, for instance, have studied the
network of citations between US patents, which is similar in some respects
to citations between academic papers.  A number of authors have looked at
peer-to-peer networks~\cite{ALPH01,ALH03,IRF02}, which are virtual networks
of computers that allow sharing of files between computer users over local-
or wide-area networks.  The network of relations between word classes in a
thesaurus has been studied by Knuth~\cite{Knuth93} and more recently by
various other authors~\cite{MDLD02,Kinouchi02,ST02}.  This network can be
looked upon as an information network---users of a thesaurus ``surf'' the
network from one word to another looking for the particular word that
perfectly captures the idea they have in mind.  However, it can also be
looked at as a conceptual network representing the structure of the
language, or possibly even the mental constructs used to represent the
language.  A number of other semantic word networks have also been
investigated~\cite{FS01a,DM01b,SC02,ST02}.

Preference networks provide an example of a bipartite information network.
A preference network is a network with two kinds of vertices representing
individuals and the objects of their preference, such as books or films,
with an edge connecting each individual to the books or films they like.
(Preference networks can also be weighted to indicate strength of likes or
dislikes.)  A widely studied example of a preference network is the
\textit{EachMovie} database of film
preferences.\footnote{\texttt{http://research.compaq.com/SRC/eachmovie/}}
Networks of this kind form the basis for \defn{collaborative filtering}
algorithms and \defn{recommender systems}, which are techniques for
predicting new likes or dislikes based on comparison of individuals'
preferences with those of others~\cite{GNOT92,SM95,RV97}.  Collaborative
filtering has found considerable commercial success for product
recommendation and targeted advertising, particularly with online
retailers.  Preference networks can also be thought of as social networks,
linking not only people to objects, but also people to other people with
similar preferences.  This approach has been adopted occasionally in the
literature~\cite{KSS97}.

\subsection{Technological networks}
\label{techno}
Our third class of networks is technological networks, man-made networks
designed typically for distribution of some commodity or resource, such as
electricity or information.  The electric power grid is a good example.
This is a network of high-voltage three-phase transmission lines that spans
a country or a portion of a country (as opposed to the local low-voltage
a.c.\ power delivery lines that span individual neighborhoods).
Statistical studies of power grids have been made by, for example, Watts
and Strogatz~\cite{WS98,Watts99a} and Amaral~\etal~\cite{ASBS00}.  Other
distribution networks that have been studied include the network of airline
routes~\cite{ASBS00}, and networks of roads~\cite{KSM03},
railways~\cite{LM02,Sen03} and pedestrian traffic~\cite{CHE02}.  River
networks could be regarded as a naturally occurring form of distribution
network (actually a collection network)
\cite{RR97,RRR98,Maritan96,DR01abc}, as could the vascular networks
discussed in Sec.~\ref{bionets}.  The telephone network and delivery
networks such as those used by the post-office or parcel delivery companies
also fall into this general category and are presumably studied within the
relevant corporations, if not yet by academic researchers.  (We distinguish
here between the physical telephone network of wires and cables and the
network of who calls whom, discussed in Sec.~\ref{socnet}.)  Electronic
circuits~\cite{FJS01} fall somewhere between distribution and communication
networks.

Another very widely studied technological network is the Internet,
i.e.,~the network of physical connections between computers.  Since there
is a large and ever-changing number of computers on the Internet, the
structure of the network is usually examined at a coarse-grained level,
either the level of routers, special-purpose computers on the network that
control the movement of data, or ``autonomous systems,'' which are groups
of computers within which networking is handled locally, but between which
data flows over the public Internet.  The computers at a single company or
university would probably form a single autonomous system---autonomous
systems often correspond roughly with domain names.

In fact, the network of physical connections on the Internet is not easy to
discover since the infrastructure is maintained by many separate
organizations.  Typically therefore, researchers reconstruct the network by
reasoning from large samples of point-to-point data routes.  So-called
``traceroute'' programs can report the sequence of network nodes that a
data packet passes through when traveling between two points and if we
assume an edge in the network between any two consecutive nodes along such
a path then a sufficiently large sample of paths will give us a fairly
complete picture of the entire network.  There may however be some edges
that never get sampled, so the reconstruction is typically a good, but not
perfect, representation of the true physical structure of the Internet.
Studies of Internet structure have been carried out by, among others,
Faloutsos~\etal~\cite{FFF99}, Broida and Claffy~\cite{BC01} and
Chen~\etal~\cite{Chen02}.

\subsection{Biological networks}
\label{bionets}
A number of biological systems can be usefully represented as networks.
Perhaps the classic example of a biological network is the network of
metabolic pathways, which is a representation of metabolic substrates and
products with directed edges joining them if a known metabolic reaction
exists that acts on a given substrate and produces a given product.  Most
of us will probably have seen at some point the giant maps of metabolic
pathways that many molecular biologists pin to their walls.\footnote{The
standard chart of the metabolic network is somewhat misleading.  For
reasons of clarity and aesthetics, many metabolites appear in more than one
place on the chart, so that some pairs of vertices are actually the same
vertex.}  Studies of the statistical properties of metabolic networks have
been performed by, for example, Jeong~\etal~\cite{Jeong00,Podani01}, Fell
and Wagner~\cite{FW00,WF01}, and Stelling~\etal~\cite{Stelling02}.  A
separate network is the network of mechanistic physical interactions
between proteins (as opposed to chemical reactions among metabolites),
which is usually referred to as a protein interaction network.  Interaction
networks have been studied by a number of
authors~\cite{Uetz00,Ito01,Jeong01,MS02a,SP03}.

Another important class of biological network is the genetic regulatory
network.  The expression of a gene, i.e.,~the production by transcription
and translation of the protein for which the gene codes, can be controlled
by the presence of other proteins, both activators and inhibitors, so that
the genome itself forms a switching network with vertices representing the
proteins and directed edges representing dependence of protein production
on the proteins at other vertices.  The statistical structure of regulatory
networks has been studied recently by various
authors~\cite{GBBK02,SMMA02,Farkas03}.  Genetic regulatory networks were in
fact one of the first networked dynamical systems for which large-scale
modeling attempts were made.  The early work on random Boolean nets by
Kauffman~\cite{Kauffman69,Kauffman71,Kauffman93} is a classic in this
field, and anticipated recent developments by several decades.

Another much studied example of a biological network is the food web, in
which the vertices represent species in an ecosystem and a directed edge
from species~A to species~B indicates that A preys
on~B~\cite{CBN90,Pimm02}---see Fig.~\ref{examples}a.  (Sometimes the
relationship is drawn the other way around, because ecologists tend to
think in terms of energy or carbon flows through food webs; a predator-prey
interaction is thus drawn as an arrow pointing from prey to predator,
indicating energy flow from prey to predator when the prey is eaten.)
Construction of complete food webs is a laborious business, but a number of
quite extensive data sets have become available in recent
years~\cite{BU89,Martinez91,GR93,HBR96}.  Statistical studies of the
topologies of food webs have been carried out by Sol\'e and
Montoya~\cite{SM01,MS02b}, Camacho~\etal~\cite{CGA02} and
Dunne~\etal~\cite{DWM02a,DWM02b,Williams02}, among others.  A particularly
thorough study of webs of plants and herbivores has been conducted by
Jordano~\etal~\cite{JBO03}, which includes statistics for no less than 53
different networks.

Neural networks are another class of biological networks of considerable
importance.  Measuring the topology of real neural networks is extremely
difficult, but has been done successfully in a few cases.  The best known
example is the reconstruction of the 282-neuron neural network of the
nematode \textit{C.~Elegans} by White~\etal~\cite{WSTB86}.  The network
structure of the brain at larger scales than individual
neurons---functional areas and pathways---has been investigated by
Sporns~\etal~\cite{STE00,Sporns02}.

Blood vessels and the equivalent vascular networks in plants form the
foundation for one of the most successful theoretical models of the effects
of network structure on the behavior of a networked system, the theory of
biological allometry~\cite{WBE97,WBE99,BMR99}, although we are not aware of
any quantitative studies of their statistical structure.

Finally we mention two examples of networks from the physical sciences, the
network of free energy minima and saddle points in glasses~\cite{Doye02}
and the network of conformations of polymers and the transitions between
them~\cite{SAB01}, both of which appear to have some interesting structural
properties.

\section{Properties of networks}
\label{props}
Perhaps the simplest useful model of a network is the random graph, first
studied by Rapoport~\cite{SR51,Rapoport57,Rapoport68} and by Erd\H{o}s and
R\'enyi~\cite{ER59,ER60,ER61}, which we describe in Sec.~\ref{poissonrg}.
In this model, undirected edges are placed at random between a fixed
number~$n$ of vertices to create a network in which each of the $\half
n(n-1)$ possible edges is independently present with some probability~$p$,
and the number of edges connected to each vertex---the degree of the
vertex---is distributed according to a binomial distribution, or a Poisson
distribution in the limit of large~$n$.  The random graph has been well
studied by mathematicians~\cite{Karonski82,Bollobas01,JLR99} and many
results, both approximate and exact, have been proved rigorously.  Most of
the interesting features of real-world networks that have attracted the
attention of researchers in the last few years however concern the ways in
which networks are \emph{not} like random graphs.  Real networks are
non-random in some revealing ways that suggest both possible mechanisms
that could be guiding network formation, and possible ways in which we
could exploit network structure to achieve certain aims.  In this section
we describe some features that appear to be common to networks of many
different types.

\begin{sidewaystable*}
\begin{center}
\begin{tabular}{l|l|l|r|r|r|r|r|l|l|r|l}
 & network                   & type       & $n$             & $m$                & $z$     & $\ell$  & $\alpha$
 & $C^{(1)}$ & $C^{(2)}$ & $r$ & Ref(s). \\
\hline
\begin{rotate}{90}
\hbox{\hspace{-7.1em}social}
\end{rotate}
 & film actors               & undirected & $449\,913$      & $25\,516\,482$     & $113.43$& $3.48$  & $2.3$
 & $0.20$  & $0.78$  & $0.208$  & \citen{WS98,ASBS00} \\
 & company directors         & undirected & $7\,673$        & $55\,392$          & $14.44$ & $4.60$  & --
 & $0.59$  & $0.88$  & $0.276$  & \citen{DYB01,NSW01} \\
 & math coauthorship         & undirected & $253\,339$      & $496\,489$         & $3.92$  & $7.57$  & --
 & $0.15$  & $0.34$  & $0.120$  & \citen{GI95,DG99}   \\
 & physics coauthorship      & undirected & $52\,909$       & $245\,300$         & $9.27$  & $6.19$  & --
 & $0.45$  & $0.56$  & $0.363$  & \citen{Newman01a,Newman01b} \\
 & biology coauthorship      & undirected & $1\,520\,251$   & $11\,803\,064$     & $15.53$ & $4.92$  & --
 & $0.088$ & $0.60$  & $0.127$  & \citen{Newman01a,Newman01b} \\
 & telephone call graph      & undirected & $47\,000\,000$  & $80\,000\,000$     & $3.16$  &         & $2.1$
 &         &      &          & \citen{ACL00,ACL02} \\
 & email messages            & directed   & $59\,912$       & $86\,300$          & $1.44$  & $4.95$  & $1.5/2.0$
 &         & $0.16$  &          & \citen{EMB02}       \\
 & email address books       & directed   & $16\,881$       & $57\,029$          & $3.38$  & $5.22$  & --
 & $0.17$  & $0.13$  & $0.092$  & \citen{NFB02}       \\
 & student relationships     & undirected & $573$           & $477$              & $1.66$  & $16.01$ & --
 & $0.005$ & $0.001$ & $-0.029$ & \citen{BMS02}       \\
 & sexual contacts           & undirected & $2\,810$        &                    &         &         & $3.2$
 &         &         &          & \citen{Liljeros01,LEA03} \\
\hline
\begin{rotate}{90}
\hbox{\hspace{-4.9em}information}
\end{rotate}
 & WWW \texttt{nd.edu}       & directed   & $269\,504$      & $1\,497\,135$      & $5.55$  & $11.27$ & $2.1$/$2.4$
 & $0.11$  & $0.29$  & $-0.067$ & \citen{AJB99,BAJ00} \\
 & WWW Altavista             & directed   & $203\,549\,046$ & $2\,130\,000\,000$ & $10.46$ & $16.18$ & $2.1$/$2.7$
 &         &         &          & \citen{Broder00}    \\
 & citation network          & directed   & $783\,339$      & $6\,716\,198$      & $8.57$  &         & $3.0$/--
 &         &         &          & \citen{Redner98}    \\
 & Roget's Thesaurus         & directed   & $1\,022$        & $5\,103$           & $4.99$  & $4.87$  & --
 & $0.13$  & $0.15$  & $0.157$  & \citen{Knuth93}     \\
 & word co-occurrence        & undirected & $460\,902$      & $17\,000\,000$     & $70.13$ &         & $2.7$
 &         & $0.44$  &          & \citen{FS01a,DM01b} \\
\hline
\begin{rotate}{90}
\hbox{\hspace{-6.5em}technological}
\end{rotate}
 & Internet                  & undirected & $10\,697$       & $31\,992$          & $5.98$  & $3.31$  & $2.5$
 & $0.035$ & $0.39$  & $-0.189$ & \citen{FFF99,Chen02} \\
 & power grid                & undirected & $4\,941$        & $6\,594$           & $2.67$  & $18.99$ & --
 & $0.10$  & $0.080$ & $-0.003$ & \citen{WS98}        \\
 & train routes              & undirected & $587$           & $19\,603$          & $66.79$ & $2.16$  & --
 &         & $0.69$  & $-0.033$ & \citen{Sen03}       \\
 & software packages         & directed   & $1\,439$        & $1\,723$           & $1.20$  & $2.42$  & $1.6/1.4$
 & $0.070$ & $0.082$ & $-0.016$ & \citen{Newman03c}   \\
 & software classes          & directed   & $1\,377$        & $2\,213$           & $1.61$  & $1.51$  & --
 & $0.033$ & $0.012$ & $-0.119$ & \citen{VCS02} \\
 & electronic circuits       & undirected & $24\,097$       & $53\,248$          & $4.34$  & $11.05$ & $3.0$
 & $0.010$ & $0.030$ & $-0.154$ & \citen{FJS01} \\
 & peer-to-peer network      & undirected & $880$           & $1\,296$           & $1.47$  & $4.28$  & $2.1$
 & $0.012$ & $0.011$ & $-0.366$ & \citen{ALPH01,RFI02} \\
\hline
\begin{rotate}{90}
\hbox{\hspace{-4.4em}biological}
\end{rotate}
 & metabolic network         & undirected & $765$           & $3\,686$           & $9.64$  & $2.56$  & $2.2$
 & $0.090$ & $0.67$  & $-0.240$ & \citen{Jeong00}    \\
 & protein interactions      & undirected & $2\,115$        & $2\,240$           & $2.12$  & $6.80$  & $2.4$
 & $0.072$ & $0.071$ & $-0.156$ & \citen{Jeong01}    \\
 & marine food web           & directed   & $135$           & $598$              & $4.43$  & $2.05$  & --
 & $0.16$  & $0.23$  & $-0.263$ & \citen{HBR96}      \\
 & freshwater food web       & directed   & $92$            & $997$              & $10.84$ & $1.90$  & --
 & $0.20$  & $0.087$ & $-0.326$ & \citen{Martinez91} \\
 & neural network            & directed   & $307$           & $2\,359$           & $7.68$  & $3.97$  & --
 & $0.18$  & $0.28$  & $-0.226$ & \citen{WSTB86,WS98} \\
\end{tabular}
\end{center}
\caption{Basic statistics for a number of published networks.  The
properties measured are: type of graph, directed or undirected; total
number of vertices~$n$; total number of edges~$m$; mean degree~$z$; mean
vertex--vertex distance~$\ell$; exponent~$\alpha$ of degree distribution if
the distribution follows a power law (or ``--'' if not; in/out-degree
exponents are given for directed graphs); clustering coefficient~$C^{(1)}$
from Eq.~\eref{defsc1}; clustering coefficient~$C^{(2)}$ from
Eq.~\eref{defsc3}; and degree correlation coefficient~$r$,
Sec.~\ref{degcorr}.  The last column gives the citation(s) for the network
in the bibliography.  Blank entries indicate unavailable data.}
\label{summary}
\end{sidewaystable*}

\subsection{The small-world effect}
\label{sweffect}
In Sec.~\ref{socnet} we described the famous experiments carried out by
Stanley Milgram in the 1960s, in which letters passed from person to person
were able to reach a designated target individual in only a small number of
steps---around six in the published cases.  This result is one of the first
direct demonstrations of the \defn{small-world effect}, the fact that most
pairs of vertices in most networks seem to be connected by a short path
through the network.

The existence of the small-world effect had been speculated upon before
Milgram's work, notably in a remarkable 1929 short story by the Hungarian
writer Frigyes Karinthy~\cite{Karinthy29}, and more rigorously in the
mathematical work of Pool and Kochen~\cite{PK78} which, although published
after Milgram's studies, was in circulation in preprint form for a decade
before Milgram took up the problem.  Nowadays, the small-world effect has
been studied and verified directly in a large number of different networks.

Consider an undirected network, and let us define $\ell$ to be the mean
geodesic (i.e.,~shortest) distance between vertex pairs in a network:
\begin{equation}
\ell = {1\over\half n(n+1)} \sum_{i\ge j} d_{ij},
\label{ell1}
\end{equation}
where $d_{ij}$ is the geodesic distance from vertex~$i$ to vertex~$j$.
Notice that we have included the distance from each vertex to itself (which
is zero) in this average.  This is mathematically convenient for a number
of reasons, but not all authors do it.  In any case, its inclusion simply
multiplies $\ell$ by $(n-1)/(n+1)$ and hence gives a correction of order
$n^{-1}$, which is often negligible for practical purposes.

The quantity~$\ell$ can be measured for a network of $n$ vertices and $m$
edges in time $\ord(mn)$ using simple breadth-first search~\cite{AMO93},
also called a ``burning algorithm'' in the physics literature.  In
Table~\ref{summary}, we show values of $\ell$ taken from the literature for
a variety of different networks.  As the table shows, the values are in all
cases quite small---much smaller than the number~$n$ of vertices, for
instance.

The definition~\eref{ell1} of~$\ell$ is problematic in networks that have
more than one component.  In such cases, there exist vertex pairs that have
no connecting path.  Conventionally one assigns infinite geodesic distance
to such pairs, but then the value of $\ell$ also becomes infinite.  To
avoid this problem one usually defines $\ell$ on such networks to be the
mean geodesic distance between all pairs that have a connecting path.
Pairs that fall in two different components are excluded from the average.
The figures in Table~\ref{summary} were all calculated in this way.  An
alternative and perhaps more satisfactory approach is to define $\ell$ to
be the ``harmonic mean'' geodesic distance between all pairs, i.e.,~the
reciprocal of the average of the reciprocals:
\begin{equation}
\ell^{-1} = {1\over\half n(n+1)} \sum_{i\ge j} d_{ij}^{-1}.
\label{ell2}
\end{equation}
Infinite values of $d_{ij}$ then contribute nothing to the sum.  This
approach has been adopted only occasionally in network
calculations~\cite{LM01b}, but perhaps should be used more often.

The small-world effect has obvious implications for the dynamics of
processes taking place on networks.  For example, if one considers the
spread of information, or indeed anything else, across a network, the
small-world effect implies that that spread will be fast on most real-world
networks.  If it takes only six steps for a rumor to spread from any person
to any other, for instance, then the rumor will spread much faster than if
it takes a hundred steps, or a million.  This affects the number of
``hops'' a packet must make to get from one computer to another on the
Internet, the number of legs of a journey for an air or train traveler, the
time it takes for a disease to spread throughout a population, and so
forth.  The small-world effect also underlies some well-known parlor games,
particularly the calculation of Erd\H{o}s numbers~\cite{DG99} and Bacon
numbers.\footnote{\texttt{http://www.cs.virginia.edu/oracle/}}

On the other hand, the small-world effect is also mathematically obvious.
If the number of vertices within a distance $r$ of a typical central vertex
grows exponentially with~$r$---and this is true of many networks, including
the random graph (Sec.~\ref{poissonrg})---then the value of~$\ell$ will
increase as $\log n$.  In recent years the term ``small-world effect'' has
thus taken on a more precise meaning: networks are said to show the
small-world effect if the value of~$\ell$ scales logarithmically or slower
with network size for fixed mean degree.  Logarithmic scaling can be proved
for a variety of network
models~\cite{Bollobas81,Bollobas01,CL02b,FFH03,DMS03b} and has also been
observed in various real-world networks~\cite{Newman01a,Newman01c,AB02}.
Some networks have mean vertex--vertex distances that increase slower
than~$\log n$.  Bollob\'as and Riordan~\cite{BR02} have shown that networks
with power-law degree distributions (Sec.~\ref{dd}) have values of $\ell$
that increase no faster than $\log n/\log\log n$ (see also
Ref.~\citen{FFH03}), and Cohen and Havlin~\cite{CH03} have given arguments
that suggest that the actual variation may be slower even than this.

\subsection{Transitivity or clustering}
\label{transitivity}
\begin{figure}
\resizebox{3.5cm}{!}{\includegraphics{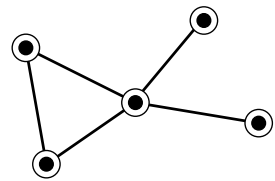}}
\caption{Illustration of the definition of the clustering coefficient~$C$,
Eq.~\eref{defsc1}.  This network has one triangle and eight connected
triples, and therefore has a clustering coefficient of
$3\times1/8=\frac38$.  The individual vertices have local clustering
coefficients, Eq.~\eref{defsci}, of 1, 1, $\frac16$, 0 and~0, for a mean
value, Eq.~\eref{defsc3}, of $C=\frac{13}{30 }$.}
\label{cc}
\end{figure}

A clear deviation from the behavior of the random graph can be seen in the
property of network transitivity, sometimes also called clustering,
although the latter term also has another meaning in the study of networks
(see Sec.~\ref{communities}) and so can be confusing.  In many networks it
is found that if vertex~A is connected to vertex~B and vertex~B to
vertex~C, then there is a heightened probability that vertex~A will also be
connected to vertex~C.  In the language of social networks, the friend of
your friend is likely also to be your friend.  In terms of network
topology, transitivity means the presence of a heightened number of
triangles in the network---sets of three vertices each of which is
connected to each of the others.  It can be quantified by defining a
clustering coefficient~$C$ thus:
\begin{equation}
C = {\mbox{$3\times$ number of triangles in the network}\over
     \mbox{number of connected triples of vertices}},
\label{defsc1}
\end{equation}
where a ``connected triple'' means a single vertex with edges running to an
unordered pair of others (see Fig.~\ref{cc}).

In effect, $C$~measures the fraction of triples that have their third edge
filled in to complete the triangle.  The factor of three in the numerator
accounts for the fact that each triangle contributes to three triples and
ensures that $C$ lies in the range $0\le C\le1$.  In simple terms, $C$~is
the mean probability that two vertices that are network neighbors of the
same other vertex will themselves be neighbors.  It can also be written in
the form
\begin{equation}
C = {\mbox{$6\times$ number of triangles in the network}\over
     \mbox{number of paths of length two}},
\label{defsc2}
\end{equation}
where a path of length two refers to a directed path starting from a
specified vertex.  This definition shows that $C$ is also the mean
probability that the friend of your friend is also your friend.

The definition of~$C$ given here has been widely used in the sociology
literature, where it is referred to as the ``fraction of transitive
triples.''\footnote{For example, the standard network analysis program
UCInet includes a function to calculate this quantity for any network.}  In
the mathematical and physical literature it seems to have been first
discussed by Barrat and Weigt~\cite{BW00}.

An alternative definition of the clustering coefficient, also widely used,
has been given by Watts and Strogatz~\cite{WS98}, who proposed defining a
local value
\begin{equation}
C_i = {\mbox{number of triangles connected to vertex~$i$}\over
       \mbox{number of triples centered on vertex~$i$}}.
\label{defsci}
\end{equation}
For vertices with degree 0 or~1, for which both numerator and denominator
are zero, we put $C_i=0$.  Then the clustering coefficient for the whole
network is the average
\begin{equation}
C = {1\over n}\sum_i C_i.
\label{defsc3}
\end{equation}
This definition effectively reverses the order of the operations of taking
the ratio of triangles to triples and of averaging over vertices---one here
calculates the mean of the ratio, rather than the ratio of the means.  It
tends to weight the contributions of low-degree vertices more heavily,
because such vertices have a small denominator in Eq.~\eref{defsci} and
hence can give quite different results from Eq.~\eref{defsc1}.  In
Table~\ref{summary} we give both measures for a number of networks (denoted
$C^{(1)}$ and $C^{(2)}$ in the table).  Normally our first
definition~\eref{defsc1} is easier to calculate analytically,
but~\eref{defsc3} is easily calculated on a computer and has found wide use
in numerical studies and data analysis.  It is important when reading (or
writing) literature in this area to be clear about which definition of the
clustering coefficient is in use.  The difference between the two is
illustrated in Fig.~\ref{cc}.

The local clustering~$C_i$ above has been used quite widely in its own
right in the sociological literature, where it is referred to as the
``network density''~\cite{Scott00}.  Its dependence on the degree~$k_i$ of
the central vertex~$i$ has been studied by Dorogovtsev~\etal~\cite{DGM02a}
and Szab\'o~\etal~\cite{SAK03}; both groups found that $C_i$ falls off with
$k_i$ approximately as $k_i^{-1}$ for certain models of scale-free networks
(Sec.~\ref{sfnet}).  Similar behavior has also been observed empirically
in real-world networks~\cite{Ravasz02,RB03,Vazquez03b}.

In general, regardless of which definition of the clustering coefficient is
used, the values tend to be considerably higher than for a random graph
with a similar number of vertices and edges.  Indeed, it is suspected that
for many types of networks the probability that the friend of your friend
is also your friend should tend to a non-zero limit as the network becomes
large, so that $C=\ord(1)$ as $n\to\infty$.\footnote{An exception is
scale-free networks with $C_i\sim k_i^{-1}$, as described above.  For such
networks Eq.~\eref{defsc1} tends to zero as $n\to\infty$, although
Eq.~\eref{defsc3} is still finite.} On the random graph, by contrast,
$C=\ord(n^{-1})$ for large~$n$ (either definition of~$C$) and hence the
real-world and random graph values can be expected to differ by a factor of
order~$n$.  This point is discussed further in Sec.~\ref{poissonrg}.

The clustering coefficient measures the density of triangles in a network.
An obvious generalization is to ask about the density of longer loops also:
loops of length four and above.  A number of authors have looked at such
higher order clustering
coefficients~\cite{GSWF01,FHJS02,Newman03a,CPV03,BC03}, although there is
so far no clean theory, similar to a cumulant expansion, that separates the
independent contributions of the various orders from one another.  If more
than one edge is permitted between a pair of vertices, then there is also a
lower order clustering coefficient that describes the density of loops of
length two.  This coefficient is particularly important in directed graphs
where the two edges in question can point in opposite directions.  The
probability that two vertices in a directed network point to each other is
called the \defn{reciprocity} and is often measured in directed social
networks~\cite{WF94,Scott00}.  It has been examined occasionally in other
contexts too, such as the World Wide Web~\cite{AA01,EM02} and email
networks~\cite{NFB02}.

\subsection{Degree distributions}
\label{dd}
Recall that the degree of a vertex in a network is the number of edges
incident on (i.e.,~connected to) that vertex.  We define $p_k$ to be the
fraction of vertices in the network that have degree~$k$.  Equivalently,
$p_k$~is the probability that a vertex chosen uniformly at random has
degree~$k$.  A plot of $p_k$ for any given network can be formed by making
a histogram of the degrees of vertices.  This histogram is the degree
distribution for the network.  In a random graph of the type studied by
Erd\H{o}s and R\'enyi~\cite{ER59,ER60,ER61}, each edge is present or absent
with equal probability, and hence the degree distribution is, as mentioned
earlier, binomial, or Poisson in the limit of large graph size.  Real-world
networks are mostly found to be very unlike the random graph in their
degree distributions.  Far from having a Poisson distribution, the degrees
of the vertices in most networks are highly right-skewed, meaning that
their distribution has a long right tail of values that are far above the
mean.

Measuring this tail is somewhat tricky.  Although in theory one just has to
construct a histogram of the degrees, in practice one rarely has enough
measurements to get good statistics in the tail, and direct histograms are
thus usually rather noisy (see the histograms in Refs.~\citen{Broder00},
\citen{FFF99} and~\citen{Price65} for example).  There are two accepted
ways to get around this problem.  One is to constructed a histogram in
which the bin sizes increase exponentially with degree.  For example the
first few bins might cover degree ranges 1, 2--3, 4--7, 8--15, and so on.
The number of samples in each bin is then divided by the width of the bin
to normalize the measurement.  This method of constructing a histogram is
often used when the histogram is to be plotted with a logarithmic degree
scale, so that the widths of the bins will appear even.  Because the bins
get wider as we get out into the tail, the problems with statistics are
reduced, although they are still present to some extent as long as $p_k$
falls off faster than~$k^{-1}$, which it must if the distribution is to be
integrable.

An alternative way of presenting degree data is to make a plot of the
cumulative distribution function
\begin{equation}
P_k=\sum_{k'=k}^\infty p_{k'},
\label{cumulative}
\end{equation}
which is the probability that the degree is greater than or equal to~$k$.
Such a plot has the advantage that all the original data are represented.
When we make a conventional histogram by binning, any differences between
the values of data points that fall in the same bin are lost.  The
cumulative distribution function does not suffer from this problem.  The
cumulative distribution also reduces the noise in the tail.  On the
downside, the plot doesn't give a direct visualization of the degree
distribution itself, and adjacent points on the plot are not statistically
independent, making correct fits to the data tricky.

\begin{figure*}
\resizebox{15cm}{!}{\includegraphics{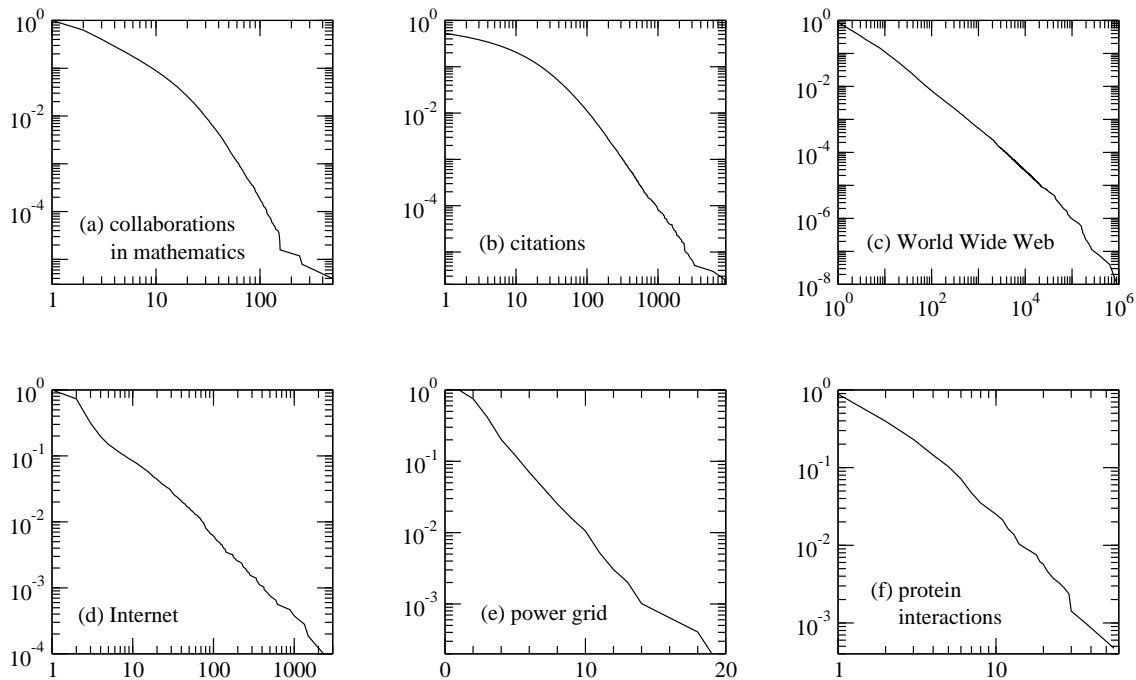}}
\caption{Cumulative degree distributions for six different networks.  The
horizontal axis for each panel is vertex degree~$k$ (or in-degree for the
citation and Web networks, which are directed) and the vertical axis is the
cumulative probability distribution of degrees, i.e.,~the fraction of
vertices that have degree greater than or equal to~$k$.  The networks shown
are: (a)~the collaboration network of mathematicians~\cite{GI95};
(b)~citations between 1981 and 1997 to all papers cataloged by the
Institute for Scientific Information~\cite{Redner98}; (c)~a 300 million
vertex subset of the World Wide Web, \textit{circa} 1999~\cite{Broder00};
(d)~the Internet at the level of autonomous systems, April
1999~\cite{Chen02}; (e)~the power grid of the western United
States~\cite{WS98}; (f)~the interaction network of proteins in the
metabolism of the yeast
\textit{S. Cerevisiae}~\cite{Jeong01}.  Of these networks, three of them,
(c), (d) and (f), appear to have power-law degree distributions, as
indicated by their approximately straight-line forms on the doubly
logarithmic scales, and one (b) has a power-law tail but deviates markedly
from power-law behavior for small degree.  Network (e) has an exponential
degree distribution (note the log-linear scales used in this panel) and
network~(a) appears to have a truncated power-law degree distribution of
some type, or possibly two separate power-law regimes with different
exponents.}
\label{degree}
\end{figure*}

In Fig.~\ref{degree} we show cumulative distributions of degree for a
number of the networks described in Sec.~\ref{data}.  As the figure shows,
the distributions are indeed all right-skewed.  Many of them follow power
laws in their tails: $p_k\sim k^{-\alpha}$ for some constant
exponent~$\alpha$.  Note that such power-law distributions show up as power
laws in the cumulative distributions also, but with exponent $\alpha-1$
rather than~$\alpha$:
\begin{equation}
P_k \sim \sum_{k'=k}^\infty {k'}^{-\alpha}
    \sim k^{-(\alpha-1)}.
\end{equation}
Some of the other distributions have exponential tails:
$p_k\sim\e^{-k/\kappa}$.  These also give exponentials in the cumulative
distribution, but with the \emph{same} exponent:
\begin{equation}
P_k = \sum_{k'=k}^\infty p_k
 \sim \sum_{k'=k}^\infty \e^{-k'/\kappa}
 \sim \e^{-k/\kappa}.
\end{equation}
This makes power-law and exponential distributions particularly easy to
spot experimentally, by plotting the corresponding cumulative distributions
on logarithmic scales (for power laws) or semi-logarithmic scales (for
exponentials).

For other types of networks degree distributions can be more complicated.
For bipartite graphs, for instance (Sec.~\ref{types}), there are two degree
distributions, one for each type of vertex.  For directed graphs each
vertex has both an in-degree and an out-degree, and the degree distribution
therefore becomes a function~$p_{jk}$ of two variables, representing the
fraction of vertices that simultaneously have in-degree~$j$ and
out-degree~$k$.  In empirical studies of directed graphs like the Web,
researchers have usually given only the individual distributions of in- and
out-degree~\cite{AJB99,BAJ00,Broder00}, i.e.,~the distributions derived by
summing $p_{jk}$ over one or other of its indices.  This however discards
much of the information present in the joint distribution.  It has been
found that in- and out-degrees are quite strongly correlated in some
networks~\cite{NFB02}, which suggests that there is more to be gleaned from
the joint distribution than is normally appreciated.

\subsubsection{Scale-free networks}
\label{sfnet}
Networks with power-law degree distributions have been the focus of a great
deal of attention in the literature~\cite{Strogatz01,AB02,DM02}.  They are
sometimes referred to as \defn{scale-free networks}~\cite{BA99b}, although
it is only their degree distributions that are scale-free;\footnote{The
term ``scale-free'' refers to any functional form~$f(x)$ that remains
unchanged to within a multiplicative factor under a rescaling of the
independent variable~$x$.  In effect this means power-law forms, since
these are the only solutions to $f(ax)=b f(x)$, and hence ``power-law'' and
``scale-free'' are, for our purposes, synonymous.} one can and usually does
have scales present in other network properties.  The earliest published
example of a scale-free network is probably Price's network of citations
between scientific papers~\cite{Price65} (see Sec.~\ref{infonets}).  He
quoted a value of $\alpha=2.5$ to~$3$ for the exponent of his network.  In
a later paper he quoted a more accurate figure of
$\alpha=3.04$~\cite{Price76}.  He also found a power-law distribution for
the out-degree of the network (number of bibliography entries in each
paper), although later work has called this into question~\cite{Vazquez01}.
More recently, power-law degree distributions have been observed in a host
of other networks, including notably other citation
networks~\cite{Seglen92,Redner98}, the World Wide
Web~\cite{AJB99,BAJ00,Broder00}, the Internet~\cite{FFF99,Chen02,VPV02},
metabolic networks~\cite{Jeong00,Jeong01}, telephone call
graphs~\cite{ACL00,ACL02}, and the network of human sexual
contacts~\cite{Liljeros01,JH03}.  The degree distributions of some of these
networks are shown in Fig.~\ref{degree}.

Other common functional forms for the degree distribution are exponentials,
such as those seen in the power grid~\cite{ASBS00} and railway
networks~\cite{Sen03}, and power laws with exponential cutoffs, such as
those seen in the network of movie actors~\cite{ASBS00} and some
collaboration networks~\cite{Newman01a}.  Note also that while a particular
form may be seen in the degree distribution for the network as a whole,
specific subnetworks within the network can have other forms.  The
World Wide Web, for instance, shows a power-law degree distribution overall
but unimodal distributions within domains~\cite{Pennock02}.

\subsubsection{Maximum degree}
\label{highest}
The maximum degree~$k_\mathrm{max}$ of a vertex in a network will in
general depend on the size of the network.  For some calculations on
networks the value of this maximum degree matters (see, for example,
Sec.~\ref{guided}).  In work on scale-free networks,
Aiello~\etal~\cite{ACL00} assumed that the maximum degree was approximately
the value above which there is less than one vertex of that degree in the
graph on average, i.e.,~the point where $np_k=1$.  This means, for
instance, that $k_{\rm max}\sim n^{1/\alpha}$ for the power-law degree
distribution $p_k\sim k^{-\alpha}$.  This assumption however can give
misleading results; in many cases there will be vertices in the network
with significantly higher degree than this, as discussed by
Adamic~\etal~\cite{ALPH01}.

Given a particular degree distribution (and assuming all degrees to be
sampled independently from it, which may not be true for networks in the
real world), the probability of there being exactly $m$ vertices of
degree~$k$ and no vertices of higher degree is ${n\choose m} p_k^m
(1-P_k)^{n-m}$, where $P_k$ is the cumulative probability distribution,
Eq.~\eref{cumulative}.  Hence the probability~$h_k$ that the highest degree
on the graph is~$k$ is
\begin{eqnarray}
h_k &=& \sum_{m=1}^n {n\choose m} p_k^m (1-P_k)^{n-m}\nonumber\\
    &=& (p_k+1-P_k)^n - (1-P_k)^n,
\end{eqnarray}
and the expected value of the highest degree is $k_\mathrm{max} = \sum_k k
h_k$.

For both small and large values of $k$, $h_k$~tends to zero, and the sum
over $k$ is dominated by the terms close to the maximum.  Thus, in most
cases, a good approximation to the expected value of the maximum degree is
given by the modal value.  Differentiating and observing that $\d P_k/\d
k=p_k$, we find that the maximum of~$h_k$ occurs when
\begin{equation}
\biggl( {\d p_k\over\d k} - p_k \biggr) (p_k+1-P_k)^{n-1}
  + p_k (1-P_k)^{n-1} = 0,
\end{equation}
or $k_\mathrm{max}$ is a solution of
\begin{equation}
{\d p_k\over\d k} \simeq -np_k^2,
\end{equation}
where we have made the (fairly safe) assumption that $p_k$ is sufficiently
small for $k\gtrsim k_\mathrm{max}$ that $np_k\ll1$ and $P_k\ll1$.

For example, if $p_k\sim k^{-\alpha}$ in its tail, then we find that
\begin{equation}
k_\mathrm{max} \sim n^{1/(\alpha-1)}.
\label{kmax}
\end{equation}
As shown by Cohen~\etal~\cite{CEBH00}, a simple rule of thumb that leads to
the same result is that the maximum degree is roughly the value of $k$ that
solves $nP_k=1$.  Note however that, as shown by Dorogovtsev and
Samukhin~\cite{DS03}, the fluctuations in the tail of the degree
distribution are very large for the power-law case.

Dorogovtsev~\etal~\cite{DMS01c} have also shown that Eq.~\eref{kmax} holds
for networks generated using the ``preferential attachment'' procedure of
Barab\'asi and Albert~\cite{BA99b} described in Sec.~\ref{bamodel}, and a
detailed numerical study of this case has been carried out by
Moreira~\etal~\cite{MAA03}.

\subsection{Network resilience}
\label{resilience}
Related to degree distributions is the property of resilience of networks
to the removal of their vertices, which has been the subject of a good deal
of attention in the literature.  Most of the networks we have been
considering rely for their function on their connectivity, i.e.,~the
existence of paths leading between pairs of vertices.  If vertices are
removed from a network, the typical length of these paths will increase,
and ultimately vertex pairs will become disconnected and communication
between them through the network will become impossible.  Networks vary in
their level of resilience to such vertex removal.

There are also a variety of different ways in which vertices can be removed
and different networks show varying degrees of resilience to these also.
For example, one could remove vertices at random from a network, or one
could target some specific class of vertices, such as those with the
highest degrees.  Network resilience is of particular importance in
epidemiology, where ``removal'' of vertices in a contact network might
correspond for example to vaccination of individuals against a disease.
Because vaccination not only prevents the vaccinated individuals from
catching the disease but may also destroy paths between other individuals
by which the disease might have spread, it can have a wider reaching effect
than one might at first think, and careful consideration of the efficacy of
different vaccination strategies could lead to substantial advantages for
public health.

\begin{figure}
\resizebox{6cm}{!}{\includegraphics{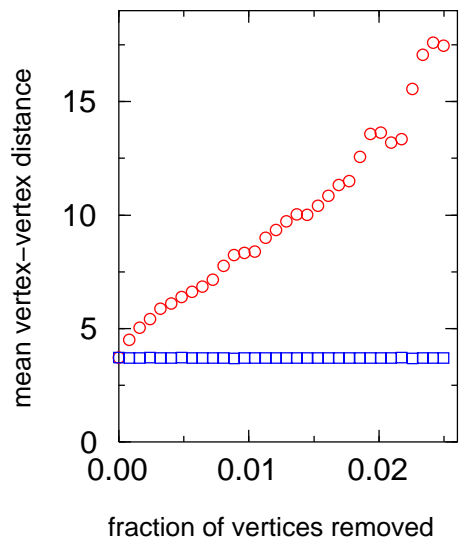}}
\caption{Mean vertex--vertex distance on a graph representation
of the Internet at the autonomous system level, as vertices are removed one
by one.  If vertices are removed in random order (squares), distance
increases only very slightly, but if they are removed in order of their
degrees, starting with the highest degree vertices (circles), then distance
increases sharply.  After Albert~\etal~\cite{AJB00}.}
\label{attack}
\end{figure}

Recent interest in network resilience has been sparked by the work of
Albert~\etal~\cite{AJB00}, who studied the effect of vertex deletion in two
example networks, a 6000-vertex network representing the topology of the
Internet at the level of autonomous systems (see Sec.~\ref{techno}), and a
$326\,000$-page subset of the World Wide Web.  Both of the Internet and the
Web have been observed to have degree distributions that are approximately
power-law in form~\cite{FFF99,Chen02,AJB99,Broder00,VPV02}
(Sec.~\ref{sfnet}).  The authors measured average vertex--vertex distances
as a function of number of vertices removed, both for random removal and
for progressive removal of the vertices with the highest
degrees.\footnote{In removing the vertices with the highest degrees,
Albert~\etal\ recalculated degrees following the removal of each vertex.
Most other authors who have studied this issue have adopted a slightly
different strategy of removing vertices in order of their \emph{initial}
degree in the network before any removal.}  In Fig.~\ref{attack} we show
their results for the Internet.  They found for both networks that distance
was almost entirely unaffected by random vertex removal, i.e.,~the networks
studied were highly resilient to this type of removal.  This is intuitively
reasonable, since most of the vertices in these networks have low degree
and therefore lie on few paths between others; thus their removal rarely
affects communications substantially.  On the other hand, when removal is
targeted at the highest degree vertices, it is found to have devastating
effect.  Mean vertex--vertex distance increases very sharply with the
fraction of vertices removed, and typically only a few percent of vertices
need be removed before essentially all communication through the network is
destroyed.  Albert~\etal\ expressed their results in terms of failure or
sabotage of network nodes.  The Internet (and the Web) they suggest, is
highly resilient against the random failure of vertices in the network, but
highly vulnerable to deliberate attack on its highest-degree vertices.

Similar results to those of Albert~\etal\ were found independently by
Broder~\etal~\cite{Broder00} for a much larger subset of the Web graph.
Interestingly, however, Broder~\etal\ gave an entirely opposite
interpretation of their results.  They found that in order to destroy
connectivity in the Web one has to remove all vertices with degree greater
than five, which seems like a drastic attack on the network, given that
some vertices have degrees in the thousands.  They thus concluded that the
network was very \emph{resilient} against targeted attack.  In fact however
there is not such a conflict between these results as at first appears.
Because of the highly skewed degree distribution of the Web, the fraction
of vertices with degree greater than five is only a small fraction of all
vertices.

Following these studies, many authors have looked into the question of
resilience for other networks.  In general the picture seems to be
consistent with that seen in the Internet and Web.  Most networks are
robust against random vertex removal but considerably less robust to
targeted removal of the highest-degree vertices.
Jeong~\etal~\cite{Jeong01} have looked at metabolic networks,
Dunne~\etal~\cite{DWM02a,DWM02b} at food webs, Newman~\etal~\cite{NFB02} at
email networks, and a variety of authors at resilience of model
networks~\cite{AJB00,CEBH00,CEBH01,CNSW00,Holme02a}, which we discuss in
more detail in later sections of the review.  A particularly thorough study
of the resilience of both real-world and model networks has been conducted
by Holme~\etal~\cite{Holme02a}, who looked not only at vertex removal but
also at removal of edges, and considered some additional strategies for
selecting vertices based on so-called ``betweenness'' (see
Secs.~\ref{communities} and~\ref{otherprop}).

\subsection{Mixing patterns}
\label{mixing}
Delving a little deeper into the statistics of network structure, one can
ask about which vertices pair up with which others.  In most kinds of
networks there are at least a few different types of vertices, and the
probabilities of connection between vertices often depends on types.  For
example, in a food web representing which species eat which in an ecosystem
(Sec.~\ref{bionets}) one sees vertices representing plants, herbivores, and
carnivores.  Many edges link the plants and herbivores, and many more the
herbivores and carnivores.  But there are few edges linking herbivores to
other herbivores, or carnivores to plants.  For the Internet,
Maslov~\etal~\cite{MSZ03} have proposed that the structure of the network
reflects the existence of three broad categories of nodes: high-level
connectivity providers who run the Internet backbone and trunk lines,
consumers who are end users of Internet service, and ISPs who join the two.
Again there are many links between end users and ISPs, and many between
ISPs and backbone operators, but few between ISPs and other ISPs, or
between backbone operators and end users.

In social networks this kind of selective linking is called
\defn{assortative mixing} or \defn{homophily} and has been widely studied,
as it has also in epidemiology.  (The term ``assortative matching'' is also
seen in the ecology literature, particularly in reference to mate choice
among animals.)  A classic example of assortative mixing in social networks
is mixing by race.  Table~\ref{sanfran} for example reproduces results from
a study of $1\,958$ couples in the city of San Francisco, California.
Among other things, the study recorded the race (self-identified) of study
participants in each couple.  As the table shows, participants appear to
draw their partners preferentially from those of their own race, and this
is believed to be a common phenomenon in many social networks: we tend to
associate preferentially with people who are similar to ourselves in some
way.

\begin{table}[t]
\begin{tabular}{l|r|rrrr}
\multicolumn{2}{c|}{} & \multicolumn{4}{c}{women}               \\
\cline{3-6}
\multicolumn{2}{c|}{} & black & hispanic & white & other \\
\hline
\begin{rotate}{90}
\hbox{\hspace{-25pt}men}
\end{rotate}
& black    & 506   & 32    & 69    & 26    \\
& hispanic & 23    & 308   & 114   & 38    \\
& white    & 26    & 46    & 599   & 68    \\
& other    & 10    & 14    & 47    & 32    \\
\end{tabular}
\caption{Couples in the study of Catania~\etal~\cite{CCKF92} tabulated by
race of either partner.  After Morris~\cite{Morris95}.}
\label{sanfran}
\end{table}

Assortative mixing can be quantified by an ``assortativity coefficient,''
which can be defined in a couple of different ways.  Let $E_{ij}$ be the
number of edges in a network that connect vertices of types~$i$ and~$j$,
with $i,j=1\ldots N$, and let $\vE$ be the matrix with elements~$E_{ij}$,
as depicted in Table~\ref{sanfran}.  We define a normalized mixing matrix
by
\begin{equation}
\ve = {\vE\over\norm{\vE}},
\end{equation}
where $\norm{\vx}$ means the sum of all the elements of the matrix~$\vx$.
The elements $e_{ij}$ measure the \emph{fraction} of edges that fall
between vertices of types $i$ and~$j$.  One can also ask about the
conditional probability $P(j|i)$ that my network neighbor is of type~$j$
given that I am of type~$i$, which is given by $P(j|i)=e_{ij}/\sum_j
e_{ij}$.  These quantities satisfy the normalization conditions
\begin{equation}
\sum_{ij} e_{ij} = 1,\qquad \sum_j P(j|i) = 1.
\end{equation}

Gupta~\etal~\cite{GAM89} have suggested that assortative mixing be
quantified by the coefficient
\begin{equation}
Q = {\sum_i P(i|i) - 1\over N-1}.
\end{equation}
This quantity has the desirable properties that it is~1 for a perfectly
assortative network (every edge falls between vertices of the same type),
and 0 for randomly mixed networks, and it has been quite widely used in the
literature.  But it suffers from two shortcomings~\cite{Newman03c}: (1)~for
an asymmetric matrix like the one in Table~\ref{sanfran}, $Q$~has two
different values, depending on whether we put the men or the women along
the horizontal axis, and it is unclear which of these two values is the
``correct'' one for the network; (2)~the measure weights each vertex type
equally, regardless of how many vertices there are of each type, which can
give rise to misleading figures for~$Q$ in cases where community size is
heterogeneous, as it often is.

An alternative assortativity coefficient that remedies these problems is
defined by~\cite{Newman03c}
\begin{equation}
r = {\Tr\ve - \norm{\ve^2}\over1 - \norm{\ve^2}}.
\end{equation}
This quantity is also~0 in a randomly mixed network and~1 in a perfectly
assortative one.  But its value is not altered by transposition of the
matrix and it weights vertices equally rather than communities, so that
small communities make an appropriately small contribution to~$r$.  For the
data of Table~\ref{sanfran} we find $r=0.621$.

Another type of assortative mixing is mixing by scalar characteristics such
as age or income.  Again it is usually found that people prefer to
associate with others of similar age and income to themselves, although of
course age and income, like race, may be proxies for other driving forces,
such as cultural differences.  Garfinkel~\etal~\cite{GGM02} and
Newman~\cite{Newman03c}, for example, have analyzed data for unmarried and
married couples respectively to show that there is strong correlation
between the ages of partners.  Mixing by scalar characteristics can be
quantified by calculating a correlation coefficient for the characteristic
in question.

In theory assortative mixing according to vector characteristics should
also be possible.  For example, geographic location probably affects
individuals' propensity to become acquainted.  Location could be viewed as
a two-vector, with the probability of connection between pairs of
individuals being assortative on the values of these vectors.

\subsection{Degree correlations}
\label{degcorr}
A special case of assortative mixing according to a scalar vertex property
is mixing according to vertex degree, also commonly referred to simply as
degree correlation.  Do the high-degree vertices in a network associate
preferentially with other high-degree vertices?  Or do they prefer to
attach to low-degree ones?  Both situations are seen in some networks, as
it turns out.  The case of assortative mixing by degree is of particular
interest because, since degree is itself a property of the graph topology,
degree correlations can give rise to some interesting network structure
effects.

Several different ways of quantifying degree correlations have been
proposed.  Maslov~\etal~\cite{MS02a,MSZ03} have simply plotted the
two-dimensional histogram of the degrees of vertices at either ends of an
edge.  They have shown results for protein interaction networks and the
Internet.  A more compact representation of the situation is that proposed
by Pastor-Satorras~\etal~\cite{PVV01,VPV02}, who in studies of the Internet
calculated the mean degree of the network neighbors of a vertex as a
function of the degree~$k$ of that vertex.  This gives a one-parameter
curve which increases with~$k$ if the network is assortatively mixed.  For
the Internet in fact it is found to decrease with~$k$, a situation we call
disassortativity.  Newman~\cite{Newman02f,Newman03c} reduced the
measurement still further to a single number by calculating the Pearson
correlation coefficient of the degrees at either ends of an edge.  This
gives a single number that should be positive for assortatively mixed
networks and negative for disassortative ones.  In Table~\ref{summary} we
show results for a number of different networks.  An interesting
observation is that essentially all social networks measured appear to be
assortative, but other types of networks (information networks,
technological networks, biological networks) appear to be disassortative.
It is not clear what the explanation for this result is, or even if there
is any one single explanation.  (Probably there is not.)

\begin{figure*}
\resizebox{15cm}{!}{\includegraphics{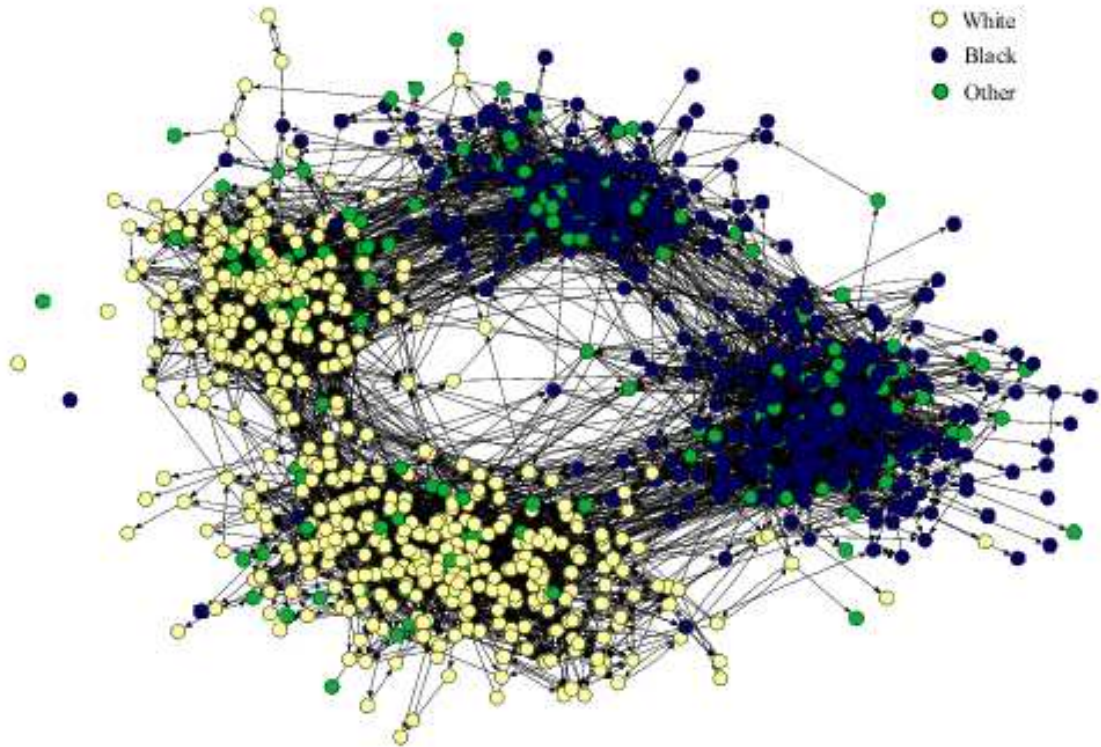}}
\caption{Friendship network of children in a US school.  Friendships are
determined by asking the participants, and hence are directed, since A may
say that B is their friend but not \textit{vice versa}.  Vertices are color
coded according to race, as marked, and the split from left to right in the
figure is clearly primarily along lines of race.  The split from top to
bottom is between middle school and high school, i.e.,~between younger and
older children.  Picture courtesy of James Moody.}
\label{school}
\end{figure*}

\subsection{Community structure}
\label{communities}
It is widely assumed~\cite{WF94,Scott00} that most social networks show
``community structure,'' i.e.,~groups of vertices that have a high density
of edges within them, with a lower density of edges between groups.  It is
a matter of common experience that people do divide into groups along lines
of interest, occupation, age, and so forth, and the phenomenon of
assortativity discussed in Sec.~\ref{mixing} certainly suggests that this
might be the case.  (It is possible for a network to have assortative
mixing but no community structure.  This can occur, for example, when there
is assortative mixing by age or other scalar quantities.  Networks with
this type of structure are sometimes said to be ``stratified.'')

In Fig.~\ref{school} we show a visualization of the friendship network of
children in a US school taken from a study by
Moody~\cite{Moody01}.\footnote{This image does not appear in the paper
cited, but it and a number of other images from the same study can be found
on the Web at \texttt{http://www.sociology.ohio-state.edu/jwm/}.}  The
figure was created using a ``spring embedding'' algorithm, in which linear
springs are placed between vertices and the system is relaxed using a
first-order energy minimization.  We have no special reason to suppose that
this very simple algorithm would reveal anything particularly useful about
the network, but the network appears to have strong enough community
structure that in fact the communities appear clearly in the figure.
Moreover, when Moody colors the vertices according to the race of the
individuals they represent, as shown in the figure, it becomes immediately
clear that one of the principal divisions in the network is by individuals'
race, and this is presumably what is driving the formation of communities
in this case.  (The other principal division visible in the figure is
between middle school and high school, which are age divisions in the
American education system.)

It would be of some interest, and indeed practical importance, were we to
find that other types of networks, such as those those listed in
Table~\ref{summary}, show similar group structure also.  One might well
imagine for example that citation networks would divide into groups
representing particular areas of research interest, and a good deal of
energy has been invested in studies of this phenomenon~\cite{Crane72,ER90}.
Similarly communities in the World Wide Web might reflect the subject
matter of pages, communities in metabolic, neural, or software networks
might reflect functional units, communities in food webs might reflect
subsystems within ecosystems, and so on.

\begin{figure}
\resizebox{5.4cm}{!}{\includegraphics{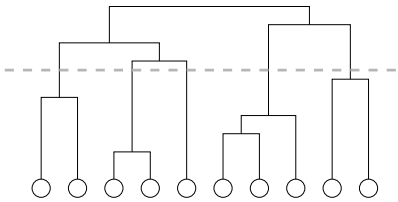}}
\caption{An example of a dendrogram showing the hierarchical clustering of
ten vertices.  A horizontal cut through the dendrogram, such as that
denoted by the dotted line, splits the vertices into a set of communities,
five in this case.}
\label{dendrogram}
\end{figure}

The traditional method for extracting community structure from a network is
\defn{cluster analysis}~\cite{Everitt74}, sometimes also called
\defn{hierarchical clustering}.\footnote{Not to be confused with the
entirely different use of the word clustering introduced in
Sec.~\ref{transitivity}.}  In this method, one assigns a ``connection
strength'' to vertex pairs in the network of interest.  In general each of
the $\half n(n-1)$ possible pairs in a network of $n$ vertices is assigned
such a strength, not just those that are connected by an edge, although
there are versions of the method where not all pairs are assigned a
strength; in that case one can assume the remaining pairs to have a
connection strength of zero.  Then, starting with $n$ vertices with no
edges between any of them, one adds edges in order of decreasing
vertex--vertex connection strength.  One can pause at any point in this
process and examine the component structure formed by the edges added so
far; these components are taken to be the communities (or ``clusters'') at
that stage in the process.  When all edges have been added, all vertices
are connected to all others, and there is only one community.  The entire
process can be represented by a tree or \defn{dendrogram} of union
operations between vertex sets in which the communities at any level
correspond to a horizontal cut through the tree---see
Fig.~\ref{dendrogram}.\footnote{For some reason such trees are
conventionally depicted with their ``root'' at the top and their ``leaves''
at the bottom, which is not the natural order of things for most trees.}

Clustering is possible according to many different definitions of the
connection strength.  Reasonable choices include various weighted
vertex--vertex distance measures, the sizes of minimum cut-sets
(i.e.,~maximum flow)~\cite{AMO93}, and weighted path counts between
vertices.  Recently a number of authors have had success with methods based
on ``edge betweenness,'' which is the count of how many geodesic paths
between vertices run along each edge in the
network~\cite{GN02,HHJ02,WH03,Guimera03}.  Results appear to show that, for
social and biological networks at least, community structure is a common
network property, although some food webs are found not to break up into
communities in any simple way.  (Food webs may be different from other
networks in that they appear to be dense: mean vertex degree increases
roughly linearly with network size, rather than remaining constant as it
does in most networks~\cite{Martinez92,DWM02a}.  The same may be true of
metabolic networks also [P.~Holme, personal communication].)

Network clustering should not be confused with the technique of data
clustering, which is a way of detecting groupings of data-points in
high-dimensional data spaces~\cite{JMF99}.  The two problems do have some
common features however, and algorithms for one can be adapted for the
other, and \textit{vice versa}.  For example, high-dimensional data can be
converted into a network by placing edges between closely spaced data
points, and then network clustering algorithms can be applied to the
result.  On balance, however, one normally finds that algorithms specially
devised for data clustering work better than such borrowed methods, and the
same is true in reverse.

In the social networks literature, network clustering has been discussed to
a great extent in the context of so-called \defn{block
models},~\cite{WBB76,BBA75} which are essentially just divisions of
networks into communities or blocks according to one criterion or another.
Sociologists have concentrated particularly on \defn{structural
equivalence}.  Two vertices in a network are said to be structurally
equivalent if they have all of the same neighbors.  Exact structural
equivalence is rare, but approximate equivalence can be used as the basis
for a hierarchical clustering method such as that described above.

Another slightly different question about community structure, but related
to the one discussed here, has been studied by Flake~\etal~\cite{FLGC02}:
if one is given an example vertex drawn from a known network, can one
identify the community to which it belongs?  Algorithmic methods for
answering this question would clearly be of some practical value for
searching networks such as the World Wide Web and citation networks.
Flake~\etal\ give what appears to be a very successful algorithm, at least
in the context of the Web, based on a maximum flow method.

\subsection{Network navigation}
\label{navigation}
Stanley Milgram's famous small-world experiment (Sec.~\ref{socnet}), in
which letters were passed from person to person in an attempt to get them
to a desired target individual, showed that there exist short paths through
social networks between apparently distant individuals.  However, there is
another conclusion that can be drawn from this experiment which Milgram
apparently failed to notice; it was pointed out in 2000 by
Kleinberg~\cite{Kleinberg00,Kleinberg00proc}.  Milgram's results
demonstrate that there exist short paths in the network, but they also
demonstrate that ordinary people are good at finding them.  This is, upon
reflection, perhaps an even more surprising result than the existence of
the paths in the first place.  The participants in Milgram's study had no
special knowledge of the network connecting them to the target person.
Most people know only who their friends are and perhaps a few of their
friends' friends.  Nonetheless it proved possible to get a message to a
distant target in only a small number of steps.  This indicates that there
is something quite special about the structure of the network.  On a random
graph for instance, as Kleinberg pointed out, short paths between vertices
exist but no one would be able to find them given only the kind of
information that people have in realistic situations.  If it were possible
to construct artificial networks that were easy to navigate in the same way
that social networks appear to be, it has been suggested they could be used
to build efficient database structures or better peer-to-peer computer
networks~\cite{ALPH01,ALH03,WDN02} (see Sec.~\ref{greedy}).

\subsection{Other network properties}
\label{otherprop}
In addition to the heavily studied network properties of the preceding
sections, a number of others have received some attention.  In some
networks the size of the largest component is an important quantity.  For
example, in a communication network like the Internet the size of the
largest component represents the largest fraction of the network within
which communication is possible and hence is a measure of the effectiveness
of the network at doing its
job~\cite{Broder00,CEBH00,CEBH01,CNSW00,NSW01,DMS01a}.  The size of the
largest component is often equated with the graph theoretical concept of
the ``giant component'' (see Sec.~\ref{poissonrg}), although technically
the two are only the same in the limit of large graph size.  The size of
the second-largest component in a network is also measured sometimes.  In
networks well above the density at which a giant component first forms, the
largest component is expected to be much larger than the second largest
(Sec.~\ref{poissonrg}).

Goh~\etal~\cite{Goh02} have made a statistical study of the distribution of
the ``betweenness centrality'' of vertices in networks.  The betweenness
centrality of a vertex~$i$ is the number of geodesic paths between other
vertices that run through~$i$~\cite{Freeman77,WF94,Scott00}.  Goh~\etal\
show that betweenness appears to follow a power law for many networks and
propose a classification of networks into two kinds based on the exponent
of this power law.  Betweenness centrality can also be viewed as a measure
of network resilience~\cite{Newman01c,Holme02a}---it tells us how many
geodesic paths will get longer when a vertex is removed from the network.
Latora and Marchiori~\cite{LM01b,LM03} have considered the harmonic mean
distance between a vertex and all others, which they call the
``efficiency'' of the vertex.  This, like betweenness centrality, can be
viewed as a measure of network resilience, indicating how much effect on
path length the removal of a vertex will have.  A number of authors have
looked at the eigenvalue spectra and eigenvectors of the graph Laplacian
(or equivalently the adjacency matrix) of a
network~\cite{BP01,Farkas02,ESMS03}, which tells us about diffusion or
vibration modes of the network, and about vertex
centrality~\cite{Bonacich72,Bonacich87} (see also the discussion of network
search strategies in Sec.~\ref{exhaustive}).

Milo~\etal~\cite{Milo02,SMMA02} have presented a novel analysis that picks
out recurrent motifs---small subgraphs---from complete networks.  They
apply their method to genetic regulatory networks, food webs, neural
networks and the World Wide Web, finding different motifs in each case.
They have also made suggestions about the possible function of these motifs
within the networks.  In regulatory networks, for instance, they identify
common subgraphs with particular switching functions in the system, such as
gates and other feed-forward logical operations.

\section{Random graphs}
\label{rg}
The remainder of this review is devoted to our primary topic of study, the
mathematics of model networks of various kinds.  Recent work has focused on
models of four general types, which we treat in four following sections.
In this section we look at random graph models, starting with the classic
Poisson random graph of Rapoport~\cite{SR51,Rapoport57} and Erd\H{o}s and
R\'enyi~\cite{ER59,ER60}, and concentrating particularly on the generalized
random graphs studied by Molloy and Reed~\cite{MR95,MR98} and others.  In
Sec.~\ref{markov} we look at the somewhat neglected but potentially very
useful Markov graphs and their more general forms, exponential random
graphs and p$^*$~models.  In Section~\ref{sw} we look at the ``small-world
model'' of Watts and Strogatz~\cite{WS98} and its generalizations.  Then in
Section~\ref{growing} we look at models of growing networks, particularly
the models of Price~\cite{Price76} and Barab\'asi and Albert~\cite{BA99b},
and generalizations.  Finally, in Section~\ref{onnets} we look at a number
of models of processes occurring on networks, such as search and navigation
processes, and network transmission and epidemiology.

The first serious attempt at constructing a model for large and
(apparently) random networks was the ``random net'' of Rapoport and
collaborators~\cite{SR51,Rapoport57}, which was independently rediscovered
a decade later by Erd\H{o}s and R\'enyi~\cite{ER59}, who studied it
exhaustively and rigorously, and who gave it the name ``random graph'' by
which it is most often known today.  Where necessary, we will here refer to
it as the ``Poisson random graph,'' to avoid confusion with other random
graph models.  It is also sometimes called the ``Bernoulli graph.''  As we
will see in this section, the random graph, while illuminating, is
inadequate to describe some important properties of real-world networks,
and so has been extended in a variety of ways.  In particular, the random
graph's Poisson degree distribution is quite unlike the highly skewed
distributions of Section~\ref{dd} and Fig.~\ref{degree}.  Extensions of the
model to allow for other degree distributions lead to the class of models
known as ``generalized random graphs,'' ``random graphs with arbitrary
degree distributions'' and the ``configuration model.''

We here look first at the Poisson random graph, and then at its
generalizations.  Our treatment of the Poisson case is brief.  A much more
thorough treatment can be found in the books by
Bollob\'as~\cite{Bollobas01} and Janson~\etal~\cite{JLR99} and the review
by Karo\'nski~\cite{Karonski82}.

\subsection{Poisson random graphs}
\label{poissonrg}
Solomonoff and Rapoport~\cite{SR51} and independently Erd\H{o}s and
R\'enyi~\cite{ER59} proposed the following extremely simple model of a
network.  Take some number $n$ of vertices and connect each pair (or not)
with probability~$p$ (or $1-p$).\footnote{Slight variations on the model
are possible depending one whether one allows self-edges or not
(i.e.,~edges that connect a vertex to itself), but this distinction makes a
negligible difference to the average behavior of the model in the limit of
large~$n$.}  This defines the model that Erd\H{o}s and R\'enyi
called~$G_{n,p}$.  In fact, technically, $G_{n,p}$~is the \emph{ensemble}
of all such graphs in which a graph having $m$ edges appears with
probability $p^m(1-p)^{M-m}$, where $M=\half n(n-1)$ is the maximum
possible number of edges.  Erd\H{o}s and R\'enyi also defined another,
related model, which they called $G_{n,m}$, which is the ensemble of all
graphs having $n$ vertices and exactly $m$ edges, each possible graph
appearing with equal probability.\footnote{Those familiar with statistical
mechanics will notice a similarity between these two models and the
so-called canonical and grand canonical ensembles.  In fact, the analogy is
exact, and one can define equivalents of the Helmholtz and Gibbs free
energies, which are generating functions for moments of graph properties
over the distribution of graphs and which are related by a Lagrange
transform with respect to the ``field''~$p$ and the ``order
parameter''~$m$.}  Here we will discuss~$G_{n,p}$, but most of the results
carry over to $G_{n,m}$ in a straightforward fashion.

Many properties of the random graph are exactly solvable in the limit of
large graph size, as was shown by Erd\H{o}s and R\'enyi in a series of
papers in the 1960s~\cite{ER59,ER60,ER61}.  Typically the limit of large
$n$ is taken holding the mean degree $z=p(n-1)$ constant, in which case the
model clearly has a Poisson degree distribution, since the presence or
absence of edges is independent, and hence the probability of a vertex
having degree~$k$ is
\begin{equation}
p_k = {n\choose k} p^k (1-p)^{n-k} \simeq {z^k \e^{-z}\over k!},
\label{poissonpk}
\end{equation}
with the last approximate equality becoming exact in the limit of large~$n$
and fixed~$k$.  This is the reason for the name ``Poisson random graph.''

The expected structure of the random graph varies with the value of~$p$.
The edges join vertices together to form components, i.e.,~(maximal)
subsets of vertices that are connected by paths through the network.  Both
Solomonoff and Rapoport and also Erd\H{o}s and R\'enyi demonstrated what is
for our purposes the most important property of the random graph, that it
possesses what we would now call a phase transition, from a low-density,
low-$p$ state in which there are few edges and all components are small,
having an exponential size distribution and finite mean size, to a
high-density, high-$p$ state in which an extensive (i.e.,~$\ord(n)$)
fraction of all vertices are joined together in a single \defn{giant
component}, the remainder of the vertices occupying smaller components with
again an exponential size distribution and finite mean size.

\begin{figure}
\resizebox{8cm}{!}{\includegraphics{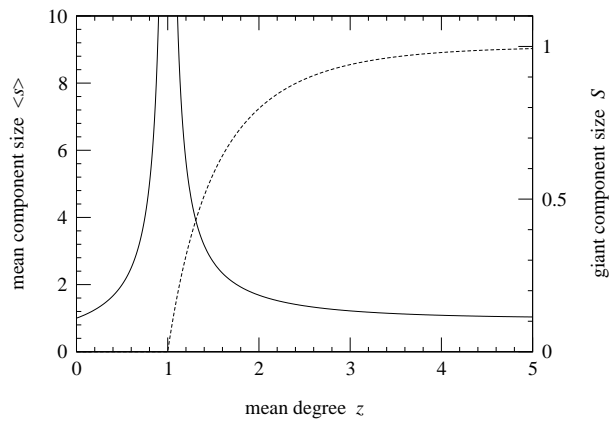}}
\caption{The mean component size (solid line), excluding the giant
component if there is one, and the giant component size (dotted line), for
the Poisson random graph, Eqs.~\eref{eqgc} and~\eref{eqavs}.}
\label{avsgc}
\end{figure}

We can calculate the expected size of the giant component from the
following simple heuristic argument.  Let $u$ be the fraction of vertices
on the graph that do not belong to the giant component, which is also the
probability that a vertex chosen uniformly at random from the graph is not
in the giant component.  The probability of a vertex not belonging to the
giant component is also equal to the probability that none of the vertex's
network neighbors belong to the giant component, which is just $u^k$ if
the vertex has degree~$k$.  Averaging this expression over the probability
distribution of~$k$, Eq.~\eref{poissonpk}, we then find the following
self-consistency relation for~$u$ in the limit of large graph size:
\begin{equation}
u = \sum_{k=0}^\infty p_k u^k = \e^{-z} \sum_{k=0}^\infty {(zu)^k\over k!}
  = \e^{z(u-1)}.
\end{equation}
The fraction~$S$ of the graph occupied by the giant component is $S=1-u$
and hence
\begin{equation}
S = 1 - \e^{-zS}.
\label{eqgc}
\end{equation}
By an argument only slightly more complex, which we give in the following
section, we can show that the mean size~$\av{s}$ of the component to which
a randomly chosen vertex belongs (for non-giant components) is
\begin{equation}
\av{s} = {1\over1-z+zS}.
\label{eqavs}
\end{equation}
The form of these two quantities is shown in Fig.~\ref{avsgc}.
Equation~\eref{eqgc} is transcendental and has no closed-form solution, but
it is easy to see that for $z<1$ its only non-negative solution is $S=0$,
while for $z>1$ there is also a non-zero solution, which is the size of the
giant component.  The phase transition occurs at $z=1$.  This is also the
point at which $\av{s}$ diverges, a behavior that will be recognized by
those familiar with the theory of phase transitions: $S$~plays the role of
the order parameter in this transition and $\av{s}$ the role of the
order-parameter fluctuations.  The corresponding critical exponents,
defined by $S\sim(z-1)^\beta$ and $\av{s}\sim|z-1|^{-\gamma}$, take the
values $\beta=1$ and $\gamma=1$.  Precisely at the transition, $z=1$, there
is a ``double jump''---the mean size of the largest component in the graph
goes as~$\ord(n^{2/3})$ for $z=1$, rather than $\ord(n)$ as it does above
the transition.  The components at the transition have a power-law size
distribution with exponent~$\tau=\frac52$ (or $\frac32$ if one asks about
the component to which a randomly chosen vertex belongs).  We look at these
results in more detail in the next section for the more general
``configuration model.''

The random graph reproduces well one of the principal features of
real-world networks discussed in Section~\ref{props}, namely the
small-world effect.  The mean number of neighbors a distance~$\ell$ away
from a vertex in a random graph is~$z^d$, and hence the value of $d$ needed
to encompass the entire network is $z^\ell\simeq n$.  Thus a typical
distance through the network is $\ell=\log n/\log z$, which satisfies the
definition of the small-world effect given in Sec.~\ref{sweffect}.
Rigorous results to this effect can be found in, for instance,
Refs.~\citen{Bollobas81} and~\citen{Bollobas01}.  However in almost all
other respects, the properties of the random graph do not match those of
networks in the real world.  It has a low clustering coefficient: the
probability of connection of two vertices is~$p$ regardless of whether they
have a common neighbor, and hence $C=p$, which tends to zero as $n^{-1}$
in the limit of large system size~\cite{WS98}.  The model also has a
Poisson degree distribution, quite unlike the distributions in
Fig.~\ref{degree}.  It has entirely random mixing patterns, no correlation
between degrees of adjacent vertices, no community structure, and
navigation is impossible on a random graph using local
algorithms~\cite{VPV02,Newman02f,Newman03c,Kleinberg00,Kleinberg00proc}.
In short it makes a good straw man but is rarely taken seriously in the
modeling of real systems.

Nonetheless, much of our basic intuition about the way networks behave
comes from the study of the random graph.  In particular, the presence of
the phase transition and the existence of a giant component are ideas that
underlie much of the work described in this review.  One often talks about
the giant component of a network, meaning in fact the largest component;
one looks at the sizes of smaller components, often finding them to be much
smaller than the largest component; one sees a giant component transition
in many of the more sophisticated models that we will look at in the coming
sections.  All of these are ideas that started with the Poisson random
graph.

\subsection{Generalized random graphs}
\label{genrg}
Random graphs can be extended in a variety of ways to make them more
realistic.  The property of real graphs that is simplest to incorporate is
the property of non-Poisson degree distributions, which leads us to the
so-called ``configuration model.''  Here we examine this model in detail;
in Sec.~\ref{dirrg}--\ref{dcrg} we describe further generalizations of the
random graph to add other features.

\subsubsection{The configuration model}
\label{confmodel}
Consider the model defined in the following way.  We specify a degree
distribution~$p_k$, such that $p_k$ is the fraction of vertices in the
network having degree~$k$.  We choose a \defn{degree sequence}, which is a
set of $n$~values of the degrees $k_i$ of vertices $i=1\ldots n$, from this
distribution.  We can think of this as giving each vertex~$i$ in our graph
$k_i$ ``stubs'' or ``spokes'' sticking out of it, which are the ends of
edges-to-be.  Then we choose pairs of stubs at random from the network and
connect them together.  It is straightforward to demonstrate~\cite{MR95}
that this process generates every possible topology of a graph with the
given degree sequence with equal probability.\footnote{Each possible graph
can be generated $\prod_i k_i!$ different ways, since the stubs around each
vertex are indistinguishable.  This factor is a constant for a given degree
sequence and hence each graph appears with equal probability.}  The
\defn{configuration model} is defined as the ensemble of graphs so
produced, with each having equal weight.\footnote{An alternative model has
recently been proposed by Chung and Lu~\cite{CL02a,CL02b}.  In their model,
each vertex~$i$ is assigned a \emph{desired} degree~$k_i$ chosen from the
distribution of interest, and then $m=\half\sum_i k_i$ edges are placed
between vertex pairs $(i,j)$ with probability proportional to $k_ik_j$.
This model has the disadvantage that the final degree sequence is not in
general precisely equal to the desired degree sequence, but it has some
significant calculational advantages that make the derivation of rigorous
results easier.  It is also a logical generalization of the Poisson random
graph, in a way that the configuration model is not.  Similar approaches
have also been taken by a number of other
authors~\cite{GKK01b,DMS03a,CCDM02}.}

Since the 1970s the configuration model has been studied by a number of
authors~\cite{BBK72,BC78,Bollobas80,Wormald81,Luczak92,MR95,MR98,NSW01,CL02a,CL02b}.
An exact condition is known in terms of $p_k$ for the model to possess a
giant component~\cite{MR95}, the expected size of that component is
known~\cite{MR98}, and the average size of non-giant components both above
and below the transition is known~\cite{NSW01}, along with a variety of
other properties, such as mean numbers of vertices a given distance away
from a central vertex and typical vertex--vertex distances~\cite{CL02b}.
Here we give a brief derivation of the main results using the generating
function formalism of Newman~\etal~\cite{NSW01}.  More rigorous treatments
of the same results can be found in Refs.~\citen{MR95,MR98,CL02a,CL02b}.

There are two important points to grasp about the configuration model.
First, $p_k$~is, in the limit of large graph size, the distribution of
degrees of vertices in our graph, but the degree of the vertex we reach by
following a randomly chosen edge on the graph is not given by~$p_k$.  Since
there are $k$ edges that arrive at a vertex of degree~$k$, we are $k$ times
as likely to arrive at that vertex as we are at some other vertex that has
degree~1.  Thus the degree distribution of the vertex at the end of a
randomly chosen edge is proportional to~$kp_k$.  In most case, we are
interested in how many edges there are leaving such a vertex other than the
one we arrived along, i.e.,~in the so-called \defn{excess degree}, which is
one less than the total degree of the vertex.  In the configuration model,
the excess degree has a distribution~$q_k$ given by
\begin{equation}
q_k = {(k+1)p_{k+1}\over\sum_k k p_k} = {(k+1)p_{k+1}\over z},
\label{defsqk}
\end{equation}
where $z=\sum_k kp_k$ is, as before, the mean degree in the network.

The second important point about the model is that the chance of finding a
loop in a small component of the graph goes as~$n^{-1}$.  The number of
vertices in a non-giant component is~$\ord(n^{-1})$, and hence the
probability of there being more than one path between any pair of vertices
is also $\ord(n^{-1})$ for suitably well-behaved degree
distributions.\footnote{Using arguments similar to those leading to
Eq.~\eref{configcc}, we can show that the density of loops in small
components will tend to zero as graph size becomes large provided that $z$
is finite and $\av{k^2}$ grows slower than~$n^{1/2}$.  See also
footnote~\ref{fn1}.}  This property is crucial to the solution of the
configuration model, but is definitely not true of most real-world networks
(see Sec.~\ref{transitivity}).  It is an open question how much the
predictions of the model would change if we were able to incorporate the
true loop structure of real networks into it.

We now proceed by defining two generating functions for the distributions
$p_k$ and~$q_k$:\footnote{Traditionally, the independent variable in a
generating function is denoted~$z$, but here we use $x$ to avoid confusion
with the mean degree~$z$.}
\begin{equation}
G_0(x) = \sum_{k=0}^\infty p_k x^k,\qquad
G_1(x) = \sum_{k=0}^\infty q_k x^k.
\label{defsg0g1}
\end{equation}
Note that, using Eq.~\eref{defsqk}, we also find that $G_1(x)=G_0'(x)/z$,
which is occasionally convenient.  Then the generating function $H_1(x)$
for the total number of vertices reachable by following an edge satisfies
the self-consistency condition
\begin{equation}
H_1(x) = xG_1(H_1(x)).
\label{defsh1}
\end{equation}
This equation says that when we follow an edge, we find at least one vertex
at the other end (the factor of~$x$ on the right-hand side), plus some
other clusters of vertices (each represented by~$H_1$) which are reachable
by following other edges attached to that one vertex.  The number of these
other clusters is distributed according to~$q_k$, hence the appearance
of~$G_1$.  A detailed derivation of Eq.~\eref{defsh1} is given in
Ref.~\citen{NSW01}.

The total number of vertices reachable from a randomly chosen vertex,
i.e.,~the size of the component to which such a vertex belongs, is
generated by $H_0(x)$ where
\begin{equation}
H_0(x) = xG_0(H_1(x)).
\label{defsh0}
\end{equation}
The solution of Eqs.~\eref{defsh1} and~\eref{defsh0} gives us the entire
distribution of component sizes.  Mean component size below the phase
transition in the region where there is no giant component is given by
\begin{equation}
\av{s} = H_0'(1) = 1 + {G_0'(1)\over1-G_1'(1)} = 1 + {z_1^2\over z_1-z_2},
\label{avs1}
\end{equation}
where $z_1=z=\av{k}=G_0'(1)$ is the average number of neighbors of a
vertex and $z_2=\av{k^2}-\av{k}=G_0'(1)G_1'(1)$ is the average number of
second neighbors.  We see that this diverges when $z_1=z_2$, or
equivalently when
\begin{equation}
G_1'(1) = 1.
\label{perctrans}
\end{equation}
This point marks the phase transition at which a giant component first
appears.  Substituting Eq.~\eref{defsg0g1} into Eq.~\eref{perctrans}, we
can also write the condition for the phase transition as
\begin{equation}
\sum_k k(k-2) p_k = 0.
\label{mrresult1}
\end{equation}
Indeed, since this sum increases monotonically as edges are added to the
graph, it follows that the giant component exists if and only if this sum
is positive.  A more rigorous derivation of this result has been given by
Molloy and Reed~\cite{MR95}.

Above the transition there is a giant component which occupies a
fraction~$S$ of the graph.  If we define $u$ to be the probability that a
randomly chosen edge leads to a vertex that is not a part of this giant
component, then, by an argument precisely analogous to the one preceding
Eq.~\eref{eqgc}, this probability must satisfy the self-consistency
condition $u=G_1(u)$ and $S$ is given by the solution of
\begin{equation}
S = 1 - G_0(u),\qquad u = G_1(u).
\label{mrresult2}
\end{equation}
An equivalent result is derived in Ref.~\citen{MR98}.  Normally the
equation for~$u$ cannot be solved in closed form, but once the generating
functions are known a solution can be found to any desired level of
accuracy by numerical iteration.  And once the value of $S$ is known, the
mean size of small components above the transition can be found by
subtracting off the giant component and applying the arguments that led to
Eq.~\eref{avs1} again, giving
\begin{equation}
\av{s} = 1 + {zu^2\over [1-S][1-G_1'(u)]}.
\label{avsrg}
\end{equation}
The result is a behavior qualitatively similar to that of the Poisson
random graph, with a continuous phase transition at a point defined by
Eq.~\eref{mrresult1}, characterized by the appearance of a giant component
and the divergence of the mean size of non-giant components.  The ratio
$z_2/z_1$ of the mean number of vertices two steps away to the number one
step away plays the role of the independent parameter governing the
transition, as the mean degree~$z$ does in the Poisson case, and one can
again define critical exponents for the transition, which take the same
values as for the Poisson case, $\beta=\gamma=1$, $\tau=\frac52$.

We can also find an expression for the clustering coefficient,
Eq.~\eref{defsc1}, of the configuration model.  A simple calculation shows
that~\cite{EMB02,Newman03b}
\begin{equation}
C = {1\over nz_1}\,\biggl[ {z_2\over z_1} \biggr]^2
  = {z\over n}\,\biggl[ {\av{k^2} - \av{k}\over\av{k}^2} \biggr]^2,
\label{configcc}
\end{equation}
which is the value $C=z/n$ for the Poisson random graph times an extra
factor that depends on $z$ and on the ratio $\av{k^2}/\av{k}^2$.  Thus $C$
will normally go to zero as $n^{-1}$ for large graphs, but for highly
skewed degree distributions, like some of those in Fig.~\ref{degree}, the
factor of $\av{k^2}/\av{k}^2$ can be quite large, so that $C$ is not
necessarily negligible for the graph sizes seen in empirical studies of
networks (see below).

\subsubsection{Example: power-law degree distribution}
As an example of the application of these results, consider the much
studied case of a network with a power-law degree distribution:
\begin{equation}
p_k = \biggl\lbrace\begin{array}{ll}
        0                               & \mbox{for $k=0$}    \\
        k^{-\alpha}/\zeta(\alpha)\qquad & \mbox{for $k\ge1$,}
      \end{array}
\label{powerlaw}
\end{equation}
for given constant~$\alpha$.  Here $\zeta(\alpha)$ is the Riemann
$\zeta$-function, which functions as a normalizing constant.  Substituting
into Eq.~\eref{defsg0g1} we find that
\begin{equation}
G_0(x) = {\Li_\alpha(x)\over\zeta(\alpha)},\qquad
G_1(x) = {\Li_{\alpha-1}(x)\over x\zeta(\alpha-1)},
\end{equation}
where $\Li_n(x)$ is the $n$th polylogarithm of~$x$.  Then
Eq.~\eref{perctrans} tells us that the phase transition occurs at the point
\begin{equation}
\zeta(\alpha-2) = 2\zeta(\alpha-1),
\end{equation}
which gives a critical value for~$\alpha$ of $\alpha_c=3.4788\ldots$\ \
Below this value a giant component exists; above it there is no giant
component.  For
$\alpha<\alpha_c$, the value of the variable~$u$ of Eq.~\eref{mrresult2} is
\begin{equation}
u = {\Li_{\alpha-1}(u)\over u\zeta(\alpha-1)},
\end{equation}
which gives $u=0$ below $\alpha=2$ and hence $S=1$.  Thus the giant
component occupies the entire graph below this point, or more strictly, a
randomly chosen vertex belongs to the giant component with probability~1 in
the limit of large graph size (but see the following discussion of the
clustering coefficient and footnote~\ref{fn1}).  In the range
$2<\alpha<\alpha_c$ we have a non-zero giant component whose size is given
by Eq.~\eref{mrresult2}.  All of these results were first shown by
Aiello~\etal~\cite{ACL00}.

We can also calculate the clustering coefficient for the power-law case
using Eq.~\eref{configcc}.  For $\alpha<3$ we have $\av{k^2}\sim
k_\mathrm{max}^{3-\alpha}$, where $k_\mathrm{max}$ is the maximum degree in
the network.  Using Eq.~\eref{kmax} for $k_\mathrm{max}$,
Eq.~\eref{configcc} then gives
\begin{equation}
C \sim n^{-\beta},\qquad \beta = {3\alpha-7\over\alpha-1}.
\end{equation}
This gives interesting behavior for the typical values $2\le\alpha\le3$ of
the exponent~$\alpha$ seen in most networks (see Table~\ref{summary}).  If
$\alpha>\frac73$, then $C$~tends to zero as the graph becomes large,
although it does so slower than the $C\sim n^{-1}$ of the Poisson random
graph provided $\alpha<3$.  At $\alpha=\frac73$, $C$~becomes constant (or
logarithmic) in the graph size, and for $\alpha<\frac73$ it actually
increases with increasing system size.\footnote{For sufficiently large
networks this implies that the clustering coefficient will be greater
than~1.  Physically this means that there will be more than one edge on
average between two vertices that share a common neighbor.}  Thus for
scale-free networks with smaller exponents~$\alpha$, we would not be
surprised to see quite substantial values of the clustering coefficient,
even if the pattern of connections were completely random.\footnote{This
means in fact that the generating function formalism breaks down for
$\alpha<\frac73$, invalidating some of the preceding results for the
power-law graph, since a fundamental assumption of the method is that there
are no short loops in the network.  Aiello~\etal~\cite{ACL00} get around
this problem by assuming that the degree distribution is cut off at
$k_\mathrm{max}\sim n^{1/\alpha}$ (see Sec.~\ref{highest}), which gives
$C\to0$ as $n\to\infty$ for all $\alpha>2$.  This however is somewhat
artificial; in real power-law networks there is normally no such
cutoff.\label{fn1}} This mechanism can, for instance, account for much of
the clustering seen in the World Wide Web~\cite{Newman03b}.

\subsubsection{Directed graphs}
\label{dirrg}
Substantially more sophisticated extensions of random graph models are
possible than the simple first example given above.  In this and the next
few sections we list some of the many possibilities, starting with directed
graphs.

Each vertex in a directed graph has both an in-degree~$j$ and an
out-degree~$k$, and the degree distribution therefore becomes, in general,
a double distribution~$p_{jk}$ over both degrees, as discussed in
Sec.~\ref{dd}.  The generating function for such a distribution is a
function of two variables
\begin{equation}
\cG(x,y) = \sum_{jk} p_{jk} x^j y^k.
\end{equation}
Each vertex~A also belongs to an \defn{in-component} and an
\defn{out-component}, which are, respectively, the set of vertices from
which A can be reached, and the set that can be reached from~A, by
following directed edges only in their forward direction.  There is also
the \defn{strongly connected component}, which is the set of vertices which
can both reach and be reached from~A.  In a random directed graph with a
given degree distribution, the giant in, out, and strongly connected
components can all be shown~\cite{NSW01} to form at a single transition
that takes place when
\begin{equation}
\sum_{jk} (2jk-j-k) p_{jk} = 0.
\end{equation}
Defining generating functions for in- and out-degree separately and their
excess-degree counterparts,
\begin{subequations}
\begin{eqnarray}
F_0(x) &=& \cG(x,1),\qquad
F_1(x) = 
{1\over z} {\partial\cG\over\partial y}\biggr|_{y=1},\\
G_0(y) &=& \cG(1,y),\qquad
G_1(y) = 
{1\over z} {\partial\cG\over\partial x}\biggr|_{x=1},
\end{eqnarray}
\end{subequations}
the sizes of the giant out-, in-, and strongly connected components are
given by~\cite{NSW01,DMS01a}
\begin{subequations}
\begin{eqnarray}
S_\mathrm{out} &=& 1 - F_0(u),\\
S_\mathrm{in}  &=& 1 - G_0(v),\\
S_\mathrm{str} &=& 1 - \cG(u,1) - \cG(1,v) + \cG(u,v),
\end{eqnarray}
\end{subequations}
where
\begin{equation}
u = F_1(u),\qquad v = G_1(v).
\end{equation}

\subsubsection{Bipartite graphs}
\label{bipartite}
Another class of generalizations of random graph models is to networks with
more than one type of vertex.  One of the simplest and most important
examples of such a network is the bipartite graph, which has two types of
vertices and edges running only between vertices of unlike types.  As
discussed in Sec.~\ref{types}, many social networks are bipartite, forming
what the sociologists call \defn{affiliation networks}, i.e.,~networks of
individuals joined by common membership of groups.  In such networks the
individuals and the groups are represented by the two vertex types with
edges between them representing group membership.  Networks of
CEOs~\cite{GM78,Galaskiewicz85}, boards of
directors~\cite{Mariolis75,DG97,DYB01}, and collaborations of
scientists~\cite{Newman01a} and film actors~\cite{WS98} are all examples of
affiliation networks.  Some other networks, such as the railway network
studied by Sen~\etal~\cite{Sen03}, are also bipartite, and bipartite graphs
have been used as the basis for models of sexual contact
networks~\cite{Ergun02,Newman02c}.

Bipartite graphs have two degree distributions, one each for the two types
of vertices.  Since the total number of edges attached to each type of
vertex is the same, the means $\mu$ and $\nu$ of the two distributions are
related to the numbers $M$ and $N$ of the types of vertices by
$\mu/M=\nu/N$.  One can define generating functions as before for the two
types of vertices, generating both the degree distribution and the excess
degree distribution, and denoted $f_0(x)$, $f_1(x)$, $g_0(x)$,
and~$g_1(x)$.  Then for example we can show that there is a phase
transition at which a giant component appears when $f_1'(1)g_1'(1)=1$.
Expressions for the expected size of giant and non-giant components can
easily be derived~\cite{NSW01}.

In many cases, graphs that are fundamentally bipartite are actually studied
by projecting them down onto one set of vertices or the other---so called
``one-mode'' projections.  For example, in the study of boards of directors
of companies, it has become standard to look at board ``interlocks.''  Two
boards are said to be interlocked if they share one or more common members,
and the graph of board interlocks is the one-mode projection of the full
board graph onto the vertices representing just the boards.  Many results
for these one-mode projections can also be extracted from the generating
function formalism.  To give one example, the projected networks do not
have a vanishing clustering coefficient~$C$ in the limit of large system
size, but instead can be shown to obey~\cite{NSW01}
\begin{equation}
{1\over C} - 1 =
{(\mu_2-\mu_1)(\nu_2-\nu_1)^2\over\mu_1\nu_1(2\nu_1-3\nu_2+\nu_3)},
\end{equation}
where $\mu_n$ and $\nu_n$ are the $n$th moments of the degree distributions
of the two vertex types.

More complicated types of network structure can be introduced by increasing
the number of different types of vertices beyond two, and by relaxing the
patterns of connection between vertex types.  For example, one can define a
model with the type of mixing matrix shown in Table~\ref{sanfran}, and
solve exactly for many of the standard
properties~\cite{Soderberg02,Newman03c}.

\subsubsection{Degree correlations}
\label{dcrg}
The type of degree correlations discussed in Sec.~\ref{degcorr} can also be
introduced into a random graph model~\cite{Newman02f}.  Extending the
formalism of Sec.~\ref{mixing}, we can define the probability
distribution~$e_{jk}$ to be the probability that a randomly chosen edge on
a graph connects vertices of excess degrees~$j$ and~$k$.  On an undirected
graph, this quantity is symmetric and satisfies
\begin{equation}
\sum_{jk} e_{jk} = 1,\qquad \sum_j e_{jk} = q_k.
\label{sumrules}
\end{equation}
Then the equivalent of Eq.~\eref{mrresult2} is
\begin{equation}
S = 1 - p_0 - \sum_{k=1}^\infty p_k u_{k-1}^k,\qquad
u_j = {\sum_k e_{jk} u_k^k\over\sum_k e_{jk}},
\label{gcmix}
\end{equation}
which must be solved self-consistently for the entire set $\set{u_k}$ of
quantities, one for each possible value of the excess degree.  The phase
transition at which a giant component appears takes place when
$\det(\vI-\vm)=0$, where $\vm$ is the matrix with elements $m_{jk} =
ke_{jk}/q_j$.  Matrix conditions of this form appear to be the typical
generalization of the criterion for the appearance of a giant component to
graphs with non-trivial mixing patterns~\cite{BPV02,VM03,Newman03c}.

Two other random graph models for degree correlations are also worth
mentioning.  One is the exponential random graph, which we study in more
detail in the following section.  This is a general model, which has been
applied to the particular problem of degree correlations by Berg and
L\"assig~\cite{BL02}.

A more specialized model that aims to explain the degree anticorrelations
seen in the Internet has been put forward by Maslov~\etal~\cite{MSZ03}.
They suggest that these anticorrelations are a simple result of the fact
that the Internet graph has at most one edge between any vertex pair.  Thus
they are led to consider the ensemble of all networks with a given degree
sequence and no double edges.  (The configuration model, by contrast,
allows double edges, and typical graphs usually have at least a few such
edges, which would disqualify them from membership in the ensemble of
Maslov~\etal)\ \ The ensemble with no duplicate edges, it turns out, is
hard to treat analytically~\cite{BC78,WZ98}, so Maslov~\etal\ instead
investigate it numerically, sampling the ensemble at random using a Monte
Carlo algorithm.  Their results appear to indicate that anticorrelations of
the type seen in the Internet do indeed arise as a finite-size effect
within this model.  (An alternative explanation of the same observations
has been put forward by Capocci~\etal~\cite{CCD03}, who use a modified
version of the model of Barab\'asi and Albert discussed in
Sec.~\ref{bamodel} to show that correlations can arise through network
growth processes.)

\section{Exponential random graphs and Markov graphs}
\label{markov}
The generalized random graph models of the previous sections effectively
address one of the principal shortcomings of early network models such as
the Poisson random graph, their unrealistic degree distribution.  However,
they have a serious shortcoming in that they fail to capture the common
phenomenon of transitivity described in Sec.~\ref{transitivity}.  The only
solvable random graph models that currently incorporate transitivity are
the bipartite and community-structured models of Sec.~\ref{bipartite} and
certain dual-graph models~\cite{RKM03}, and these cover rather special
cases.  For general networks we currently have no idea how to incorporate
transitivity into random graph models; the crucial property of independence
between the neighbors of a vertex is destroyed by the presence of short
loops in a network, invalidating all the techniques used to derive
solutions.  Some approximate methods may be useful in limited
ways~\cite{Newman03a} or perhaps some sort of perturbative analysis will
prove possible, but no progress has yet been made in this direction.

The main hope for progress in understanding the effects of transitivity,
which are certainly substantial, seems to lie in formulating a completely
different model or models, based around some alternative ensemble of graph
structures.  In this and the following section we describe two candidate
models, the Markov graphs of Holland and Leinhardt~\cite{HL81} and
Strauss~\cite{Strauss86,FS86} and the small-world model of Watts and
Strogatz~\cite{WS98}.

Strauss~\cite{Strauss86} considers \defn{exponential random graphs}, also
(in a slightly generalized form) called
\defn{p$^*$~models}~\cite{WP96,AWC99}, which are a class of graph ensembles
of fixed vertex number~$n$ defined by analogy with the Boltzmann ensemble
of statistical mechanics.\footnote{Indeed, in a development typical of this
highly interdisciplinary field, exponential random graphs have recently
been rediscovered, apparently quite independently, by
physicists~\cite{BCK01,BL02}.}  Let $\set{\epsilon_i}$ be a set of
measurable properties of a single graph, such as the number of edges, the
number of vertices of given degree, or the number of triangles of edges in
the graph.  These quantities play a role similar to energy in statistical
mechanics.  And let $\set{\beta_i}$ be a set of inverse-temperature or
field parameters, whose values we are free to choose.  We then define the
exponential random graph model to be the set of all possible graphs
(undirected in the simplest case) of $n$ vertices in which each graph~$G$
appears with probability
\begin{equation}
P(G) = {1\over Z}\,\exp\biggl(-{\sum_i \beta_i\epsilon_i}\biggr),
\label{defspg}
\end{equation}
where the partition function~$Z$ is
\begin{equation}
Z = \sum_G \exp\biggl(-{\sum_i \beta_i\epsilon_i}\biggr).
\end{equation}
For a sufficiently large set of temperature parameters $\set{\beta_i}$,
this definition can encompass any probability distribution over graphs that
we desire, although its practical application requires that the size of the
set be limited to a reasonably small number.

The calculation of the ensemble average of a graph observable $\epsilon_i$
is then found by taking a suitable derivative of the (reduced) free energy
$f=-\log Z$:
\begin{eqnarray}
\av{\epsilon_i} &=& \sum_G \epsilon_i(G) P(G)
                 =  {1\over Z} \sum_G \epsilon_i
                    \exp\biggl(-{\sum_i \beta_i\epsilon_i}\biggr)\nonumber\\
                &=& {\partial f\over\partial\beta_i}.
\end{eqnarray}
Thus, the free energy is a generating function for the expectation values
of the observables, in a manner familiar from statistical field theory.  If
a particular observable of interest does not appear in the exponent
of~\eref{defspg} (the ``graph Hamiltonian''), then one can simply introduce
it, with a corresponding temperature~$\beta_i$ which is set to zero.

While these preliminary developments appear elegant in principle, little
real progress has been made.  One would like to find the appropriate
Gaussian field theory for which $f$ can be expressed in closed form, and
then perturb around it to derive a diagrammatic expansion for the effects
of higher-order graph operators.  In fact, one can show that the Feynman
diagrams for the expansion \emph{are} the networks themselves.
Unfortunately, carrying through the entire field-theoretic program has not
proved easy.  The general approach one should take is
clear~\cite{BCK01,BL02}, but the mechanics appear intractable for most
cases of interest.  Some progress can be made by restricting ourselves to
\defn{Markov graphs}, which are the subset of graphs in which the presence
or absence of an edge between two vertices in the graph is correlated only
with those edges that share one of the same two vertices---edge pairs that
are disjoint (have no vertices in common) are uncorrelated.  Overall
however, the question of how to carry out calculations in exponential
random graph ensembles is an open one.

In the absence of analytic progress on the model, therefore, researchers
have turned to Monte Carlo simulation, a technique to which the exponential
random graph lends itself admirably.  Once the values of the parameters
$\set{\beta_i}$ are specified, the form~\eref{defspg} of $P(G)$ makes
generation of graphs correctly sampled from the ensemble straightforward
using a Metropolis--Hastings type Markov chain method.  One defines an
ergodic move-set in the space of graphs with given~$n$, and then repeatedly
generates moves from this set, accepting them with probability
\begin{equation}
p = \biggl\lbrace\begin{array}{ll}
        1          & \qquad\mbox{if $P(G')>P(G)$} \\
        P(G')/P(G) & \qquad\mbox{otherwise,}
      \end{array}
\end{equation}
and rejecting them with probability $1-p$, where $G'$ is the graph after
performance of the move.  Because of the particular form,
Eq.~\eref{defspg}, assumed for $P(G)$, this acceptance probability is
particularly simple to calculate:
\begin{equation}
{P(G')\over P(G)} = \exp\biggl(-{\sum_i
\beta_i[\epsilon_i'-\epsilon_i]}\biggr).
\end{equation}
This expression is independent of the value of the partition function and
its evaluation involves calculating only the differences
$\epsilon_i'-\epsilon_i$ of the energy-like graph properties~$\epsilon_i$,
which for local move-sets and local properties can often be accomplished in
time independent of graph size.  Suitable move-sets are: (a)~addition and
removal of edges between randomly chosen vertex pairs for the case of
variable edge numbers; (b)~movement of edges randomly from one place to
another for the case of fixed edge numbers but variable degree sequence;
(c)~edge swaps of the form $\set{(v_1,w_1), (v_2,w_2)}\to\set{(v_1,v_2),
(w_1,w_2)}$ for the case of fixed degree sequence, where $(v_1,w_1)$
denotes an edge from vertex $v_1$ to vertex $w_1$.  Monte Carlo algorithms
of this type are straightforward to implement and appear to converge
quickly allowing us to study quite large graphs.

There is however, one unfortunate pathology of the exponential random graph
that plagues numerical work, and particularly affects Markov graphs as they
are used to model transitivity.  If, for example, we include a term in the
graph Hamiltonian that is linear in the number of triangles in the graph,
with an accompanying positive temperature favoring these triangles, then
the model has a tendency to ``condense,'' forming regions of the graph that
are essentially complete cliques---subsets of vertices within which every
possible edge exists.  It is easy to see why the model shows this
behavior: cliques have the largest number of triangles for the number of
edges they contain, and are therefore highly energetically favored, while
costing the system a minimum in entropy by virtue of leaving the largest
possible number of other edges free to contribute to the (presumably
extensive) entropy of the rest of the graph.  Networks in the real world
however do not seem to have this sort of ``clumpy'' transitivity---regions
of cliquishness contributing heavily to the clustering coefficient,
separated by other regions with few triangles.  It is not clear how this
problem is to be circumvented, although for higher temperatures (lower
values of the parameters $\set{\beta_i}$) it is less problematic, since
higher temperatures favor entropy over energy.

Another area in which some progress has been made is in techniques for
extracting appropriate values for the temperature parameters in the model
from real-world network data.  Procedures for doing this have been
particularly important for social network applications.  Parameters so
extracted can be fed back into the Monte Carlo graph generation methods
described above to generate model graphs which have similar statistical
properties to their real-world counterparts and which can be used for
hypothesis testing or as a substrate for further network simulations.
Reviews of parameter extraction techniques can be found in
Refs.~\citen{AWC99} and~\citen{Snijders02}.

\section{The small-world model}
\label{sw}
A less sophisticated but more tractable model of a network with high
transitivity is the \defn{small-world model} proposed by Watts and
Strogatz~\cite{WS98,Watts99a,Watts99b}.\footnote{An equivalent model was
proposed by Ball~\etal~\cite{BMS97} some years earlier, as a model of the
spread of disease between households, but appears not to have been widely
adopted.}  As touched upon in Sec.~\ref{mixing}, networks may have a
geographical component to them; the vertices of the network have positions
in space and in many cases it is reasonable to assume that geographical
proximity will play a role in deciding which vertices are connected to
which others.  The small-world model starts from this idea by positing a
network built on a low-dimensional regular lattice and then adding or
moving edges to create a low density of ``shortcuts'' that join remote
parts of the lattice to one another.

\begin{figure*}
\resizebox{15cm}{!}{\includegraphics{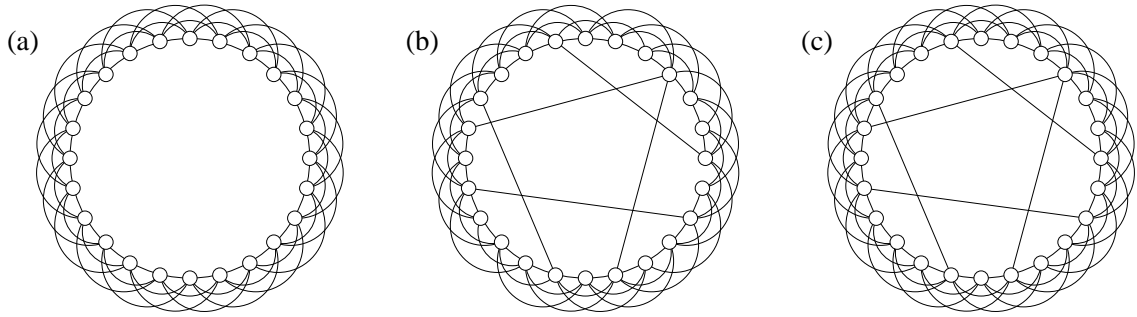}}
\caption{(a) A one-dimensional lattice with connections between all vertex
pairs separated by $k$ or fewer lattice spacing, with $k=3$ in this case.
(b)~The small-world model~\cite{WS98,Watts99a} is created by choosing at
random a fraction~$p$ of the edges in the graph and moving one end of each
to a new location, also chosen uniformly at random.  (c)~A slight variation
on the model~\cite{NW99a,Monasson99} in which shortcuts are added randomly
between vertices, but no edges are removed from the underlying
one-dimensional lattice.}
\label{swfig}
\end{figure*}

Small-world models can be built on lattices of any dimension or topology,
but the best studied case by far is one-dimensional one.  If we take a
one-dimensional lattice of $L$ vertices with periodic boundary conditions,
i.e.,~a ring, and join each vertex to its neighbors $k$ or fewer lattice
spacings away, we get a system like Fig.~\ref{swfig}a, with $Lk$ edges.
The small-world model is then created by taking a small fraction of the
edges in this graph and ``rewiring'' them.  The rewiring procedure involves
going through each edge in turn and, with probability~$p$, moving one end
of that edge to a new location chosen uniformly at random from the lattice,
except that no double edges or self-edges are ever created.  This process
is illustrated in Fig.~\ref{swfig}b.

\begin{figure}[b]
\resizebox{8cm}{!}{\includegraphics{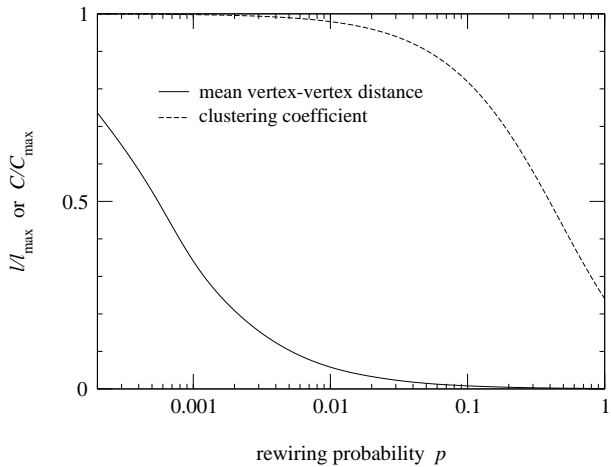}}
\caption{The clustering coefficient~$C$ and mean vertex--vertex
distance~$\ell$ in the small-world model of Watts and Strogatz~\cite{WS98}
as a function of the rewiring probability~$p$.  For convenience, both $C$
and $\ell$ are divided by their maximum values, which they assume when
$p=0$.  Between the extremes $p=0$ and $p=1$, there is a region in which
clustering is high and mean vertex--vertex distance is simultaneously low.}
\label{swlc}
\end{figure}

The rewiring process allows the small-world model to interpolate between a
regular lattice and something which is similar, though not identical (see
below), to a random graph.  When $p=0$, we have a regular lattice.  It is
not hard to show that the clustering coefficient of this regular lattice is
$C=(3k-3)/(4k-2)$, which tends to $\frac34$ for large~$k$.  The regular
lattice, however, does not show the small-world effect.  Mean geodesic
distances between vertices tend to $L/4k$ for large~$L$.  When $p=1$, every
edge is rewired to a new random location and the graph is almost a random
graph, with typical geodesic distances on the order of $\log L/\log k$, but
very low clustering $C\simeq 2k/L$ (see Sec.~\ref{poissonrg}).  As Watts
and Strogatz showed by numerical simulation, however, there exists a
sizable region in between these two extremes for which the model has both
low path lengths and high transitivity---see Fig.~\ref{swlc}.

The original model proposed by Watts and Strogatz is somewhat baroque.  The
fact that only one end of each chosen edge is rewired, not both, that no
vertex is ever connected to itself, and that an edge is never added between
vertex pairs where there is already one, makes it quite difficult to
enumerate or average over the ensemble of graphs.  For the purposes of
mathematical treatment, the model can be simplified considerably by
rewiring both ends of each chosen edge, and by allowing both double and
self edges.  This results in a system that genuinely interpolates between a
regular lattice and a random graph.  Another variant of the model that has
become popular was proposed independently by Monasson~\cite{Monasson99} and
by Newman and Watts~\cite{NW99a}.  In this variant, no edges are rewired.
Instead ``shortcuts'' joining randomly chosen vertex pairs are added to the
low-dimensional lattice---see Fig.~\ref{swfig}c.  The parameter~$p$
governing the density of these shortcuts is defined so as to make it as
similar as possible to the parameter~$p$ in the first version of the model:
$p$~is defined as the probability per edge on the underlying lattice, of
there being a shortcut anywhere in the graph.  Thus the mean total number
of shortcuts is $Lkp$ and the mean degree is $2Lk(1+p)$.  This version of
the model has the desirable property that no vertices ever become
disconnected from the rest of the network, and hence the mean
vertex--vertex distance is always formally finite.  Both this version and
the original have been studied at some length in the mathematical and
physical literature~\cite{Newman00b}.

\subsection{Clustering coefficient}
The clustering coefficient for both versions of the small-world model can
be calculated relatively easily.  For the original version, Barrat and
Weigt~\cite{BW00} showed that
\begin{equation}
C = {3(k-1)\over2(2k-1)}(1-p)^3,
\label{swc1}
\end{equation}
while for the version without rewiring, Newman~\cite{Newman02e}
showed that
\begin{equation}
C = {3(k-1)\over2(2k-1)+4kp(p+2)}.
\label{swc2}
\end{equation}

\subsection{Degree distribution}
The degree distribution of the small-world model does not match most
real-world networks very well, although this is not surprising, since this
was not a goal of the model in the first place.  For the version without
rewiring, each vertex has degree at least~$2k$, for the edges of the
underlying regular lattice, plus a binomially distributed number of
shortcuts.  Hence the probability $p_j$ of having degree~$j$ is
\begin{equation}
p_j = {L\choose j-2k} \biggl[ {2kp\over L} \biggr]^{j-2k}
        \biggl[ 1 - {2kp\over L} \biggr]^{L-j+2k}
\end{equation}
for $k\ge2k$, and $p_j=0$ for $j<2k$.  For the rewired version of the
model, the distribution has a lower cutoff at $k$ rather than~$2k$, and is
rather more complicated.  The full expression is~\cite{BW00}
\begin{equation}
p_j = \hspace{-0.6em} \sum_{n=0}^{\min(j-k,k)} {k\choose n} (1-p)^n p^{k-n}
          {(pk)^{j-k-n}\over(j-k-n)!}\,\e^{-pk}
\end{equation}
for $j\ge k$, and $p_j=0$ for~$j<k$.

\subsection{Average path length}
By far the most attention has been focused on the average geodesic path
length of the small-world model.  We denote this quantity~$\ell$.  We do
not have any exact solution for the value of~$\ell$ yet, but a number of
partial exact results are known, including scaling forms, as well as some
approximate solutions for its behavior as a function of the model's
parameters.

In the limit~$p\to0$, the model is a ``large world''---the typical path
length tends to $\ell=L/4k$, as discussed above.  Small-world behavior, by
contrast, is typically characterized by logarithmic scaling $\ell\sim\log
L$ (see Sec.~\ref{sweffect}), which we see for large~$p$, where the model
becomes like a random graph.  In between these two limits there is
presumably some sort of crossover from large- to small-world behavior.
Barth\'el\'emy and Amaral~\cite{BA99a} conjectured that $\ell$ satisfies a
scaling relation of the form
\begin{equation}
\ell = \xi g(L/\xi),
\label{baconject}
\end{equation}
where $\xi$ is a correlation length that depends on~$p$, and $g(x)$ an
unknown but universal scaling function that depends only on system
dimension and lattice geometry, but not on $L$, $\xi$ or~$p$.  The
variation of~$\xi$ defines the crossover from large- to small-world
behavior; the known behavior of~$\ell$ for small and large~$L$, can be
reproduced by having $\xi$ diverge as $p\to0$ and
\begin{equation}
g(x) \sim \biggl\lbrace\begin{array}{ll}
            x                & \mbox{for $x\gg1$}    \\
            \log x \qquad    & \mbox{for $x\ll1$}.
      \end{array}
\end{equation}
Barth\'el\'emy and Amaral conjectured that $\xi$ diverges as $\xi\sim
p^{-\tau}$ for small~$p$, where $\tau$ is a constant exponent.  These
conjectures have all turned out to be correct.  Barth\'el\'emy and Amaral
also conjectured on the basis of numerical results that $\tau=\frac23$,
which turned out not to be correct~\cite{Barrat99,NW99a,BA99c}.

Equation~\eref{baconject} has been shown to be correct by a renormalization
group treatment of the model~\cite{NW99a}.  From this treatment one can
derive a scaling form for $\ell$ of
\begin{equation}
\ell = {L\over k} f(Lkp),
\label{nwscale}
\end{equation}
which is equivalent to~\eref{baconject}, except for a factor of~$k$, if
$\xi=1/kp$ and $g(x)=x f(x)$.  Thus we immediately conclude that the
exponent $\tau$ defined by Barth\'el\'emy and Amaral is~1, as was also
argued by Barrat~\cite{Barrat99} using a mixture of scaling ideas and
numerical simulation.

The scaling form~\eref{nwscale} shows that we can go from the large-world
regime to the small-world one either by increasing~$p$ or by increasing the
system size~$L$.  Indeed, the crucial scaling variable $Lkp$ that appears
as the argument of the scaling function is simply equal to the mean number
of shortcuts in the model, and hence $\ell$ as a fraction of system size
depends only on how many shortcuts there are, for given~$k$.

Making any further progress has proved difficult.  We would like to be able
to calculate the scaling function~$f(x)$, but this turns out not to be
easy.  The calculation is possible, though complicated, for a variant model
in which there are no short cuts but random sites are connected to a single
central ``hub'' vertex~\cite{DM00a}.  But for the normal small-world model
no exact solution is known, although some additional exact scaling forms
have been found~\cite{KAS00,AKS02}.  Accurate numerical measurements have
been carried out for system sizes up to about
$L=10^7$~\cite{BA99a,Barrat99,NW99a,NW99b,Moukarzel99,MMP00} and quite good
results can be derived using series expansions~\cite{NW99b}.  A mean-field
treatment of the model has been given by Newman~\etal~\cite{NMW00}, which
shows that $f(x)$ is approximately
\begin{equation}
f(x) = {1\over2\sqrt{x^2+2x}} \tanh^{-1} \sqrt{x\over x+2},
\label{fscale}
\end{equation}
and Barbour and Reinert~\cite{BR01} have further shown that this result is
the leading order term in an expansion for~$\ell$ that can be used to
derive more accurate results for~$f(x)$.

The primary use of the small-world model has been as a substrate for the
investigation of various processes taking place on graphs, such as
percolation~\cite{NW99b,MN00b,Ozana01,Sander02},
coloring~\cite{Walsh99,Svenson01}, coupled
oscillators~\cite{WS98,HCK02,BP02b}, iterated
games~\cite{WS98,AK01,Kim02,EB02}, diffusion
processes~\cite{Monasson99,JSB00,GKK01a,FDBV01,PA01,LKK01,LKK02}, epidemic
processes~\cite{BMS97,KG99,MN00a,KA01,Zanette01,ZC01}, and spin
models~\cite{BW00,Pekalski01,HKC02,KZ02,Herrero02,ZZ03}.  Some of this work
is discussed further in Section~\ref{onnets}.

A few of variations of the small-world model have been proposed.  Several
authors have studied the model in dimension higher than
one~\cite{NW99a,NW99b,Moukarzel99,MMP00,Ozana01}---the results are
qualitatively similar to the one-dimensional case and follow the expected
scaling laws.  Various authors have also studied models in which shortcuts
preferentially join vertices that are close together on the underlying
lattice~\cite{Kleinberg00,Kleinberg00proc,JB00,SC01,MD02}.  Of particular
note is the work by Kleinberg~\cite{Kleinberg00,Kleinberg00proc}, which is
discussed in Sec.~\ref{greedy}.  Rozenfeld~\etal~\cite{RCBH02} and
independently Warren~\etal~\cite{WSS02} have studied models in which there
are \emph{only} shortcuts and no underlying lattice, but the signature of
the lattice still remains, guiding shortcuts to fall with higher
probability between more closely spaced vertices (see Sec.~\ref{siteperc}).

\section{Models of network growth}
\label{growing}
All of the models discussed so far take observed properties of real-world
networks, such as degree sequences or transitivity, and attempt to create
networks that incorporate those properties.  The models do not however help
us to understand how networks come to have those properties in the first
place.  In this section we examine a class of models whose primary goal is
to explain network properties.  In these models, the networks typically
grow by the gradual addition of vertices and edges in some manner intended
to reflect growth processes that might be taking place on the real
networks, and it is these growth processes that lead to the characteristic
structural features of the network.\footnote{An alternative and intriguing
idea, which has so far not been investigated in much depth, is that
features such as power-law degree distributions may arise through network
optimization.  See, for instance,
Refs.~\citen{WBE97,WBE99,BMR99,GLS00,FS01b,VCS02}.}  For example, a number
of
authors~\cite{BC96,Watts99a,Watts99b,JGN01,DEB02,KE02,JJ02,HK02b,Vazquez03a,Vazquez03b}
have studied models of network transitivity that make use of ``triadic
closure'' processes.  In these models, edges are added to the network
preferentially between pairs of vertices that have another third vertex as
a common neighbor.  In other words, edges are added so as to complete
triangles, thereby increasing the denominator in Eq.~\eref{defsc1} and so
increasing the amount of transitivity in the network.  (There is some
empirical evidence from collaboration networks in support of this
mechanism~\cite{Newman01d}.)

But the best studied class of network growth models by far, and the class
on which we concentrate primarily in this section, is the class of models
aimed at explaining the origin of the highly skewed degree distributions
discussed in Sec.~\ref{dd}.  Indeed these models are some of the best
studied in the whole of the networks literature, having been the subject of
an extraordinary number of papers in the last few years.  In this section
we describe first the archetypal model of Price~\cite{Price76}, which was
based in turn on previous work by Simon~\cite{Simon55}.  Then we describe
the highly influential model of Barab\'asi and Albert~\cite{BA99b}, which
has been the driving force behind much of the recent work in this area.  We
also describe a number of variations and generalizations of these models
due to a variety of authors.

\subsection{Price's model}
As discussed in Sec.~\ref{dd}, the physicist-turned-historian-of-science
Derek de Solla Price described in 1965 probably the first example of what
would now be called a scale-free network; he studied the network of
citations between scientific papers and found that both in- and out-degrees
(number of times a paper has been cited and number of other papers a paper
cites) have power-law distributions~\cite{Price65}.  Apparently intrigued
by the appearance of these power laws, Price published another paper some
years later~\cite{Price76} in which he offered what is now the accepted
explanation for power-law degree distributions.  Like many after him, his
work built on ideas developed in the 1950s by Herbert
Simon~\cite{Simon55,BE01}, who showed that power laws arise when ``the rich
get richer,'' when the amount you get goes up with the amount you already
have.  In sociology this is referred to as the \defn{Matthew
effect}~\cite{Merton68}, after the biblical edict, ``For to every one that
hath shall be given\dots''\ (Matthew~25:29).\footnote{In fact, this is
really only a half of the Matthew effect, since the same verse continues,
``\dots but from him that hath not, that also which he seemeth to have
shall be taken away.''  In the processes studied by Simon and Price nothing
is taken away from anyone.  The full Matthew effect, with both the giving
and the taking away, corresponds more closely to the Polya urn process than
to Price's cumulative advantage.  Price points out this distinction in his
paper~\cite{Price76}.}  Price called it
\defn{cumulative advantage}.  Today it is usually known under the name
\defn{preferential attachment}, coined by Barab\'asi and
Albert~\cite{BA99b}.

The important contribution of Price's work was to take the ideas of Simon
and apply them to the growth of a network.  Simon was thinking of wealth
distributions in his early work, and although he later gave other
applications of his ideas, none of them were to networked systems.  Price
appears to have been the first to discuss cumulative advantage specifically
in the context of networks, and in particular in the context of the network
of citations between papers and its in-degree distribution.  His idea was
that the rate at which a paper gets new citations should be proportional to
the number that it already has.  This is easy to justify in a qualitative
way.  The probability that one comes across a particular paper whilst
reading the literature will presumably increase with the number of other
papers that cite it, and hence the probability that you cite it yourself in
a paper that you write will increase similarly.  The same argument can be
applied to other networks also, such as the Web.  It is not clear that the
dependence of citation probability on previous citations need be strictly
linear, but certainly this is the simplest assumption one could make and it
is the one that Price, following Simon, adopts.  We now describe in detail
Price's model and his exact solution of it, which uses what we would now
call a \defn{master-equation} or \defn{rate-equation} method.

Consider a directed graph of $n$ vertices, such as a citation network.  Let
$p_k$ be the fraction of vertices in the network with in-degree~$k$, so
that $\sum_k p_k=1$.  New vertices are continually added to the network,
though not necessarily at a constant rate.  Each added vertex has a certain
out-degree---the number of papers that it cites---and this out-degree is
fixed permanently at the creation of the vertex.  The out-degree may vary
from one vertex to another, but the mean out-degree, which is denoted~$m$,
is a constant over time.\footnote{Elsewhere in this review we have used the
letter~$z$ to denote mean degree.  While it would make sense in many ways
to use the same notation here, we have opted instead to change notation and
use~$m$ because this is the notation used in most of the recent papers on
growing networks.  The reader should bear in mind therefore that $m$ is
not, as previously, the total number of edges in the graph.}  (Certain
conditions on the distribution of $m$ about the mean must hold; see for
instance Ref.~\citen{Durrett03}.)  The value~$m$ is also the mean in-degree
of the network: $\sum_k kp_k=m$.  Since the out-degree can vary between
vertices, $m$~can take non-integer values, including values less than~1.

In the simplest form of cumulative advantage process the probability of
attachment of one of our new edges to an old vertex---i.e.,~the probability
that a newly appearing paper cites a previous paper---is simply
proportional to the in-degree~$k$ of the old vertex.  This however
immediately gives us a problem, since each vertex starts with in-degree
zero, and hence would forever have zero probability of gaining new edges.
To circumvent this problem, Price suggests that the probability of
attachment to a vertex should be proportional to $k+k_0$, where $k_0$ is a
constant.  Although he discusses the case of general~$k_0$, all his
mathematical developments are for $k_0=1$, which he justifies for the
citation network by saying that one can consider the initial publication of
a paper to be its first citation (of itself by itself).  Thus the
probability of a new citation is proportional to $k+1$.

The probability that a new edge attaches to \emph{any} of the vertices with
degree~$k$ is thus
\begin{equation}
{(k+1)p_k\over\sum_k(k+1)p_k} = {(k+1)p_k\over m+1}.
\label{priceattach}
\end{equation}
The mean number of new citations per vertex added is simply~$m$, and hence
the mean number of new citations to vertices with current in-degree~$k$ is
$(k+1)p_k m/(m+1)$.  The number $np_k$ of vertices with in-degree~$k$
decreases by this amount, since the vertices that get new citations become
vertices of degree~$k+1$.  However, the number of vertices of in-degree~$k$
increases because of influx from the vertices previously of degree~$k-1$
that have also just acquired a new citation, except for vertices of degree
zero, which have an influx of exactly~1.  If we denote by $p_{k,n}$ the
value of $p_k$ when the graph has $n$ vertices, then the net change in
$np_k$ per vertex added is
\begin{equation}
(n+1)p_{k,n+1} - np_{k,n} = 
    \bigl[ kp_{k-1,n} - (k+1)p_{k,n} \bigr] {m\over m+1},
\end{equation}
for $k\ge1$, or
\begin{equation}
(n+1)p_{0,n+1} - np_{0,n} = 1 - p_{0,n} {m\over m+1},
\end{equation}
for $k=0$.  Looking for stationary solutions $p_{k,n+1}=p_{k,n}=p_k$, we
then find
\begin{equation}
p_k = \biggl\lbrace\begin{array}{ll}
    \bigl[ kp_{k-1} - (k+1)p_k \bigr] m/(m+1) \qquad
                                        & \mbox{for $k\ge1$,}\\
    1 - p_0 m/(m+1)                     & \mbox{for $k=0$.}
  \end{array}
\label{priceeq}
\end{equation}
Rearranging, we find $p_0=(m+1)/(2m+1)$ and $p_k = p_{k-1} k/(k+2+1/m)$ or
\begin{eqnarray}
p_k &=& {k(k-1)\ldots 1\over(k+2+1/m)\ldots(3+1/m)} \,p_0 \nonumber\\
    &=& (1+1/m) \Beta(k+1,2+1/m),
\end{eqnarray}
where $\Beta(a,b)=\Gamma(a)\Gamma(b)/\Gamma(a+b)$ is Legendre's
beta-function, which goes asymptotically as $a^{-b}$ for large~$a$ and
fixed~$b$, and hence
\begin{equation}
p_k \sim k^{-(2+1/m)}.
\end{equation}
In other words, in the limit of large~$n$, the degree distribution has a
power-law tail with exponent $\alpha=2+1/m$.  This will typically give
exponents in the interval between 2 and~3, which is in agreement with the
values seen in real-world networks---see Table~\ref{summary}.  (Bear in
mind that the mean degree $m$ need not take an integer value, and can be
less than~1.)  Price gives a comparison between his model and citation
network data from the Science Citation Index, making a plausible case that
the parameter~$m$ has about the right value to give the observed power-law
citation distribution.

Note that Price's assumption that the offset parameter $k_0=1$ can be
justified \textit{a posteriori} because the value of the exponent does not
depend on~$k_0$.  (This contrasts with the behavior of the model of
Barab\'asi and Albert~\cite{BA99b}, which is discussed in
Sec.~\ref{bagen}.)  The argument above is easily generalized to the case
$k_0\ne1$, and we find that
\begin{equation}
p_k = {m+1\over m(k_0+1)+1}\,{\Beta(k+k_0,2+1/m)\over\Beta(k_0,2+1/m)},
\end{equation}
and hence $\alpha=2+1/m$ again for large $k$ and fixed~$k_0$.  See
Sec.~\ref{bagen} and Refs.~\citen{DMS00} and~\citen{KR01} for further
discussion of the effects of offset parameters.  Thorough reviews of
master-equation methods for grown graph models have been given by
Dorogovtsev and Mendes~\cite{DM02} and Krapivsky and Redner~\cite{KR03}.

The analytic solution above was the extent of the progress Price was able
to make in understanding his model network.  Unlike present-day authors,
for instance, he did not have computational resources available to simulate
the model, and so could give no numerical results.  In recent years, a
great deal more progress has been made in understanding cumulative
advantage processes and the growth of networks.  Most of this work has been
carried out using a slightly different model, however, the model of
Barab\'asi and Albert, which we now describe.

\subsection{The model of Barab\'asi and Albert}
\label{bamodel}
The mechanism of cumulative advantage proposed by Price~\cite{Price76} is
now widely accepted as the probable explanation for the power-law degree
distribution observed not only in citation networks but in a wide variety
of other networks also, including the World Wide Web, collaboration
networks, and possibly the Internet and other technological networks also.
The work of Price himself, however, is largely unknown in the scientific
community, and cumulative advantage did not achieve currency as a model of
network growth until its rediscovery some decades later by Barab\'asi and
Albert~\cite{BA99b}, who gave it the new name of \defn{preferential
attachment}.  In a highly influential paper published---like Price's first
paper on citation networks---in the journal \textit{Science}, they proposed
a network growth model of the Web that is very similar to Price's, but with
one important difference.

The model of Barab\'asi and Albert~\cite{BA99b,BAJ99} is the same as
Price's in having vertices that are added to the network with degree~$m$,
which is never changed thereafter, the other end of each edge being
attached to (``citing'') another vertex with probability proportional to
the degree of that vertex.  The difference between the two models is that
in the model of Barab\'asi and Albert edges are undirected, so there is no
distinction between in- and out-degree.  This has pros and cons.  On the
one hand, both citation networks and the Web are in reality directed
graphs, so any undirected graph model is missing a crucial feature of these
networks.  On the other hand, by ignoring the directed nature of the
network, the model of Barab\'asi and Albert gets around Price's problem of
how a paper gets its first citation or a Web site gets its first link.
Each vertex in the graph appears with initial degree~$m$, and hence
automatically has a non-zero probability of receiving new links.  (Note
that for the model to be solvable using the master-equation approach as
demonstrated below, the number of edges added with each vertex must be
exactly~$m$---it cannot vary around the mean value as in the model of
Price.  Hence it must also be an integer and must always have a value
$m\ge1$.)

Another way of looking at the model of Barab\'asi and Albert is to say the
network \emph{is} directed, with edges going from the vertex just added to
the vertex that it is citing or linking to, but that the probability of
attachment of a new edge is proportional to the sum of the in- and
out-degrees of the vertex.  This however is perhaps a less satisfactory
viewpoint, since it is difficult to conjure up a mechanism, either for
citation networks or the Web, which would give rise to such an attachment
process.  Overall, perhaps the best way to look at the model of Barab\'asi
and Albert is as a model that sacrifices some of the realism of Price's
model in favor of simplicity.  As we will see, the main result of this
sacrifice is that the model produces only a single value $\alpha=3$ for the
exponent governing the degree distribution, although this has been remedied
in later generalizations of the model, which we discuss in
Sec.~\ref{bagen}.

The model of Barab\'asi and Albert can be solved exactly in the limit of
large graph size\footnote{The behavior of the model at finite system sizes
has been investigated by Krapivsky and Redner~\cite{KR02b}.} using the
master-equation method and such a solution has been given by
Krapivsky~\etal~\cite{KRL00} and independently by
Dorogovtsev~\etal~\cite{DMS00}.  (Barab\'asi and Albert themselves gave an
approximate solution based on the assumption that all vertices of the same
age have the same degree~\cite{BA99b,BAJ99}.  The method of
Krapivsky~\etal\ and Dorogovtsev~\etal\ does not make this assumption.)

The probability that a new edge attaches to a vertex of degree~$k$---the
equivalent of Eq.~\eref{priceattach}---is
\begin{equation}
{kp_k\over\sum_k kp_k} = {kp_k\over2m}.
\end{equation}
The sum in the denominator is equal to the mean degree of the network,
which is $2m$, since there are $m$ edges for each vertex added, and each
edge, being now undirected, contributes two ends to the degrees of network
vertices.  Now the mean number of vertices of degree~$k$ that gain an edge
when a single new vertex with $m$ edges is added is $m\times kp_k/2m=\half
kp_k$, independent of~$m$.  The number $np_k$ of vertices with degree~$k$
thus decreases by this same amount, since the vertices that get new edges
become vertices of degree~$k+1$.  The number of vertices of degree~$k$ also
increases because of influx from vertices previously of degree~$k-1$ that
have also just acquired a new edge, except for vertices of degree~$m$,
which have an influx of exactly~1.  If we denote by $p_{k,n}$ the
value of $p_k$ when the graph has $n$ vertices, then the net change in
$np_k$ per vertex added is
\begin{equation}
(n+1)p_{k,n+1} - np_{k,n} = 
    \half (k-1)p_{k-1,n} - \half kp_{k,n},
\end{equation}
for $k>m$, or
\begin{equation}
(n+1)p_{m,n+1} - np_{m,n} = 1 - \half m p_{m,n},
\end{equation}
for $k=m$, and there are no vertices with $k<m$.

Looking for stationary solutions $p_{k,n+1}=p_{k,n}=p_k$ as before, the
equations equivalent to Eq.~\eref{priceeq} for the model are
\begin{equation}
p_k = \biggl\lbrace\begin{array}{ll}
        \half (k-1) p_{k-1} - \half k p_k \qquad & \mbox{for $k>m$,}\\
        1 - \half m p_m                         & \mbox{for $k=m$.}
      \end{array}
\label{krldms}
\end{equation}
Rearranging for $p_k$ once again, we find $p_m=2/(m+2)$ and
$p_k=p_{k-1}(k-1)/(k+2)$, or~\cite{KRL00,DMS00}
\begin{equation}
p_k = {(k-1)(k-2)\ldots m\over (k+2)(k+1)\ldots(m+3)}\,p_m
    = {2m(m+1)\over (k+2)(k+1)k}.
\end{equation}
In the limit of large~$k$ this gives a power law degree distribution
$p_k\sim k^{-3}$, with only the single fixed exponent~$\alpha=3$.  A more
rigorous derivation of this result has been given by
Bollob\'as~\etal~\cite{BRST01}.

In addition to the basic solution of the model for its degree distribution,
many other results are now known about the model of Barab\'asi and Albert.
Krapivsky and Redner~\cite{KR01} have conducted a thorough analytic study
of the model, showing among other things that the model has two important
types of correlations.  First, there is a correlation between the age of
vertices and their degrees, with older vertices having higher mean degree.
For the case $m=1$, for instance, they find that the probability
distribution of the degree of a vertex~$i$ with age~$a$, measured as the
number of vertices added after vertex~$i$, is
\begin{equation}
p_k(a) = \sqrt{1-\frac{a}{n}} \biggl( 1 - \sqrt{1-\frac{a}{n}} \biggr)^k.
\end{equation}
Thus for specified age~$a$ the distribution is exponential, with a
characteristic degree scale that diverges as $(1-a/n)^{-1/2}$ as $a\to n$;
the earliest vertices added have substantially higher expected degree than
those added later, and the overall power-law degree distribution of the
whole graph is a result primarily of the influence of these earliest
vertices.

This correlation between degree and age has been used by Adamic and
Huberman~\cite{AH00} to argue against the model as a model of the World
Wide Web---they show using actual Web data that there is no such
correlation in the real Web.  This does not mean that preferential
attachment is not the explanation for power-law degree distributions in the
Web, only that the dynamics of the Web must be more complicated than this
simple model to account also for the observed age
distribution~\cite{BAJB00}.  An extension of the model that may explain why
age and degree are not correlated has been given by Bianconi and
Barab\'asi~\cite{BB01a,BB01b} and is discussed in Sec.~\ref{bagen}.

Second, Krapivsky and Redner~\cite{KR01} show that there are correlations
between the degrees of adjacent vertices in the model, of the type
discussed in Sec.~\ref{degcorr}.  Looking again at the special case $m=1$,
they show that the quantity $e_{jk}$ defined in Sec.~\ref{dcrg}, which is
the number of edges that connect vertex pairs with (excess) degrees $j$
and~$k$, is
\begin{eqnarray}
e_{jk} &=& {4j\over (k+1)(k+2)(j+k+2)(j+k+3)(j+k+4)}\nonumber\\
       & & \hspace{-2.5em}
           {} + {12j\over (k+1)(j+k+1)(j+k+2)(j+k+3)(j+k+4)}.\nonumber\\
\label{krcorr}
\end{eqnarray}
Note that this quantity is asymmetric.  This is because Krapivsky and
Redner regard the network as being directed, with edges leading from the
vertex just added to the pre-existing vertex to which they attach.  In the
expression above, however, $j$~and $k$ are total degrees of vertices, not
in- and out-degree.

Although~\eref{krcorr} shows that the vertices of the model have
non-trivial correlations, the correlation coefficient of the degrees of
adjacent vertices in the network is asymptotically zero as
$n\to\infty$~\cite{Newman02f}.  This is because the correlation coefficient
measures correlations relative to a linear model, and no such correlations
are present in this case.

One of the main advantages that we have today over early workers such as
Price is the widespread availability of powerful computer resources.  Quite
a number of numerical studies have been performed of the model of
Barab\'asi and Albert, which would have been entirely impossible thirty
years earlier.  It is worth mentioning here how simulations of these types
of models are conducted.  We consider the Barab\'asi--Albert model.  The
exact same ideas can be applied to Price's model also.

A naive simulation of the preferential attachment process is quite
inefficient.  In order to attach to a vertex in proportion to its degree we
normally need to examine the degrees of all vertices in turn, a process
that takes $\ord(n)$ time for each step of the algorithm.  Thus the
generation of a graph of size $n$ would take $\ord(n^2)$ steps overall.  A
much better procedure, which works in $\ord(1)$ time per step and $\ord(n)$
time overall, is the following.  We maintain a list, in an integer array
for instance, that includes $k_i$ entries of value $i$ for each vertex~$i$.
Thus, for example, a network of four vertices labeled 1, 2, 3, and 4 with
degrees 2, 1, 1, and~3, respectively could be represented by the array
$(1,1,2,3,4,4,4)$.  Then in order to choose a target vertex for a new edge
with the correct preferential attachment, one simply chooses a number at
random from this list.  Of course, the list must be updated as new vertices
and edges are added, but this is simple.  Notice that there is no
requirement that the items in the list be in any particular order.  If we
add a new vertex~5 to our network above, for example, with degree~1 and one
edge that connects it to vertex~2, the list can be updated by adding new
items to the end, so that it reads $(1,1,1,2,3,4,4,4,5,2)$.  And so forth.
Models such as Price's, in which there is an offset~$k_0$ in the
probability of selecting a vertex (so that the total probability goes as
$k+k_0$), can be treated with the same method---the offset merely means
that with some probability one chooses a vertex with preferential
attachment and otherwise one chooses it uniformly from the set of all
vertices.

An alternative method for simulating the model of Barab\'asi and Albert has
been described by Krapivsky and Redner~\cite{KR01}.  Their method uses the
network structure itself in place of the list of vertices above and works
as follows.  The model is regarded as a directed network in which there are
exactly $m$ edges running out of each vertex, pointing to others.  We first
pick a vertex at random from the graph and then with some probability we
either keep that vertex or we ``redirect'' to one of its neighbors,
meaning that we pick at random one of the vertices it points to.  Since
each vertex has exactly $m$ outgoing edges, the latter operation is
equivalent to choosing an edge at random from the graph and following it,
and hence alights on a target vertex with probability proportional to the
in-degree~$j$ of that target (because there are $j$ ways to arrive at a
vertex of in-degree~$j$---see Sec.~\ref{confmodel}).  Thus the total
probability of selecting any given vertex is proportional to $j+c$, where
$c$ is some constant.  However, since the out-degree of all vertices is
simply~$m$, the total degree is $k=j+m$ and the selection probability is
therefore also proportional to $k+c-m$.  By choosing the probability of
redirection appropriately, we can arrange for the constant~$c$ to be equal
to~$m$, and hence for the probability of selecting a vertex to be simply
proportional to~$k$.  Since it does not require an extra array for the
vertex list, this method of simulation is more memory efficient than the
previous method, although it is slightly more complicated to implement.

In their original paper on their model, Barab\'asi and Albert~\cite{BA99b}
gave simulations showing the power-law distribution of degrees.  A number
of authors have subsequently published more extensive simulation results.
Of particular note is the work by Dorogovtsev and Mendes~\cite{DM00b,DM00c}
and by Krapivsky and Redner~\cite{KR02b}.

A crucial element of both the models of Price and of Barab\'asi and Albert
is the assumption of linear preferential attachment.  It is worth asking
whether there is any empirical evidence in support of this assumption.  (We
discuss in the next section some work on models that relax the linearity
assumption.)  Two studies indicate that it may be a reasonable
approximation to the truth.  Jeong~\etal~\cite{JNB03} looked at the time
evolution of citation networks, the Internet, and actor and scientist
collaboration networks, and measured the number of new edges a vertex
acquires in a single year as a function of the number of previously
existing edges.  They found that the one quantity was roughly proportional
to the other, and hence concluded that linear preferential attachment was
at work in these networks.  Newman~\cite{Newman01d} performed a similar
study for scientific collaboration networks, but with finer time
resolution, measured by the publication of individual papers, and came to
similar conclusions.

\subsection{Generalizations of the Barab\'asi--Albert model}
\label{bagen}
The model of Barab\'asi and Albert~\cite{BA99b} has attracted an
exceptional amount of attention in the literature.  In addition to analytic
and numerical studies of the model itself, many authors have suggested
extensions or modifications of the model that alter its behavior or make
it a more realistic representation of processes taking place in real-world
networks.  We discuss a few of these here.  A more extensive review of
developments in this area has been given by Albert and
Barab\'asi~\cite{AB02} (see particularly Table~III in that paper).

Dorogovtsev~\etal~\cite{DMS00} and Krapivsky and Redner~\cite{KR01} have
examined the model in which the probability of attachment to a vertex of
degree~$k$ is proportional to $k+k_0$, where the offset~$k_0$ is a
constant.  Note that $k_0$ is allowed to be negative---it can fall anywhere
in the range $-m<k_0<\infty$ and the probability of attachment will be
positive.  The equations for the stationary state of the degree
distribution of this model, analogous to Eq.~\eref{krldms}, are
\begin{equation}
p_k = \biggl\lbrace\begin{array}{ll}
        \bigl[ (k-1) p_{k-1} - k p_k \bigr] m/(2m+k_0) \qquad
                                              & \mbox{for $k>m$,}\\
        1 - p_m m^2/(2m+k_0)                  & \mbox{for $k=m$,}
    \end{array}
\end{equation}
which gives $p_m=(2m+k_0)/(m^2+2m+k_0)$ and
\begin{eqnarray}
p_k &=& {(k-1)\ldots m\over(k+2+k_0/m)\ldots(m+3+k_0/m)}\,p_m\nonumber\\
    &=& {\Beta(k,3+k_0/m)\over\Beta(m,2+k_0/m)},
\end{eqnarray}
where $\Beta(a,b)=\Gamma(a)\Gamma(b)/\Gamma(a+b)$ is again the Legendre
beta-function.  This gives a power law for large~$k$ once more, with
exponent $\alpha=3+k_0/m$.  It is proposed that negative values of $k_0$
could be the explanation for the values $\alpha<3$ seen in real-world
networks.\footnote{Price's result $\alpha=2+1/m$~\cite{Price76} corresponds
to $k_0=-(m-1)$ so that the ``attractiveness'' of a new vertex is~1.  The
model of Barab\'asi and Albert corresponds to $k_0=0$, so that $\alpha=3$.}
A longer discussion of the effects of offset parameters is given in
Ref.~\citen{KR01}.

Krapivsky~\etal~\cite{KRL00,KR01} also consider another important
generalization of the model, to the case where the probability of
attachment to a vertex is not linear in the degree~$k$ of the vertex, but
goes instead as some general power of degree~$k^\gamma$.  Again this model
is solvable using methods similar to those above, and the authors find
three general classes of behavior.  For $\gamma=1$ exactly, we recover the
normal linear preferential attachment and power-law degree sequences.  For
$\gamma<1$, the degree distribution is a power law multiplied by a
stretched exponential, whose exponent is a complicated function
of~$\gamma$.  (In fact, in most cases there is no known analytic solution
for the equations governing the exponent; they must be solved numerically.)
For $\gamma>1$ there is a ``condensation'' phenomenon, in which a single
vertex gets a finite fraction of all the connections in the network, and
for $\gamma>2$ there is a non-zero probability that this ``gel node'' will
be connected to every other vertex on the graph.  The remainder of the
vertices have an exponentially decaying degree distribution.

Another variation on the basic growing network theme is to make the mean
degree change over time.  There is evidence to suggest that in the
World Wide Web the average degree of a vertex is increasing with time,
i.e.,~the parameter~$m$ appearing in the models is increasing.  Dorogovtsev
and Mendes~\cite{DM01a,DM03a} have studied a variation of the
Barab\'asi--Albert model that incorporates this process.  They assume that
the number $m$ of new edges added per new vertex increases with network
size~$n$ as $n^a$ for some constant~$a$, and that the probability of
attaching to a given vertex goes as $k+Bn^a$ for constant~$B$.  They show
that the resulting degree distribution follows a power law with exponent
$\alpha=2+B(1+a)/(1-Ba)$.  (Note that when $a=0$, this model reduces to the
model studied previously by Dorogovtsev~\etal~\cite{DMS00}, but the
expression for~$\alpha$ given here is not valid in this limit.)  Thus this
process offers another possible mechanism by which the exponent of the
degree distribution can be tuned to match that observed in real-world
networks.

In Price's model of citation networks, no new out-going edges are added to
a vertex after its first appearance, and edges once added to the graph
remain where they are forever.  This makes sense for citation networks.
But the model of Barab\'asi and Albert is intended to be a model of the
World Wide Web, in which new links are often added to pre-existing Web
sites, and old links are frequently moved or removed.  A number of authors
have proposed models that incorporate processes like these.  In particular,
Dorogovtsev and Mendes~\cite{DM00b} have proposed a model that adds to the
standard Barab\'asi--Albert model an extra mechanism whereby edges appear
and disappear between pre-existing vertices with stochastically constant
but possibly different rates.  They find that over a wide range of values
of the rates the power-law degree distribution is maintained, although
again the exponent varies from the value~$-3$ seen in the original model.
Krapivsky and Redner~\cite{KR02a} have also proposed a model that allows
edges to be added after vertices are created, which we discuss in the next
section.  Albert and Barab\'asi~\cite{AB00a} and
Tadi\'c~\cite{Tadic01,Tadic02} have studied models in which edges can move
around the network after they are added.  These models can show both
power-law and exponential degree distributions depending on the model
parameters.

As discussed in Sec.~\ref{bamodel}, Adamic and Huberman~\cite{AH00} have
shown that the real World Wide Web does not have the correlations between
age and degree of vertices that are found in the model of Barab\'asi and
Albert.  Adamic and Huberman suggest that this is because the degree of
vertices is also a function of their intrinsic worth; some Web sites are
useful to more people than others and so gain links at a higher rate.
Bianconi and Barab\'asi~\cite{BB01a,BB01b} have proposed an extension of
the Barab\'asi--Albert model that mimics this process.  In their model each
newly appearing vertex~$i$ is given a ``fitness''~$\eta_i$ that represents
its attractiveness and hence its propensity to accrue new links.  Fitnesses
are chosen from some distribution~$\rho(\eta)$ and links attach to vertices
with probability proportional now not just to the degree $k_i$ of
vertex~$i$ but to the product~$\eta_ik_i$.

Depending on the form of the distribution~$\rho(\eta)$ this model shows two
regimes of behavior~\cite{BB01a,KR02a}.  If the distribution has finite
support, then the network shows a power-law degree distribution, as in the
original Barab\'asi--Albert model.  However, if the distribution has
infinite support, then the one vertex with the highest fitness accrues a
finite fraction of all the edges in the network---a sort of ``winner takes
all'' phenomenon, which Bianconi and Barab\'asi liken to monopoly dominance
of a market.

A number of variations on the fitness theme have been studied by Erg\"un
and Rodgers~\cite{ER02}, who looked at a directed version of the
Bianconi--Barab\'asi model and at models where instead of multiplying the
attachment probability, the fitness~$\eta_i$ contributes additively to the
probability of attaching a new edge to vertex~$i$.  Treating the models
analytically, they found in each case that for suitable parameter values
the power-law degree distribution is preserved, although again the exponent
may be affected by the distribution of fitnesses, and in some cases there
are also logarithmic corrections to the degree distribution.  A model with
vertex fitness but no preferential attachment has been studied by
Caldarelli~\etal~\cite{CCDM02}, and also gives power-law degree
distributions under some circumstances.

\subsection{Other growth models}
The model of Barab\'asi and Albert~\cite{BA99b} is elegant and simple, but
lacks a number of features that are present in the real World Wide Web:
\begin{itemize}
\item The model is a model of an undirected network, where the real Web
is directed.
\item As mentioned previously one can regard the model as a model of a
directed network, but in that case attachment is in proportion to the sum
of in- and out-degrees of a vertex, which is unrealistic---presumably
attachment should be in proportion to in-degree only, as in the model of
Price.
\item If we regard the model as producing a directed network, then it
generates acyclic graphs (see Sec.~\ref{types}), which are a poor
representation of the Web.
\item All vertices in the model belong to a single connected component (a
weakly connected component if the graph is regarded as directed---the graph
has no strongly connected components because it is acyclic).  In the real
Web there are many separate components (and strongly connected components).
\item The out-degree distribution of the Web follows a power law, whereas
out-degree is a constant in the model.\footnote{What's more, although it is
rarely pointed out, it is clearly the case that a different mechanism must
be responsible for the out-degree distribution from the one responsible for
the in-degree distribution.  We can justify preferential attachment for
in-degree by saying that Web sites are easier to find if they have more
links to them, and hence they get more new links because people find them.
No such argument applies for out-degree.  It is usually assumed that
out-degree is subject to preferential attachment nonetheless.  One can
certainly argue that sites with many out-going links are more likely to add
new ones in the future than sites with few, but it's far from clear that
this must be the case.}
\end{itemize}
Many of these criticisms are also true of Price's model, but Price's model
is intended to be a model of a citation network and citation networks
really are directed, acyclic, and to a good approximation all vertices
belong to a single component, unless they cite and are cited by no one else
at all.  Thus Price's model is, within its own limited sphere, a reasonable
one.  For the World Wide Web a number of authors have suggested new growth
models that address one or more of the concerns above.  Here we describe a
number of these models, starting with some very simple ones and working up
to the more complex.

Consider first the issue of the component structure of the network.  In the
models of Price and of Barab\'asi and Albert each vertex joins to at least
one other when it first appears.  It follows trivially then that, so long
as no edges are ever removed, all vertices belong to a single
(weakly-connected) component.  This is not true in the real Web.  How can
we get around it?  To address this question
Callaway~\etal~\cite{Callaway01} proposed the following extremely simple
model of a growing network.  Vertices are added to the network one by one
as before, and a mean number $m$ of undirected edges are added with each
vertex.  As with Price's model, the value of~$m$ is only an average---the
actual number of edges added per step can vary---and so $m$ is not
restricted to integer values, and indeed we will see that the interesting
behavior of the model takes place at values $m<1$.

The important difference between this model and the previous models is that
edges are not, in general, attached to the vertex that has just been added.
Instead, both ends of each edge are attached to vertices chosen uniformly
at random from the whole graph, without preferential attachment.  Vertices
therefore normally have degree zero when they are first added to the graph.
Because of the lack of preferential attachment this model does not show
power-law degree distributions---in fact the degree distribution can be
show to be exponential---but it does have an interesting component
structure.  A related model has been studied, albeit to somewhat different
purpose, by Aldous and Pittel~\cite{AP00}.  Their model is equivalent to
the model of Callaway~\etal\ in the case $m=1$.  Also Bauer and
collaborators~\cite{BB02,CB02} have investigated a directed-graph version
of the model.

Initially, one might imagine that the model of Callaway~\etal\ generated an
ordinary Poisson random graph of the Erd\"{o}s--R\'enyi type.  Further
reflection reveals however that this is not the case; older vertices in the
network will tend to be connected to one another, so the network has a
cliquish core of old-timers surrounded by a sea of younger vertices.
Nonetheless, like the Poisson random graph, the model does have many
separate components, with a phase transition at a finite value of~$m$ at
which a giant component appears that occupies a fixed fraction of the
volume of the network as $n\to\infty$.  To demonstrate this,
Callaway~\etal\ used a master-equation approach similar to that used for
degree distributions in the preceding sections.  One defines $p_s$ to be
the probability that a randomly chosen vertex belongs to a component of~$s$
vertices, and writes difference equations that give the change in $p_s$
when a single vertex and $m$ edges are added to the graph.  Looking for
stationary solutions, one then finds in the limit of large graph size that
\begin{equation}
p_s = \biggl\lbrace\begin{array}{ll}
         ms \sum_{j=1}^{s-1} p_j p_{s-j} - 2ms p_s \qquad
                                            & \mbox{for $s>1$} \\
         1 - 2m p_1                         & \mbox{for $s=1$.}
\end{array}
\label{defsps}
\end{equation}
Being nonlinear in $p_s$, these equations are harder to solve than those
for the degree distributions in previous sections, and indeed no exact
solution has been found.  Nonetheless, we can see that a giant component
must form by defining a generating function for the component size
distribution similar to that of Eq.~\eref{defsh0}: $H(x)=\sum_{s=0}^\infty
p_s x^s$.  Then~\eref{defsps} implies that
\begin{equation}
{\d H\over\d x} = {1\over2m} \biggl[ {1 - H(x)/x\over 1 - H(x)}
\biggl].
\label{eqdh}
\end{equation}
If there is no giant component, then $H(1)=1$ and the average component
size is~$\av{s}=H'(1)$.  Taking the limit $x\to1$ in Eq.~\eref{eqdh}, we
find that $\av{s}$ is a solution of the quadratic equation
$2m\av{s}^2-\av{s}+1=0$, or
\begin{equation}
\av{s} = {1-\sqrt{1-8m}\over4m}.
\end{equation}
(The other solution to the quadratic gives a non-physical value.)  This
solution exists only up to $m=\frac18$ however, and hence above this point
there must be a giant component.  This doesn't tell us where in the
interval $0\le m\le\frac18$ the giant component appears, but a proof that
the transition in fact falls precisely at $m=\frac18$ was later given by
Durrett~\cite{Durrett03}.

The model of Callaway~\etal\ has been generalized to include preferential
attachment by Dorogovtsev~\etal~\cite{DMS01b}.  In their version of the
model both ends of each edge are attached in proportion to the degrees of
vertices plus a constant offset to ensure that vertices of degree zero have
a chance of receiving an edge.  Again they find many components and a phase
transition at nonzero~$m$, and in addition the power-law degree
distribution is now restored.

Taking the process a step further, Krapivsky and Redner~\cite{KR02a}
studied a full directed-graph model in which both vertices and directed
edges are added at stochastically constant rates and the out-going end of
each edge is attached to vertices in proportion to their out-degree and the
in-going end in proportion to in-degree, plus appropriate constant offsets.
This appears to be quite a reasonable model for the growth of the Web.  It
produces a directed graph, it allows edges to be added after the creation
of a vertex, it allows for separate components in the graph, and, as
Krapivsky and Redner showed, it gives power laws in both the in- and
out-degree distributions, just as observed in the real Web.  By varying the
offset parameters for the in- and out-degree attachment mechanisms, one can
even tune the exponents of the two distributions to agree with those
observed in the wild.  (Krapivsky and Redner's model is a development of an
earlier model that they proposed~\cite{KRR01} that had all the same
features, but gave rise to only a single weakly connected component because
each added vertex came with one edge that attached it to the rest of the
network from the outset.  In their later paper, they abandoned this
feature.  A similar model has also been studied by Rodgers and
Darby-Dowman~\cite{RD01}.)  A slight variation on the model of Krapivsky
and Redner has been proposed independently by Aiello~\etal~\cite{ACL02},
who give rigorous proofs of some of its properties.

\subsection{Vertex copying models}
There are some networks that appear to have power-law degree distributions,
but for which preferential attachment is clearly not an appropriate model.
Good examples are biochemical interaction networks of various
kinds~\cite{Jeong00,Jeong01,FW00,WF01,SP03,Stelling02}.  A number of
studies have been performed, for instance, of the interaction networks of
proteins (see Sec.~\ref{bionets}) in which the vertices are proteins and
the edges represent reactions.  These networks do change on very long
time-scales because of biological evolution, but there is no reason to
suppose that protein networks grow according to a simple cumulative
advantage or preferential attachment process.  Nonetheless, it appears that
the degree distribution of these networks obeys a power law, at least
roughly.

A possible explanation for this observation has been suggested by
Kleinberg~\etal~\cite{Kleinberg99b,Kumar00}, who proposed that these
networks grow, at least in part, by the copying of vertices.
Kleinberg~\etal\ were interested in the growth of the Web, for which their
model is as follows.  The graph grows by stochastically constant addition
of vertices and addition of directed edges either randomly or by copying
them from another vertex.  Specifically, one chooses an existing vertex and
a number~$m$ of edges to add to it, and one then decides the targets of
those edges, by choosing at random another vertex and copying targets from
$m$ of its edges, randomly chosen.  If the chosen vertex has less than $m$
outgoing edges, then its $m$ edges are copied and one moves on to another
vertex and copies its edges, and so forth until $m$ edges in total have
been copied.  In its most general form, the model of Kleinberg~\etal\ also
incorporates mechanisms for the removal of edges and vertices, which we do
not describe here.

It is straightforward to see that the copying mechanism will give rise to
power-law distributions.  The mean probability that an edge from a randomly
chosen vertex will lead to a particular other vertex with in-degree~$k$ is
proportional to~$k$ (see Sec.~\ref{confmodel}), and hence the rate of
increase of a vertex's degree is proportional to its current degree.  As
with the model of Price, this mechanism will never add new edges to
vertices that currently have degree zero, so Kleinberg~\etal\ also include
a finite probability that the target of a newly added edge will be chosen
at random, so that vertices with degree zero have a chance to gain edges.
In their original paper, Kleinberg~\etal\ present only numerical evidence
that their model results in a power law degree distribution, but in a later
paper a subset of the same authors~\cite{Kumar00} proved that the degree
distribution is a power law with exponent $\alpha=(2-a)/(1-a)$, where $a$
is the ratio of the number of edges added whose targets are chosen at
random to the number whose targets are copied from other vertices.  For
small values of~$a$, between 0 and~$\half$, i.e.,~for models in which most
target selection is by copying, this produces exponents $2\le\alpha\le3$,
which is the range observed in most real-world networks---see
Table~\ref{summary}.  Some further analytic results for copying models have
been given by Chung~\etal~\cite{CLDG03}.

It is not clear whether the copying mechanism really is at work in the
growth of the World Wide Web, but there has been considerable interest in
its application as a model of the evolution of protein interaction networks
of one sort or another.  The argument here is that the genes that code for
proteins can and do, in the course of their evolutionary development,
duplicate.  That is, upon reproduction of an organism, two copies of a gene
are erroneously made where only one existed before.  Since the proteins
coded for by each copy are the same, their interactions are also the same,
i.e.,~the new gene copies its edges in the interaction network from the
old.  Subsequently, the two genes may develop differences because of
evolutionary drift or selection~\cite{Wagner01}.  Models of protein
networks that make use of copying mechanisms have been proposed by a number
of authors~\cite{SPSK02,KKKR02,VFMV03,BLW03}.

A variation on the idea of vertex copying appears in the autocatalytic
network models of Jain and Krishna~\cite{JK98,JK01}, in which a network of
interacting chemical species evolves by reproduction and mutation, giving
rise ultimately to self-sustaining autocatalytic loops reminiscent of the
``hypercycles'' of Eigen and Schuster~\cite{ES79}, which have been proposed
as a possible explanation of the origin of life.

\section{Processes taking place on networks}
\label{onnets}
As discussed in the introduction, the ultimate goal of the study of the
structure of networks is to understand and explain the workings of systems
built upon those networks.  We would like, for instance, to understand how
the topology of the World Wide Web affects Web surfing and search engines,
how the structure of social networks affects the spread of information, how
the structure of a food web affects population dynamics, and so forth.
Thus, the next logical step after developing models of network structure,
such as those described in the previous sections of this review, is to look
at the behavior of models of physical (or biological or social) processes
going on on those networks.  Progress on this front has been slower than
progress on understanding network structure, perhaps because without a
thorough understanding of structure an understanding of the effects of that
structure is hard to come by.  However, there have been some important
advances made, particularly in the study of network failure, epidemic
processes on networks, and constraint satisfaction problems.  In this
section we review what has been learned so far.

\subsection{Percolation theory and network resilience}
\label{siteperc}
\begin{figure}
\resizebox{7cm}{!}{\includegraphics{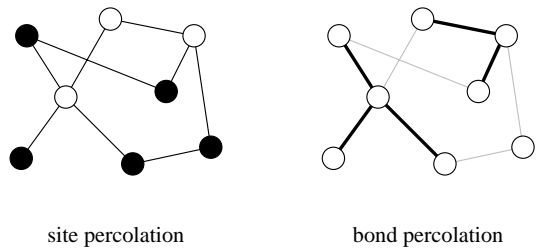}}
\caption{Site and bond percolation on a network.  In site percolation,
vertices (``sites'' in the physics parlance) are either occupied (solid
circles) or unoccupied (open circles) and studies focus on the shape and
size of the contiguous clusters of occupied sites, of which there are three
in this small example.  In bond percolation, it is the edges (``bonds'' in
physics) that are occupied or not (black or gray lines) and the vertices
that are connected together by occupied edges that form the clusters of
interest.}
\label{perc}
\end{figure}

One of the first examples to be studied thoroughly of a process taking
place on a network has been percolation processes, mostly simple site and
bond percolation---see Fig.~\ref{perc}---although a number of variants have
been studied also.  A percolation process is one in which vertices or edges
on a graph are randomly designated either ``occupied'' or ``unoccupied''
and one asks about various properties of the resulting patterns of
vertices.  One of the main motivations for the percolation model when it
was first proposed in the 1950s was the modeling of the spread of
disease~\cite{BH57,Hammersley57}, and it is in this context also that it
was first studied in the current wave of interest in real-world
networks~\cite{NW99b}.  We consider epidemiological applications of
percolation theory in Sec.~\ref{episec}.  Here however, we depart from the
order of historical developments to discuss first a simpler application to
the question of network resilience.

As discussed in Sec.~\ref{resilience}, real-world networks are found often
to be highly resilient to the random deletion of their vertices.
Resilience can be measured in different ways, but perhaps the simplest
indicator of resilience in a network is the variation (or lack of
variation) in the fraction of vertices in the largest component of the
network, which we equate with the giant component in our models (see
Sec.~\ref{poissonrg}).  If one is thinking of a communication network, for
example, in which the existence of a connecting path between two vertices
means that those two can communicate with one another, then the vertices in
the giant component can communicate with an extensive fraction of the
entire network, while those in the small components can communicate with
only a few others at most.  Following the numerical studies of
Broder~\etal~\cite{Broder00} and Albert~\etal~\cite{AJB00} on subsets of
the Web graph, it was quickly realized~\cite{CEBH00,CNSW00} that the
problem of resilience to random failure of vertices in a network is
equivalent to a site percolation process on the network.  Vertices are
randomly occupied (working) or unoccupied (failed), and the number of
vertices remaining that can successfully communicate is precisely the giant
component of the corresponding percolation model.

A number of analytic results have been derived for percolation on networks
with the structure of the configuration model of Sec.~\ref{confmodel},
i.e.,~a random graph with a given degree sequence.
Cohen~\etal~\cite{CEBH00} made the following simple argument.  Suppose we
have a configuration model with degree distribution~$p_k$.  That is, a
randomly chosen vertex has degree~$k$ with probability $p_k$ in the limit
of large number~$n$ of vertices.  Now suppose that only a fraction $q$ of
the vertices are ``occupied,'' or functional, that fraction chosen
uniformly at random from the entire graph.  For a vertex with degree~$k$,
the number~$k'$ of occupied vertices to which it is connected is
distributed binomially so that the probability of having a particular value
of $k'$ is ${k\choose k'} q^{k'} (1-q)^{k-k'}$, and hence the total
probability that a randomly chosen vertex is connected to $k'$ other
occupied vertices is
\begin{equation}
p_{k'} = \sum_{k=k'}^\infty p_k {k\choose k'} q^{k'} (1-q)^{k-k'}.
\end{equation}
Since vertex failure is random and uncorrelated, the subset of all vertices
that are occupied forms another another configuration model with this
degree distribution.  Cohen~\etal\ then applied the criterion of Molloy and
Reed, Eq.~\eref{mrresult1}, to determine whether this network has a giant
component.  (One could also apply Eqs.~\eref{mrresult2} and~\eref{avsrg} to
determine the size of the giant and non-giant components, although this is
not done in Ref.~\citen{CEBH00}.)

One of the most interesting conclusions of the work of Cohen~\etal\ is for
the case of networks with power-law degree distributions $p_k\sim
k^{-\alpha}$ for some constant~$\alpha$.  When $\alpha\le3$, they find that
the critical value~$q_c$ of~$q$ where the transition takes place at which a
giant component forms is zero or negative, indicating that the network
always has a giant component, or in the language of physics, the network
always percolates.  This echos the numerical results of
Albert~\etal~\cite{AJB00}, who found that the connectivity of power-law
networks was highly robust to the random removal of vertices.  In general,
the method of Cohen~\etal\ indicates that $q_c\le0$ for any degree
distribution with a diverging second moment.

An alternative and more general approach to the percolation problem on the
configuration model has been put forward by Callaway~\etal~\cite{CNSW00},
using a generalization of the generating function formalism discussed in
Sec.~\ref{confmodel}.  In their method, the probability of occupation of a
vertex can be any function of the degree $k$ of that vertex.  Thus the
constant~$q$ of the approach of Cohen~\etal\ is generalized to~$q_k$, the
probability that a vertex having degree~$k$ is occupied.  One defines
generating functions
\begin{equation}
F_0(x) = \sum_{k=0}^\infty p_k q_k x^k,\qquad
F_1(x) = {\sum_k kp_kq_kx^{k-1}\over\sum_k kp_k},
\label{defsf0f1}
\end{equation}
and it can then be shown that the probability distribution of the size of
the component of occupied vertices to which a randomly chosen vertex
belongs is generated by $H_0(x)$ where
\begin{subequations}
\begin{eqnarray}
H_0(x) &=& 1 - F_0(1) + x F_0(H_1(x)),\\
H_1(x) &=& 1 - F_1(1) + x F_1(H_1(x)).
\end{eqnarray}
\end{subequations}
(Note that $F_0$ is not a properly normalized generating function in the
sense that $F_0(1)\ne1$.)  From this one can derive an expression for the
mean component size:
\begin{equation}
\av{s} = F_0(1) + {F_0'(1) F_1(1)\over1-F_1'(1)},
\end{equation}
which immediately tells us that the phase transition at which a giant
component forms takes place at $F_1'(1)=1$.  The size of the giant
component is given by
\begin{equation}
S = F_0(1) - F_0(u),\qquad u = 1 - F_1(1) + F_1(u).
\end{equation}

For instance, in the case studied by Cohen~\etal~\cite{CEBH00} of uniform
occupation probability~$q_k=q$, this gives a critical occupation
probability of $q_c=1/G_1'(1)$, where $G_1(x)$ is the generating function
for the degree distribution itself, as defined in Eq.~\eref{defsg0g1}.
Taking the example of a power-law degree distribution
$p_k=k^{-\alpha}/\zeta(\alpha)$, Eq.~\eref{powerlaw}, we find
\begin{equation}
q_c = {\zeta(\alpha-1)\over\zeta(\alpha-2)-\zeta(\alpha-1)}.
\end{equation}
This is negative (and hence unphysical) for $\alpha<3$, confirming the
finding that the system always percolates in this regime.  Note that
$q_c>1$ for sufficiently large~$\alpha$, which is also unphysical.  One
finds that the system \emph{never} percolates for $\alpha>\alpha_c$, where
$\alpha_c$ is the solution of $\zeta(\alpha-2)=2\zeta(\alpha-1)$, which
gives $\alpha_c=3.4788\ldots$\ \ This corresponds to the point at which the
underlying network itself ceases to have a giant component, as shown by
Aiello~\etal~\cite{ACL00} and discussed in Sec.~\ref{confmodel}.

The main advantage of the approach of Callaway~\etal\ is that it allows us
to remove vertices from the network in an order that depends on their
degree.  If, for instance, we set $q_k=\theta(k-k_\mathrm{max})$, where
$\theta(x)$ is the Heaviside step function, then we remove all vertices
with degrees greater than $k_\mathrm{max}$.  This corresponds precisely to
the experiment of Broder~\etal~\cite{Broder00} who looked at the behavior
of the World Wide Web graph as vertices were removed in order of decreasing
degree.  (Similar but not identical calculations were also performed by
Albert~\etal~\cite{AJB00}.)  In agreement with the numerical calculations
(see Sec.~\ref{resilience}), Callaway~\etal\ find that networks with
power-law degree distributions are highly susceptible to this type of
targeted attack; one need only remove a small percentage of vertices to
destroy the giant component entirely.  Similar results were also found
independently by Cohen~\etal~\cite{CEBH01}, using a closely similar method,
and in a later paper~\cite{Schwartz02} some of the same authors extended
their calculations to directed networks also, which show a considerably
richer component structure, as described in Sec.~\ref{dirrg}.

\begin{figure}
\resizebox{8cm}{!}{\includegraphics{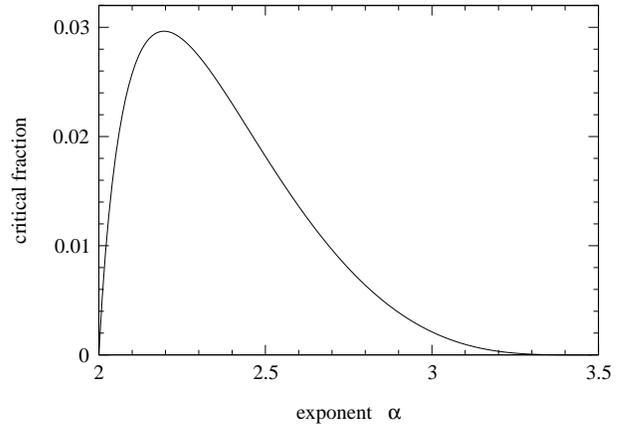}}
\caption{The fraction of vertices that must be removed from a network to
destroy the giant component, if the network has the form of a configuration
model with a power-law degree distribution of exponent~$\alpha$, and
vertices are removed in decreasing order of their degrees.}
\label{fc}
\end{figure}

As an example, consider Fig.~\ref{fc}, which shows the fraction of the
highest degree vertices that must be removed from a network with a
power-law degree distribution to destroy the giant component, as a function
of the exponent~$\alpha$ of the power law~\cite{DM01d,Newman03b}.  As the
figure shows, the maximum fraction is less than three percent, and for most
values of~$\alpha$ the fraction is significantly less than this.  This
appears to imply that networks like the Internet and the Web that have
power-law degree distributions are highly susceptible to such
attacks~\cite{Broder00,AJB00,CEBH01}.

These results are for the configuration model.  Other models offer some
further insights.  The finding by Cohen~\etal~\cite{CEBH00} that the
threshold value~$q_c$ at which percolation sets in for the configuration
model is zero for degree distributions with a divergent second moment has
attracted particular interest.  Vazquez and Moreno~\cite{VM03}, for
example, have shown that the threshold may be zero even for finite second
moment if the degrees of adjacent vertices in the network are positively
correlated (see Secs.~\ref{degcorr} and~\ref{dcrg}).  Conversely, if the
second moment does diverge there may still be a non-zero threshold if there
are negative degree correlations.  Warren~\etal~\cite{WSS02} have shown
that there can also be a non-zero threshold for a network incorporating
geographical effects, in which each vertex occupies a position in a
low-dimensional space (typically two-dimensional) and probability of
connection is higher for vertex pairs that are close together in that
space.  A similar spatial model has been studied by
Rozenfeld~\etal~\cite{RCBH02}, and both models are closely related to
continuum percolation~\cite{MR96}.

An issue related to resilience to vertex deletion, is the issue of
cascading failures.  In some networks, such as electrical power networks,
that carry load or distribute a resource, the operation of the network is
such that the failure of one vertex or edge results in the redistribution
of the load on that vertex or edge to other nearby vertices or edges.  If
vertices or edges fail when the load on them exceeds some maximum capacity,
then this mechanism can result in a cascading failure or avalanche in which
the redistribution of load pushes a vertex or edge over its threshold and
causes it to fail, leading to further redistribution.  Such a cascading
failure in the western United States in August 1996 resulted in the spread
of what was initially a small power outage in El Paso, Texas through six
states as far as Oregon and California, leaving several million electricity
customers without power.  Watts~\cite{Watts02} has given a simple model of
this process that can be mapped onto a type of percolation model and hence
can be solved using generating function methods similar to those for simple
vertex removal processes above.

In Watts's model, a vertex~$i$ fails if a given fraction $\phi_i$ of its
neighbors have failed, where the quantities $\set{\phi_i}$ are iid
variables drawn from a distribution~$f(\phi)$.  The model is seeded by the
initial failure of some non-zero density $\Phi_0$ of vertices, chosen
uniformly at random.  It is assumed that $\Phi_0\ll1$, so that the initial
seed consists, to leading order, of single isolated vertices.  Watts
considers networks with the topology of the configuration model
(Sec.~\ref{confmodel}), for which, because of the vanishing density of
short loops making the networks tree-like at small length-scales, each
vertex will have at most only a single failed neighboring vertex in the
initial stages of the cascade, and hence will fail itself if and only if
its threshold for failure satisfies $\phi<1/k$, where $k$ is its degree.
Watts calls vertices satisfying this criterion
\defn{vulnerable}.  The probability of a vertex being vulnerable is $q_k =
\int_0^{1/k} f(\phi) \>\d\phi$, and the cascade will spread only if such
vertices connect to form a percolating (i.e.,~extensive) cluster on the
network.  Thus the problem maps directly onto the generalized percolation
process studied by Callaway~\etal~\cite{CNSW00} above, allowing us to find
a condition for the spread of the initial seed to give a large-scale
cascade.  The percolation model applies only to the vulnerable vertices
however, so to calculate the final sizes of cascades Watts performs
numerical simulations.

Models of cascading failure have also been studied by Holme and
Kim~\cite{HK02a,Holme02b}, by Moreno~\etal~\cite{MGP02,MPVV03} and by
Motter and Lai~\cite{ML02}.  In the model of Holme and Kim, for instance,
load on a vertex is quantified by the betweenness centrality of the vertex
(see Sec.~\ref{otherprop}), and vertices fail when the betweenness exceeds
a given threshold.  Holme and Kim give simulation results for the avalanche
size distribution in their model.

\subsection{Epidemiological processes}
\label{episec}
One of the original, and still primary, reasons for studying networks is to
understand the mechanisms by which diseases and other things (information,
computer viruses, rumors) spread over them.  For instance, the main reason
for the study of networks of sexual
contact~\cite{GAM89,Klovdahl94,Morris97,FG00,Liljeros01,RBTM01,BMS02,LEA03,JH03}
(Sec.~\ref{socnet}) is to help us understand and perhaps control the spread
of sexually transmitted diseases.  Similarly one studies networks of email
contact~\cite{EMB02,NFB02} to learn how computer viruses
spread.\footnote{Computer viruses are an interesting case in that the
networks over which they spread are normally directed, unlike the contact
networks for most human diseases~\cite{KW91}.}

\subsubsection{The SIR model}
\label{secsir}
The simplest model of the spread of a disease over a network is the SIR
model of epidemic disease~\cite{Bailey75,AM91,Hethcote00}.\footnote{One
distinguishes between an \defn{epidemic} disease such as influenza, which
sweeps through the population rapidly and infects a significant fraction of
individuals in a short outbreak, and an \defn{endemic} disease such as
measles, which persists within the population at a level roughly constant
over time.  The SIR model is a model of the former.  The SIS model
discussed in Sec.~\ref{secsis} is a model of the latter.}  This model,
first formulated, though never published, by Lowell Reed and Wade Hampton
Frost in the 1920s, divides the population into three classes:
susceptible~(S), meaning they don't have the disease of interest but can
catch it if exposed to someone who does, infective\footnote{In everyday
parlance the more common word is ``infectious,'' but infective is the
standard term among epidemiologists.} (I) meaning they have the disease
and can pass it on, and recovered~(R), meaning they have recovered from the
disease and have permanent immunity, so that they can never get it again or
pass it on.  (Some authors consider the R to stand for ``removed,'' a
general term that encompasses also the possibility that people may die of
the disease and remove themselves from the infective pool in that fashion.
Others consider the R to mean ``refractory,'' which is the common term
among those who study the closely related area of reaction diffusion
processes~\cite{Winfree00,Strogatz94}.)

In traditional mathematical epidemiology~\cite{Bailey75,AM91,Hethcote00},
one then assumes that any susceptible individual has a uniform
probability~$\beta$ per unit time of catching the disease from any
infective one and that infective individuals recover and become immune at
some stochastically constant rate~$\gamma$.  The fractions $s$, $i$ and~$r$
of individuals in the states S, I and~R are then governed by the
differential equations
\begin{equation}
{\d s\over\d t} = -\beta i s,\qquad
{\d i\over\d t} = \beta i s - \gamma i,\qquad
{\d r\over\d t} = \gamma i.
\label{sir}
\end{equation}
Models of this type are called \defn{fully mixed}, and although they have
taught us much about the basic dynamics of diseases, they are obviously
unrealistic in their assumptions.  In reality diseases can only spread
between those individuals who have actual physical contact of one sort or
another, and the structure of the contact network is important to the
pattern of development of the disease.

The SIR model can be generalized in a straightforward manner to an epidemic
taking place on a network, although the resulting dynamical system is
substantially more complicated than its fully mixed counterpart.  The
important observation that allows us to make progress, first made by
Grassberger~\cite{Grassberger83}, is that the model can be mapped exactly
onto bond percolation on the same network.  Indeed, as pointed out by
Sander~\etal~\cite{Sander02}, significantly more general models can also be
mapped to percolation, in which transmission probability between pairs of
individuals and the times for which individuals remain infective both vary,
but are chosen in iid fashion from some appropriate distributions.  Let us
suppose that the distribution of infection rates~$\beta$, defined as the
probability per unit time that an infective individual will pass the
disease onto a particular susceptible network neighbor, is drawn from a
distribution~$P_i(\beta)$.  And suppose that the recovery rate~$\gamma$ is
drawn from another distribution~$P_r(\gamma)$.  Then the resulting model
can be shown~\cite{Newman02c} to be equivalent to uniform bond percolation
on the same network with edge occupation probability
\begin{equation}
T = 1 - \int_0^\infty P_i(\beta) P_r(\gamma)\,\e^{-\beta/\gamma}
    \>\d\beta\>\d\gamma.
\label{defstcont}
\end{equation}

The extraction of predictions about epidemics from the percolation model is
simple: the distribution of percolation clusters (i.e.,~components
connected by occupied edges) corresponds to the distribution of the sizes
of disease outbreaks that start with a randomly chosen initial carrier, the
percolation transition corresponds to the ``epidemic threshold'' of
epidemiology, above which an epidemic outbreak is possible (i.e.,~one that
infects a non-zero fraction of the population in the limit of large system
size), and the size of the giant component above this transition
corresponds to the size of the epidemic.  What the mapping cannot tell us,
but standard epidemiological models can, is the time progression of a
disease outbreak.  The mapping gives us results only for the ultimate
outcome of the disease in the limit of long times, in which all individuals
are in either the S or R states, and no new cases of the disease are
occurring.  Nonetheless, there is much to be learned by studying even the
non-time-varying properties of the model.

The solution of bond percolation for the configuration model was given by
Callaway~\etal~\cite{CNSW00}, who showed that, for uniform edge occupation
probability~$T$, the distribution of the sizes of clusters (i.e.,~disease
outbreaks in epidemiological language) is generated by the function
$H_0(x)$ where
\begin{subequations}
\begin{eqnarray}
H_0(x) &=& x G_0(H_1(x)),\\
H_1(x) &=& 1 - T + T x G_1(H_1(x)),
\end{eqnarray}
\end{subequations}
where $G_0(x)$ and $G_1(x)$ are defined in Eqs.~\eref{defsg0g1}.  This
gives an epidemic transition that takes place at $T_c=1/G_1'(1)$, a mean
outbreak size~$\av{s}$ given by
\begin{equation}
\av{s} = H_0'(1) = T\left[1+\frac{TG_0'(1)}{1-TG_1'(1)}\right],
\end{equation}
and an epidemic outbreak that affects a fraction~$S$ of the network, where
\begin{equation}
S = 1 - G_0(u),\qquad u = 1 - T + T G_1(u).
\end{equation}
Similar solutions can be found for a wide variety of other model networks,
including networks with correlations of various kinds between the rates of
infection or the infectivity times~\cite{Newman02c}, networks with
correlations between the degrees of vertices~\cite{MV03}, and networks
with more complex structure, such as different types of
vertices~\cite{Newman02c,ANMS02}.

One of the most important conclusions of this work is for the case of
networks with power-law degree distributions, for which, as in the case of
site percolation (Sec.~\ref{siteperc}), there is no non-zero epidemic
threshold so long as the exponent of the power law is less than~3.  Since
most power-law networks satisfy this condition, we expect diseases always
to propagate in these networks, regardless of transmission probability
between individuals, a point that was first made, in the context of models
of computer virus epidemiology, by Pastor-Satorras and
Vespignani~\cite{PV01a,PV03}, although, as pointed out by Lloyd and
May~\cite{LM01a,ML01}, precursors of the same result can be seen in earlier
work of May and Anderson~\cite{MA88}.  May and Anderson studied traditional
(fully mixed) differential equation models of epidemics, without network
structure, but they divided the population into activity classes with
different values of the infection rate~$\beta$.  They showed that the
variation of the number of infective individuals over time depends on the
variance of this rate over the classes, and in particular that the disease
always multiplies exponentially if the variance diverges---precisely the
situation in a network with a power-law degree distribution and exponent
less than~3.

The conclusion that diseases always spread on scale-free networks has been
revised somewhat in the light of later discoveries.  In particular, there
may be a non-zero percolation threshold for certain types of correlations
between vertices~\cite{BCK02,BP02a,BPV02,VM03,MV03,BPV03}, if the network
is embedded in a low-dimensional (rather than infinite-dimensional)
space~\cite{RCBH02,WSS02}, or if the network has high
transitivity~\cite{EK02} (see Sec.~\ref{transitivity}).

An interesting combination of the ideas of epidemiology with those of
network resilience explored in the preceding section arises when one
considers vaccination of a population against the spread of a disease.
Vaccination can be regarded as the removal from a network of some
particular set of vertices, and this in turn can be modeled as a site
percolation process.  Thus one is led to consideration of joint site/bond
percolation on networks, which has also been solved, in the simplest
uniformly random case, by Callaway~\etal~\cite{CNSW00}.  If the site
percolation is correlated with vertex degree (as in Eq.~\eref{defsf0f1} and
following), for example removing the vertices with highest degree, then one
has a model for targeted vaccination strategies also.  A good discussion
has been given by Pastor-Satorras and Vespignani~\cite{PV02a}.  As with the
models of Sec.~\ref{siteperc}, one finds that networks tend to be
particularly vulnerable to removal of their highest degree vertices, so
this kind of targeted vaccination is expected to be particularly effective.
(This of course is not news to the public health community, who have long
followed a policy of focusing their most aggressive disease prevention
efforts on the ``core communities'' of high-degree vertices in a network.)

Unfortunately, it is not always easy to find the highest degree vertices in
a social network.  The number of sexual contacts a person has had can
normally only be found by asking them, and perhaps not even then.  An
interesting method that circumvents this problem has been suggested by
Cohen~\etal~\cite{CBH02}.  They observe that since the probability of
reaching a particular vertex by following a randomly chosen edge in a graph
is proportional to the vertex's degree (Sec.~\ref{genrg}), one is more
likely to find high-degree vertices by following edges than by choosing
vertices at random.  They propose thus that a population can be immunized
by choosing a random person from that population and vaccinating a
\emph{friend} of that person, and then repeating the process.  They show
both by analytic calculations and by computer simulation that this strategy
is substantially more effective than random vaccination.  In a sense, in
fact, this strategy is already in use.  The ``contact tracing''
methods~\cite{KVS96} used to control sexually transmitted diseases, and the
``ring vaccination'' method~\cite{Greenhalgh86,MSK00} used to control
smallpox and foot-and-mouth disease are both examples of roughly this type
of acquaintance vaccination.

\subsubsection{The SIS model}
\label{secsis}
Not all diseases confer immunity on their survivors.  Diseases that, for
instance, are not self-limiting but can be cured by medicine, can usually
be caught again immediately by an unlucky patient.  Tuberculosis and
gonorrhea are two much-studied examples.  Computer viruses also fall into
this category; they can be ``cured'' by anti-virus software, but without a
permanent virus-checking program the computer has no way to fend off
subsequent attacks by the same virus.

With diseases of this kind carriers that are cured move from the infective
pool not to a recovered pool, but back into the susceptible one.  A model
with this type of dynamics is called an SIS model, for obvious reasons.  In
the simplest, fully mixed, single-population case, its dynamics are
described by the differential equations
\begin{equation}
{\d s\over\d t} = -\beta i s + \gamma i,\qquad
{\d i\over\d t} = \beta i s - \gamma i,
\label{sis}
\end{equation}
where $\beta$ and $\gamma$ are, as before, the infection and recovery
rates.

The SIS model is a model of endemic disease.  Since carriers can be
infected many times, it is possible, and does happen in some parameter
regimes, that the disease will persist indefinitely, circulating around the
population and never dying out.  The equivalent of the SIR epidemic
transition is the phase boundary between the parameter regimes in which the
disease persists and those in which it does not.

The SIS model cannot be solved exactly on a network as the SIR model can,
but a detailed mean-field treatment has been given by Pastor-Satorras and
Vespignani~\cite{PV01a,PV01b} for SIS epidemics on the configuration model.
Their approach is based on the differential equations, Eq.~\eref{sis}, but
they allow the rate of infection~$\beta$ to vary between members of the
population, rather than holding it constant.  (This is similar to the
approach of May and Anderson~\cite{MA88} for the SIR model, discussed in
Sec.~\ref{secsir}, but is more general, since it does not involve the
division of the population into a binned set of activity classes, as the
May--Anderson approach does.)  The calculation proceeds as follows.

The quantity $\beta i$ appearing in~\eref{sis} represents the average rate
at which susceptible individuals become infected by their neighbors.  For
a vertex of degree~$k$, Pastor-Satorras and Vespignani make the replacement
$\beta i\to k\lambda\Theta(\lambda)$, where $\lambda$ is the rate of
infection via contact with a single infective individual and
$\Theta(\lambda)$ is the probability that the neighbor at the other end of
an edge will in fact be infective.  Note that $\Theta$ is a function of
$\lambda$ since presumably the probability of being infective will increase
as the probability of passing on the disease increases.  The remaining
occurrences of the variables $s$ and $i$ Pastor-Satorras and Vespignani
replace by $s_k$ and $i_k$, which are degree-dependent generalizations
representing the fraction of vertices of degree~$k$ that are susceptible or
infective.  Then, noticing that $i_k$ and $s_k$ obey $i_k+s_k=1$, we can
rewrite~\eref{sis} as the single differential equation
\begin{equation}
{\d i_k\over\d t} = k\lambda \Theta(\lambda) (1-i_k) - i_k,
\label{ik}
\end{equation}
where we have, without loss of generality, set the recovery rate~$\gamma$
equal to~1.  There is an approximation inherent in this formulation, since
we have assumed that $\Theta(\lambda)$ is the same for all vertices, when
in general it too will be dependent on vertex degree.  This is in the
nature of a mean-field approximation, and can be expected to give a
reasonable guide to the qualitative behavior of the system, although
certain properties (particularly close to the phase transition) may be
quantitatively mispredicted.

Looking for stationary solutions, we find
\begin{equation}
i_k = {k\lambda\Theta(\lambda)\over1+k\lambda\Theta(\lambda)}.
\end{equation}
To calculate the value of $\Theta(\lambda)$, one averages the probability
$i_k$ of being infected over all vertices.  Since $\Theta(\lambda)$ is
defined as the probability that the vertex at the end of an edge is
infective, $i_k$~should be averaged over the distribution $kp_k/z$ of the
degrees of such vertices (see Sec.~\ref{confmodel}), where $z=\sum_k k p_k$
is, as usual, the mean degree.  Thus
\begin{equation}
\Theta(\lambda) = {1\over z} \sum_k k p_k i_k.
\label{theta}
\end{equation}
Eliminating $i_k$ from Eqs.~\eref{ik} and~\eref{theta} we then obtain an
implicit expression for~$\Theta(\lambda)$:
\begin{equation}
{\lambda\over z} \sum_k {k^2 p_k\over1+k\lambda\Theta(\lambda)} = 1.
\end{equation}

For particular choices of $p_k$ this equation can be solved for
$\Theta(\lambda)$ either exactly or approximately.  For instance, for a
power-law degree distribution of the form~\eref{powerlaw}, Pastor-Satorras
and Vespignani solve it by making an integral approximation, and hence show
that there is no non-zero epidemic threshold for the SIS model in the
power-law case---the disease will always persist, regardless of the value
of the infection rate parameter~$\lambda$~\cite{PV01a}.  They have also
generalized the solution to a number of other cases, including other degree
distributions~\cite{PV01b}, finite-sized networks~\cite{PV02b}, and models
that include vaccination of some fraction of individuals~\cite{PV02a,PV03}.
In the latter case, they tackle both random vaccination and vaccination
targeted at the vertices with highest degree using a method similar to that
of Cohen~\etal~\cite{CEBH00} in which they calculate the effective degree
distribution of the network after the removal of a given set of vertices
and then apply their mean-field method to the resulting network.  As we
would expect from the results of Cohen~\etal, propagation of the disease
turns out to be relatively robust against random vaccination, at least in
networks with right-skewed degree distributions, but highly susceptible to
vaccination of the highest-degree individuals.  The mean-field method has
also been applied to networks with degree correlations of the type
discussed in Sec.~\ref{degcorr}, by Bogu\~n\'a~\etal~\cite{BPV02}.  Of
particular note is their finding that for the case of power-law degree
distributions neither assortative nor disassortative mixing by degree can
produce a non-zero epidemic threshold in the SIS model, at least within the
mean-field approximation.  This contrasts with the case for the SIR model,
where it was found that disassortative mixing can produce a non-zero
threshold~\cite{VM03}.

The mean-field method can also be applied to the SIR
model~\cite{Andersson99,MPV02}.  Although we have an exact solution for the
SIR model as described in Sec.~\ref{secsir}, that solution can only tell us
about the long-time behavior of an outbreak---its expected final size and
so forth.  The mean-field method, although approximate, can tell us about
the time evolution of an outbreak, so the two methods are complementary.
The mean-field method for the SIR model can also be used to treat
approximately the effects of network
transitivity~\cite{Andersson99,Keeling99,KG99,FG00}.

\subsection{Search on networks}
\label{search}
Another example of a process taking place on a network that has important
practical applications is network search.  Suppose some resource of
interest is stored at the vertices of a network, such as information on Web
pages, or computer files on a distributed database or file-sharing network.
One would like to determine rapidly where on the network a particular item
of interest can be found (or determine that it is not on the network at
all).  One way of doing this, which is used by Web search engines, is
simply to catalog exhaustively (or ``crawl'') the entire network, creating
a distilled local map of the data found.  Such a strategy is favored in
cases where there is a heavy communication cost to searching the network in
real time, so that it makes sense to create a local index.  While
performing a network crawl is, in principle, straightforward (although in
practice it may be technically very challenging~\cite{BP98}), there are
nonetheless some interesting theoretical questions arising.

\subsubsection{Exhaustive network search}
\label{exhaustive}
One of the triumphs of recent work on networks has been the development of
effective algorithms for mining network crawl data for information of
interest, particularly in the context of the World Wide Web.  The important
trick here turns out to be to use the information contained in the edges of
the network as well as in the vertices.  Since the edges, or hyperlinks, in
the World Wide Web are created by people in order to highlight connections
between the contents of pairs of pages, their structure contains
information about page content and relevance which can help us to improve
search performance.  The good search engines therefore make a local catalog
not only of the contents of web pages, but also of which ones link to which
others.  Then when a query is made of the database, usually in the form of
a textual string of interest, the typical strategy would be to select a
subset of pages from the database by searching for that string, and then to
rank the results using the edge information.  The classic algorithm, due to
Brin and Page~\cite{BP98,PBMW98}, is essentially identical in its simplest
form to the eigenvector centrality long used in social network
analysis~\cite{Bonacich72,Bonacich87,WF94,Scott00}.  Each vertex~$i$ is
assigned a weight $x_i>0$, which is defined to be proportional to the sum
of the weights of all vertices that point to~$i$: $x_i=\lambda^{-1}\sum_j
A_{ij}x_j$ for some $\lambda>0$, or in matrix form
\begin{equation}
\vA\vx = \lambda\vx,
\label{google}
\end{equation}
where $\vA$ is the (asymmetric) adjacency matrix of the graph, whose
elements are $A_{ij}$, and $\vx$ is the vector whose elements are
the~$x_i$.  This of course means that the weights we want are an
eigenvector of the adjacency matrix with eigenvalue~$\lambda$ and, provided
the network is connected (there are no separate components), the
Perron--Frobenius theorem then tells us that there is only one eigenvector
with all weights non-negative, which is the unique eigenvector
corresponding to the largest eigenvalue.  This eigenvector can be found
trivially by repeated multiplication of the adjacency matrix into any
initial non-zero vector which is not itself an eigenvector.

This algorithm, which is implemented (along with many additional tricks) in
the widely used search engine \textit{Google}, appears to be highly
effective.  In essence the algorithm makes the assumption that a page is
important if it is pointed to by other important pages.  A more
sophisticated version of the same idea has been put forward by
Kleinberg~\cite{Kleinberg99a,KL01}, who notes that, since the Web is a
directed network, one can ask not only about which vertices point to a
vertex of interest, but also about which vertices are pointed to by that
vertex.  This then leads to two different weights $x_i$ and $y_i$ for each
vertex.  Kleinberg refers to a vertex that is pointed to by highly ranked
vertices as an authority---it is likely to contain relevant information.
Such a vertex gets a weight~$x_i$ that is large.  A vertex that points to
highly ranked vertices is referred to as a hub; while it may not contain
directly relevant information, it can tell you where to find such
information.  It gets a weight $y_i$ that is large.  (Certainly it is
possible for a vertex to have both weights large; there is no reason why
the same page cannot be both a hub and an authority.)  The appropriate
generalization of Eq.~\eref{google} for the two weights is then
\begin{equation}
\vA\vy = \lambda\vx,\qquad \vA^T\vx = \mu\vy,
\label{kleinberg}
\end{equation}
where $\vA^T$ is the transpose of~$\vA$.  Most often we are interested in
the authority weights which, eliminating $\vy$, obey
$\vA\vA^T\vx=\lambda\mu\vx$, so that the primary difference between the
method of Brin and Page~\cite{BP98} and the method of Kleinberg is the
replacement of the adjacency matrix with the symmetric product~$\vA\vA^T$.
More general forms than~\eref{kleinberg} are also possible.  One could for
example allow the authority weight of a vertex to depend on the authority
weights of the vertices that point to it (and not just their hub weights,
as in Eq.~\eref{kleinberg}).  This leads to a model that interpolates
smoothly between the Brin--Page and Kleinberg methods.  As far as we are
aware however, this has not been tried.  Neither has Kleinberg's method
been implemented yet in a commercial web search engine, to the best of our
knowledge.

The methods described here can also be used for search on other directed
information networks.  Kleinberg's method is be particularly suitable for
ranking publications in citation networks, for example.  The
\textit{Citeseer} literature search engine implements a form of article
ranking of this type.

\subsubsection{Guided network search}
\label{guided}
An alternative approach to searching a network is to perform a guided
search.  Guided search strategies may be appropriate for certain kinds of
Web search, particularly searches for specialized content that could be
missed by generic search engines (whose coverage tends to be quite poor),
and also for searching on other types of networks such as distributed
databases.  Exhaustive search of the type discussed in the preceding
section crawls a network once to create an index of the data found, which
is then stored and searched locally.  Guided searches perform small
special-purpose crawls for every search query, crawling only a small
fraction of the network, but doing so in an intelligent fashion that
deliberately seeks out the network vertices most likely to contain relevant
information.

One practical example of a guided search is the specialized Web crawler or
``spider'' of Menczer~\etal~\cite{MB00,MPRS01}.  This is a program that
performs a Web crawl to find results for a particular query.  The method
used is a type of genetic algorithm~\cite{Mitchell96} or enrichment
method~\cite{Grassberger02} that in its simplest form has a number of
``agents'' that start crawling the Web at random, looking for pages that
contain, for example, particular words or sets of words given by the user.
Agents are ranked according to their success at finding matches to the
words of interest and those that are least successful are killed off.
Those that are most successful are duplicated so that the density of agents
will be high in regions of the Web graph that contain many pages that look
promising.  After some specified amount of time has passed, the search is
halted and a list of the most promising pages found so far is presented to
the user.  The method relies for its success on the assumption that pages
that contain information on a particular topic tend to be clustered
together in local regions of the graph.  Other than this however, the
algorithm makes little use of statistical properties of the structure of
the graph.

Adamic~\etal~\cite{ALPH01,ALH03} have given a completely different
algorithm that directly exploits network structure and is designed for use
on peer-to-peer networks.  Their algorithm makes use of the skewed degree
distribution of most networks to find the desired results quickly.  It
works as follows.

Simple breadth-first search can be thought of as a query that starts from a
single source vertex on a network.  The query goes out to all neighbors of
the source vertex and says, ``Have you got the information I am looking
for?''  Each neighbor either replies ``Yes, I have it,'' in which case the
search is over, or ``No, I don't, but I have forwarded your request to all
of my neighbors.''  Each of their neighbors, when they receive the request,
either recognizes it as one they have seen before, in which case they
discard it, or they repeat the process as above.  A query of this kind
takes aggregate effort $\ord(n)$ in the network size.  Adamic~\etal\
propose to modify this algorithm as follows.  The initial source vertex
again queries each of its neighbors for the desired information.  But now
the reply is either ``Yes, I have it'' or ``No, I don't, and I have $k$
neighbors,'' where $k$ is the degree of the vertex in question.  Upon
receiving replies of the latter type from each of its neighbors, the source
vertex finds which of its neighbors has the highest value of $k$ and passes
the responsibility for the query like a runner's baton to that neighbor,
who then repeats the entire process with their neighbors.  (If the
highest-degree vertex has already handled the query in the past, then the
second highest is chosen, and so forth; complete recursive back-tracking is
used to make sure the algorithm never gets stuck in a dead end.)

The upshot of this strategy is that the baton gets passed rapidly up a
chain of increasing vertex degree until it reaches the highest degree
vertices in the network.  On networks with highly skewed degree
distributions, particularly scale-free (i.e.,~power-law) networks, the
neighbors of the high-degree vertices account for a significant fraction of
all the vertices in the network.  On average therefore, we need only go a
few steps along the chain before we find a vertex with a neighbor that has
the information we are looking for.  The maximum degree on a scale-free
network scales with network size as $n^{1/(\alpha-1)}$ (see
Sec.~\ref{highest}), and hence the number of steps required to search
$\ord(n)$ vertices is of order $n/n^{1/(\alpha-1)} =
n^{(\alpha-2)/(\alpha-1)}$, which lies between $\ord(n^{1/2})$ and
$\ord(\log n)$ for $2\le\alpha\le3$, which is the range generally observed
in power-law networks (see Table~\ref{summary}).  This is a significant
improvement over the $\ord(n)$ of the simple breadth-first search,
especially for the smaller values of~$\alpha$.

This result differs from that given by Adamic~\etal~\cite{ALPH01,ALH03},
who adopted the more conservative assumption that the maximum degree goes
as $n^{1/\alpha}$~\cite{ACL00}, which gives significantly poorer search
times between $\ord(n^{2/3})$ and $\ord(n^{1/2})$.  They point out however
that if each vertex to which the baton passes is allowed to query not only
its immediate network neighbors but also its second neighbors, then the
performance improves markedly to $\ord(n^{2(1-2/\alpha)})$.

The algorithm of Adamic~\etal\ has been tested numerically on graphs with
the structure of the configuration model~\cite{ALH03}
(Sec.~\ref{confmodel}) and the Barab\'asi--Albert preferential attachment
model~\cite{KYHJ02,ALH03} (Sec.~\ref{bamodel}), and shows behavior in
reasonable agreement with the expected scaling forms.

The reader might be forgiven for feeling that these algorithms are cheating
a little, since the running time of the algorithm is measured by the number
of hands the baton passes through.  If one measures it in terms of the
number of queries that must be responded to by network vertices, then the
algorithm is still~$\ord(n)$, just as the simple breadth-first search is.
Adamic~\etal\ suggest that each vertex therefore keep a local directory or
index of the information (such as data files) stored at neighboring
vertices, so that queries concerning those vertices can be resolved
locally.  For distributed databases and file sharing networks, where
bandwidth, in terms of communication overhead between vertices, is the
costly resource, this strategy really does improve scaling with network
size, reducing overhead per query to $\ord(\log n)$ in the best case.

\subsubsection{Network navigation}
\label{greedy}
The work of Adamic~\etal~\cite{ALPH01,ALH03} discussed in the preceding
section considers how one can design a network search algorithm to exploit
statistical features of network structure to improve performance.  A
complementary question has been considered by
Kleinberg~\cite{Kleinberg00,Kleinberg00proc}: Can one design network
structures to make a particular search algorithm perform well?  Kleinberg's
work is motivated by the observation, discussed in Sec.~\ref{navigation},
that people are able to navigate social networks efficiently with only
local information about network structure.  Furthermore, this ability does
not appear to depend on any particularly sophisticated behavior on the
part of the people.  When performing the letter-passing task of
Milgram~\cite{Milgram67,TM69}, for instance, in which participants are
asked to communicate a letter or message to a designated target person by
passing it through their acquaintance network (Sec.~\ref{socnet}), the
search for the target is performed, roughly speaking, using a simple
``greedy algorithm.''  That is, at each step along the way the letter is
passed to the person that the current holder believes to be closest to the
target.  (This in fact is precisely how participants were instructed to act
in Milgram's experiments.)  The fact that the letter often reaches the
target in only a short time then indicates that the network itself must
have some special properties, since the search algorithm clearly doesn't.

Kleinberg suggested a simple model that illustrates this behavior.  His
model is a variant of the small-world model of Watts and
Strogatz~\cite{WS98,Watts99a} (Sec.~\ref{sw}) in which shortcuts are added
between pairs of sites on a regular lattice (a square lattice in
Kleinberg's studies).  Rather than adding these shortcuts uniformly at
random as Watts and Strogatz proposed, Kleinberg adds them in a biased
fashion, with shortcuts more likely to fall between lattice sites that are
close together in the Euclidean space defined by the lattice.  The
probability of a shortcut falling between two sites goes as~$r^{-\alpha}$,
where $r$ is the distance between the sites and $\alpha$ is a constant.
Kleinberg proves a lower bound on the mean time~$t$ (i.e.,~number of steps)
taken by the greedy algorithm to find a randomly chosen target on such a
network.  His bound is $t\ge cn^\beta$ where $c$ is independent of~$n$ and
\begin{equation}
\beta = \biggl\lbrace\begin{array}{ll}
          (2-\alpha)/3                    & \mbox{for $0\le\alpha<2$} \\
          (\alpha-2)/(\alpha-1) \qquad    & \mbox{for $\alpha>2$.}
        \end{array}
\label{jkmodel}
\end{equation}
Thus the best performance of the algorithm is when $\alpha$ is close
to~$2$, and precisely at $\alpha=2$ the greedy algorithm should be capable
of finding the target in $\ord(\log n)$ steps.  Kleinberg also gave
computer simulation results confirming this result.  More generally, for
networks built on an underlying lattice in $d$ dimensions, the optimal
performance of the greedy algorithm occurs at
$\alpha=d$~\cite{Kleinberg00,Kleinberg00proc}.  (See also
Ref.~\citen{Higham02} for some rigorous results on the performance of
greedy algorithms on Watts--Strogatz type networks.)

\begin{figure}
\resizebox{7.5cm}{!}{\includegraphics{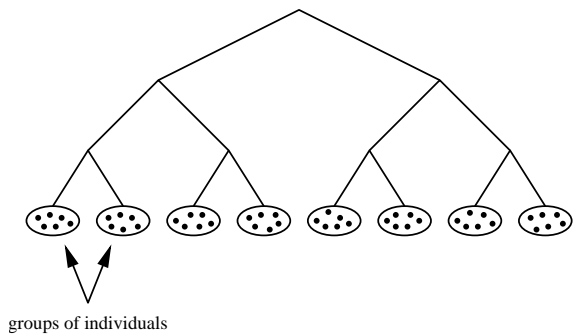}}
\caption{The hierarchical ``social distance'' tree proposed by
Watts~\etal~\cite{WDN02} and by Kleinberg~\cite{Kleinberg02}.  Individuals
are grouped together by occupation, location, interest, etc., and then
those groups are grouped together into bigger groups and so forth.  The
social distance between two individuals is measured by how far one must go
up the tree to find the lowest ``common ancestor'' of the pair.}
\label{tree}
\end{figure}

Kleinberg's work shows that many networks do not allow fast search using a
simple algorithm such as a greedy algorithm, but that it is possible to
design networks that do allow such fast search.  The particular model he
studies however is quite specialized, and certainly not a good
representation of the real social networks that inspired his
investigations.  An alternative model that shows similar behavior to
Kleinberg's, but which may shed more light on the true structure of social
networks, has been proposed by Watts~\etal~\cite{WDN02} and independently
by Kleinberg~\cite{Kleinberg02}.  The ``index'' experiments of Killworth
and Bernard~\cite{KB78,Bernard88} indicate that people in fact navigate
social networks by looking for common features between their acquaintances
and the target, such as geographical location or occupation.  This suggests
a model in which individuals are grouped (at least in the participants
minds) into categories according, for instance, to their jobs.  These
categories may then themselves be grouped in to supercategories, and so
forth, creating a tree-like hierarchy of organization that defines a
``social distance'' between any two people: the social distance between two
individuals is measured by the height of lowest level in tree at which the
two are connected---see Fig.~\ref{tree}.

The tree however is not the network, it is merely a mental construct that
affects the way the network grows.  It is assumed that the probability of
their being an edge between two vertices is greater the shorter the social
distance between those vertices, and both Watts~\etal~\cite{WDN02} and
Kleinberg~\cite{Kleinberg02} assumed that this probability falls off
exponentially with social distance.  The greedy algorithm for communicating
a message to a target person then specifies that the message should at each
step be passed to that network neighbor of the current holder who has the
shortest social distance to the target.  Watts~\etal\ showed by computer
simulation that such an algorithm performs well over a broad range of
parameters of the model, and Kleinberg showed that for appropriate
parameter choices the search can be completed in time again~$\ord(\log n)$.

While this model is primarily a model of search on social networks (or
possibly the Web~\cite{Kleinberg02}), Watts~\etal\ also suggested that it
could be used as a model for designed networks.  If one could arrange for
items in a distributed database to be grouped hierarchically according to
some identifiable characteristics, then a greedy algorithm that is aware of
those characteristics should be able to find a desired element in the
database quickly, possibly in time only logarithmic in the size of the
database.  This idea has been studied in more detail by
Iamnitchi~\etal~\cite{IRF02} and Arenas~\etal~\cite{Arenas03}.

One disadvantage of the hierarchical organizational model is that in
reality the categories into which network vertices fall almost certainly
overlap, whereas in the hierarchical model they are disjoint.  Kleinberg
has proposed a generalization of the model that allows for overlapping
categories and shows search behavior qualitatively similar to the
hierarchical model~\cite{Kleinberg02}.

\subsection{Phase transitions on networks}
\label{phasetrans}
Another group of papers has dealt with the behavior on networks of
traditional statistical mechanical models that show phase transitions.  For
example, several authors have studied spin models such as the Ising model
on networks of various kinds.  Barrat and Weigt~\cite{BW00} studied the
Ising model on networks with the topology of the small-world
model~\cite{WS98} (see Sec.~\ref{sw}) using replica methods.  They found,
unsurprisingly, that in the limit $n\to\infty$ the model has a
finite-temperature transition for all values of the shortcut density $p>0$.
Further results for Ising models on small-world networks can be found in
Refs.~\citen{Pekalski01,HKC02,KZ02,Herrero02,ZZ03}, and the model has also
been studied on random graphs~\cite{DGM02b,LVVZ02} and on networks with the
topology of the Barab\'asi--Albert growing network
model~\cite{AHS02,Bianconi02} (Sec~\ref{bamodel}).

The motivation behind studies of spin models on networks is usually either
that they can be regarded as simple models of opinion formation in social
networks~\cite{Young03} or that they provide general insight into the
effects of network topology on phase transition processes.  There are
however other more direct approaches to both of these issues.  Opinion
formation can be studied more directly using actual opinion formation
models~\cite{French56,deGroot74,SS00,Stauffer02,VKR03,CVV03}.  And
Goltsev~\etal~\cite{GDM03} have examined phase transition behavior on
networks using the general framework known as Landau theory.  They find
that the critical behavior of models on a network depends in general on
the degree distribution, and is in particular strongly affected by
power-law degree distributions.

One class of networked systems showing a phase transition that is of real
interest is the class of NP-hard computational problems such as
satisfiability and colorability that show solvability transitions.  The
simplest example of such a system is the colorability problem, which is
related to problems in operations research such as scheduling problems and
also to the Potts model of statistical mechanics.  In this problem a number
of items (vertices) are divided into a number of groups (colors).  Some
pairs of vertices cannot be in the same group.  Such a constraint is
represented by placing an edge between those vertices, so that the set of
all constraints forms a graph.  A solution to the problem of satisfying all
constraints simultaneously (if a solution exists) is then equivalent to
finding a coloring of the graph such that no two adjacent vertices have the
same color.  Problems of this type are found to show a phase transition
between a region of low graph density (low ratio of edges to vertices) in
which most graphs are colorable, to one of high density in which most are
not.  A considerable amount of work has been carried out on this and
similar problems in the computer science community~\cite{DGP97}.  However,
this work has primarily been restricted to Poisson random graphs; it is
largely an open question how the results will change when we look at more
realistic network topologies.  Walsh~\cite{Walsh99} has looked at
colorability in the Watts--Strogatz small-world model (Sec.~\ref{sw}), and
found that these networks are easily colorable for both small and large
values of the shortcut density parameter~$p$, but harder to color in
intermediate regimes.  V\'azquez and Weigt~\cite{VW03} examined the related
problem of vertex covers and found that on generalized random graphs
solutions are harder to find for networks with strong degree correlations
of the type discussed in Sec.~\ref{degcorr}.

\subsection{Other processes on networks}
Preliminary investigations, primarily numerical in nature, have been
carried out of the behavior of various other processes on networks.  A
number of authors have looked at diffusion processes.  Random walks, for
example, have been examined by Jespersen~\etal~\cite{JSB00}, Pandit and
Amritkar~\cite{PA01} and Lahtinen~\etal~\cite{LKK01,LKK02}.  Solutions of
the diffusion equation can be expressed as linear combinations of
eigenvectors of the graph Laplacian, which has led a number of authors to
investigate the Laplacian and its eigenvalue
spectrum~\cite{Monasson99,GKK01a,FDBV01}.  Discrete dynamical processes
have also attracted some attention.  One of the earliest examples of a
statistical model of a networked system falls in this category, the random
Boolean net of
Kauffman~\cite{Kauffman69,Kauffman71,Kauffman93,AB00b,CKZ01a,CKZ01b,FH01,SK03,Aldana03},
which is a model of a genetic regulatory network (see Sec.~\ref{bionets}).
Cellular automata on networks have been investigated by Watts and
Strogatz~\cite{WS98,Watts99a}, and voter models and models of opinion
formation can also be regarded as cellular
automata~\cite{KZ02,CVV03,VKR03}.  Iterated games on networks have been
investigated by several authors~\cite{WS98,AK01,Kim02,EB02}, and some
interesting differences are seen between behavior on networks and on
regular lattices.  Other topics of investigation have included weakly
coupled oscillators~\cite{WS98,HCK02,BP02b}, neural
networks~\cite{LHCS00,SADA03}, and self-organized critical
models~\cite{KAS99,DH02,MV02}.  A useful discussion of the behavior of
dynamical systems on networks has been given by Strogatz~\cite{Strogatz01}.

\section{Summary and directions for future research}
\label{concs}
In this article we have reviewed some recent work on the structure and
function of networked systems.  Work in this area has been motivated to a
high degree by empirical studies of real-world networks such as the
Internet, the World Wide Web, social networks, collaboration networks,
citation networks, and a variety of biological networks.  We have reviewed
these empirical studies in Secs.~\ref{data} and~\ref{props}, focusing on a
number of statistical properties of networks that have received particular
attention, including path lengths, degree distributions, clustering, and
resilience.  Quantitative measurements for a variety of networks are
summarized in Table~\ref{summary}.  The most important observation to come
out of studies such as these is that networks are generally very far from
random.  They have highly distinctive statistical signatures, some of
which, such as high clustering coefficients and highly skewed degree
distributions, are common to networks of a wide variety of types.

Inspired by these observations many researchers have proposed models of
networks that typically seek to explain either how networks come to have
the observed structure, or what the expected effects of that structure will
be.  The largest portion of this review has been taken up with discussion
of these models, covering random graph models and their generalizations
(Sec.~\ref{rg}), Markov graphs (Sec.~\ref{markov}), the small-world model
(Sec.~\ref{sw}), and models of network growth, particularly the
preferential attachment models (Sec.~\ref{growing}).

In the last part of this review (Sec.~\ref{onnets}) we have discussed work
on the behavior of processes that take place \emph{on} networks.  The
notable successes in this area so far have been studies of the spread of
infection over networks such as social networks or computer networks, and
studies of the effect of the failure of network nodes on performance of
communications networks.  Some progress has also be made on phase
transitions on networks and on dynamical systems on networks, particularly
discrete dynamical systems.

In looking forward to future developments in this area it is clear that
there is much to be done.  The study of complex networks is still in its
infancy.  Several general areas stand out as promising for future research.
First, while we are beginning to understand some of the patterns and
statistical regularities in the structure of real-world networks, our
techniques for analyzing networks are at present no more than a grab-bag of
miscellaneous and largely unrelated tools.  We do not yet, as we do in some
other fields, have a systematic program for characterizing network
structure.  We count triangles on networks or measure degree sequences, but
we have no idea if these are the only important quantities to measure
(almost certainly they are not) or even if they are the most important.  We
have as yet no theoretical framework to tell us if we are even looking in
the right place.  Perhaps there are other measures, so far un-thought-of,
that are more important than those we have at present.  A true
understanding of which properties of networks are the important ones to
focus on will almost certainly require us to state first what questions we
are interested in answering about a particular network.  And knowing how to
tie the answers to these questions to structural properties of the network
is therefore also an important goal.

Second, there is much to be done in developing more sophisticated models of
networks, both to help us understand network topology and to act as a
substrate for the study of processes taking place on networks.  While some
network properties, such as degree distributions, have been thoroughly
modeled and their causes and effects well understood, others such as
correlations, transitivity, and community structure have not.  It seems
certain that these properties will affect the behavior of networked
systems substantially, so our current lack of suitable techniques to handle
them leaves a large gap in our understanding.

Which leads us to our third and perhaps most important direction for future
study, the behavior of processes taking place on networks.  The work
described in Sec.~\ref{onnets} represents only a few first attempts at
answering questions about such processes, and yet this, in a sense, is our
ultimate goal in this field: to understand the behavior and function of
the networked systems we see around us.  If we can gain such understanding,
it will give us new insight into a vast array of complex and previously
poorly understood phenomena.

\onecolumngrid
\vspace{10mm}
\begin{center}
\rule{10cm}{0.5pt}
\end{center}
\vspace{6mm}
\twocolumngrid


\begin{thebibliography}{100}
\markboth{The structure and function of complex networks}{References}

\bibitem{AK01}
Abramson, G. and Kuperman, M., Social games in a social network, \textit{Phys.
  Rev. E} \textbf{63}, 030901 (2001).

\bibitem{Adamic99}
Adamic, L.~A., The small world web, in \textit{Lecture Notes in Computer
  Science}, vol. 1696, pp. 443--454, Springer, New York (1999).

\bibitem{AA01}
Adamic, L.~A. and Adar, E., Friends and neighbors on the {W}eb, \textit{Social
  Networks}  (in press).

\bibitem{AH00}
Adamic, L.~A. and Huberman, B.~A., Power-law distribution of the world wide
  web, \textit{Science} \textbf{287}, 2115 (2000).

\bibitem{ALH03}
Adamic, L.~A., Lukose, R.~M., and Huberman, B.~A., Local search in unstructured
  networks, in S.~Bornholdt and H.~G. Schuster (eds.), \textit{Handbook of
  Graphs and Networks}, Wiley-VCH, Berlin (2003).

\bibitem{ALPH01}
Adamic, L.~A., Lukose, R.~M., Puniyani, A.~R., and Huberman, B.~A., Search in
  power-law networks, \textit{Phys. Rev. E} \textbf{64}, 046135 (2001).

\bibitem{AMO93}
Ahuja, R.~K., Magnanti, T.~L., and Orlin, J.~B., \textit{Network Flows: Theory,
  Algorithms, and Applications}, Prentice Hall, Upper Saddle River, New Jersey
  (1993).

\bibitem{ACL00}
Aiello, W., Chung, F., and Lu, L., A random graph model for massive graphs, in
  \textit{Proceedings of the 32nd Annual {ACM} Symposium on Theory of
  Computing}, pp. 171--180, Association of Computing Machinery, New York
  (2000).

\bibitem{ACL02}
Aiello, W., Chung, F., and Lu, L., Random evolution of massive graphs, in
  J.~Abello, P.~M. Pardalos, and M.~G.~C. Resende (eds.), \textit{Handbook of
  Massive Data Sets}, pp. 97--122, Kluwer, Dordrecht (2002).

\bibitem{AMR02}
Alberich, R., Miro-Julia, J., and Rossello, F., {M}arvel {U}niverse looks
  almost like a real social network, Preprint cond-mat/0202174 (2002).

\bibitem{AB00b}
Albert, R. and Barab\'asi, A.-L., Dynamics of complex systems: Scaling laws for
  the period of {B}oolean networks, \textit{Phys. Rev. Lett.} \textbf{84},
  5660--5663 (2000).

\bibitem{AB00a}
Albert, R. and Barab\'asi, A.-L., Topology of evolving networks: Local events
  and universality, \textit{Phys. Rev. Lett.} \textbf{85}, 5234--5237 (2000).

\bibitem{AB02}
Albert, R. and Barab\'asi, A.-L., Statistical mechanics of complex networks,
  \textit{Rev. Mod. Phys.} \textbf{74}, 47--97 (2002).

\bibitem{AJB99}
Albert, R., Jeong, H., and Barab\'asi, A.-L., Diameter of the world-wide web,
  \textit{Nature} \textbf{401}, 130--131 (1999).

\bibitem{AJB00}
Albert, R., Jeong, H., and Barab\'asi, A.-L., Attack and error tolerance of
  complex networks, \textit{Nature} \textbf{406}, 378--382 (2000).

\bibitem{Aldana03}
Aldana, M., Dynamics of {B}oolean networks with scale-free topology, Preprint
  cond-mat/0209571 (2002).

\bibitem{AP00}
Aldous, D. and Pittel, B., On a random graph with immigrating vertices:
  Emergence of the giant component, \textit{Random Structures and Algorithms}
  \textbf{17}, 79--102 (2000).

\bibitem{AHS02}
Aleksiejuk, A., Ho{\l}yst, J.~A., and Stauffer, D., Ferromagnetic phase
  transition in {B}arab\'asi--{A}lbert networks, \textit{Physica A}
  \textbf{310}, 260--266 (2002).

\bibitem{AKS02}
Almaas, E., Kulkarni, R.~V., and Stroud, D., Characterizing the structure of
  small-world networks, \textit{Phys. Rev. Lett.} \textbf{88}, 098101 (2002).

\bibitem{ASBS00}
Amaral, L. A.~N., Scala, A., Barth\'el\'emy, M., and Stanley, H.~E., Classes of
  small-world networks, \textit{Proc. Natl. Acad. Sci. USA} \textbf{97},
  11149--11152 (2000).

\bibitem{ANMS02}
Ancel~Meyers, L., Newman, M. E.~J., Martin, M., and Schrag, S., Applying
  network theory to epidemics: Control measures for outbreaks of {M}ycoplasma
  pneumoniae, \textit{Emerging Infectious Diseases} \textbf{9}, 204--210
  (2001).

\bibitem{AWC99}
Anderson, C., Wasserman, S., and Crouch, B., A p* primer: Logit models for
  social networks, \textit{Social Networks} \textbf{21}, 37--66 (1999).

\bibitem{AM91}
Anderson, R.~M. and May, R.~M., \textit{Infectious Diseases of Humans}, Oxford
  University Press, Oxford (1991).

\bibitem{Andersson99}
Andersson, H., Epidemic models and social networks, \textit{Math. Scientist}
  \textbf{24}, 128--147 (1999).

\bibitem{Arenas03}
Arenas, A., Cabrales, A., D{\'\i}az-Guilera, A., Guimer\`a, R., and
  Vega-Redondo, F., Search and congestion in complex networks, in
  R.~Pastor-Satorras and J.~Rubi (eds.), \textit{Proceedings of the XVIII
  Sitges Conference on Statistical Mechanics}, Lecture Notes in Physics,
  Springer, Berlin (2003).

\bibitem{Bailey75}
Bailey, N. T.~J., \textit{The Mathematical Theory of Infectious Diseases and
  Its Applications}, Hafner Press, New York (1975).

\bibitem{BU89}
Baird, D. and Ulanowicz, R.~E., The seasonal dynamics of the {C}hesapeake {B}ay
  ecosystem, \textit{Ecological Monographs} \textbf{59}, 329--364 (1989).

\bibitem{BMS97}
Ball, F., Mollison, D., and Scalia-Tomba, G., Epidemics with two levels of
  mixing, \textit{Annals of Applied Probability} \textbf{7}, 46--89 (1997).

\bibitem{BMR99}
Banavar, J.~R., Maritan, A., and Rinaldo, A., Size and form in efficient
  transportation networks, \textit{Nature} \textbf{399}, 130--132 (1999).

\bibitem{BC96}
Banks, D.~L. and Carley, K.~M., Models for network evolution, \textit{Journal
  of Mathematical Sociology} \textbf{21}, 173--196 (1996).

\bibitem{Barabasi02a}
Barab\'asi, A.-L., \textit{Linked: The New Science of Networks}, Perseus,
  Cambridge, MA (2002).

\bibitem{BA99b}
Barab\'asi, A.-L. and Albert, R., Emergence of scaling in random networks,
  \textit{Science} \textbf{286}, 509--512 (1999).

\bibitem{BAJ99}
Barab\'asi, A.-L., Albert, R., and Jeong, H., Mean-field theory for scale-free
  random networks, \textit{Physica A} \textbf{272}, 173--187 (1999).

\bibitem{BAJ00}
Barab\'asi, A.-L., Albert, R., and Jeong, H., Scale-free characteristics of
  random networks: The topology of the {W}orld {W}ide {W}eb, \textit{Physica A}
  \textbf{281}, 69--77 (2000).

\bibitem{BAJB00}
Barab\'asi, A.-L., Albert, R., Jeong, H., and Bianconi, G., Power-law
  distribution of the {W}orld {W}ide {W}eb, \textit{Science} \textbf{287},
  2115a (2000).

\bibitem{Barabasi02b}
Barab\'asi, A.-L., Jeong, H., Ravasz, E., N\'eda, Z., Schuberts, A., and
  Vicsek, T., Evolution of the social network of scientific collaborations,
  \textit{Physica A} \textbf{311}, 590--614 (2002).

\bibitem{BP02b}
Barahona, M. and Pecora, L.~M., Synchronization in small-world systems,
  \textit{Phys. Rev. Lett.} \textbf{89}, 054101 (2002).

\bibitem{BR01}
Barbour, A.~D. and Reinert, G., Small worlds, \textit{Random Structures and
  Algorithms} \textbf{19}, 54--74 (2001).

\bibitem{Barrat99}
Barrat, A., Comment on `{S}mall-world networks: Evidence for crossover
  picture', Preprint cond-mat/9903323 (1999).

\bibitem{BW00}
Barrat, A. and Weigt, M., On the properties of small-world networks,
  \textit{Eur. Phys. J. B} \textbf{13}, 547--560 (2000).

\bibitem{BA99c}
Barth\'el\'emy, M. and Amaral, L. A.~N., Erratum: Small-world networks:
  Evidence for a crossover picture, \textit{Phys. Rev. Lett.} \textbf{82}, 5180
  (1999).

\bibitem{BA99a}
Barth\'el\'emy, M. and Amaral, L. A.~N., Small-world networks: Evidence for a
  crossover picture, \textit{Phys. Rev. Lett.} \textbf{82}, 3180--3183 (1999).

\bibitem{BM00}
Batagelj, V. and Mrvar, A., Some analyses of {E}rd{\H{o}}s collaboration graph,
  \textit{Social Networks} \textbf{22}, 173--186 (2000).

\bibitem{BB02}
Bauer, M. and Bernard, D., A simple asymmetric evolving random network,
  Preprint cond-mat/0203232 (2002).

\bibitem{BMS02}
Bearman, P.~S., Moody, J., and Stovel, K., Chains of affection: The structure
  of adolescent romantic and sexual networks, Preprint, Department of
  Sociology, Columbia University (2002).

\bibitem{BBK72}
Bekessy, A., Bekessy, P., and Komlos, J., Asymptotic enumeration of regular
  matrices, \textit{Stud. Sci. Math. Hungar.} \textbf{7}, 343--353 (1972).

\bibitem{BC78}
Bender, E.~A. and Canfield, E.~R., The asymptotic number of labeled graphs with
  given degree sequences, \textit{Journal of Combinatorial Theory A}
  \textbf{24}, 296--307 (1978).

\bibitem{BL02}
Berg, J. and L{\"a}ssig, M., Correlated random networks, \textit{Phys. Rev.
  Lett.} \textbf{89}, 228701 (2002).

\bibitem{BLW03}
Berg, J., L{\"a}ssig, M., and Wagner, A., Evolution dynamics of protein
  networks, Preprint cond-mat/0207711 (2002).

\bibitem{Bernard88}
Bernard, H.~R., Killworth, P.~D., Evans, M.~J., McCarty, C., and Shelley,
  G.~A., Studying social relations cross-culturally, \textit{Ethnology}
  \textbf{2}, 155--179 (1988).

\bibitem{Bianconi02}
Bianconi, G., Mean field solution of the {I}sing model on a
  {B}arab\'asi--{A}lbert network, Preprint cond-mat/0204455 (2002).

\bibitem{BB01a}
Bianconi, G. and Barab\'asi, A.-L., {B}ose--{E}instein condensation in complex
  networks, \textit{Phys. Rev. Lett.} \textbf{86}, 5632--5635 (2001).

\bibitem{BB01b}
Bianconi, G. and Barab\'asi, A.-L., Competition and multiscaling in evolving
  networks, \textit{Europhys. Lett.} \textbf{54}, 436--442 (2001).

\bibitem{BC03}
Bianconi, G. and Capocci, A., Number of loops of size h in growing scale-free
  networks, \textit{Phys. Rev. Lett.} \textbf{90}, 078701 (2003).

\bibitem{BP01}
Bilke, S. and Peterson, C., Topological properties of citation and metabolic
  networks, \textit{Phys. Rev. E} \textbf{64}, 036106 (2001).

\bibitem{BCK02}
Blanchard, P., Chang, C.-H., and Kr{\"u}ger, T., Epidemic thresholds on
  scale-free graphs: The interplay between exponent and preferential choice,
  Preprint cond-mat/0207319 (2002).

\bibitem{BP02a}
Bogu{\~n}\'a, M. and Pastor-Satorras, R., Epidemic spreading in correlated
  complex networks, \textit{Phys. Rev. E} \textbf{66}, 047104 (2002).

\bibitem{BPV02}
Bogu{\~n}\'a, M., Pastor-Satorras, R., and Vespignani, A., Absence of epidemic
  threshold in scale-free networks with connectivity correlations, Preprint
  cond-mat/0208163 (2002).

\bibitem{BPV03}
Bogu{\~n}\'a, M., Pastor-Satorras, R., and Vespignani, A., Epidemic spreading
  in complex networks with degree correlations, in R.~Pastor-Satorras and
  J.~Rubi (eds.), \textit{Proceedings of the XVIII Sitges Conference on
  Statistical Mechanics}, Lecture Notes in Physics, Springer, Berlin (2003).

\bibitem{Bollobas80}
Bollob\'as, B., A probabilistic proof of an asymptotic formula for the number
  of labelled regular graphs, \textit{European Journal on Combinatorics}
  \textbf{1}, 311--316 (1980).

\bibitem{Bollobas81}
Bollob\'as, B., The diameter of random graphs, \textit{Trans. Amer. Math. Soc.}
  \textbf{267}, 41--52 (1981).

\bibitem{Bollobas98}
Bollob\'as, B., \textit{Modern Graph Theory}, Springer, New York (1998).

\bibitem{Bollobas01}
Bollob\'as, B., \textit{Random Graphs}, Academic Press, New York, 2nd ed.
  (2001).

\bibitem{BR02}
Bollob\'as, B. and Riordan, O., The diameter of a scale-free random graph,
  Preprint, Department of Mathematical Sciences, University of Memphis (2002).

\bibitem{BRST01}
Bollob\'as, B., Riordan, O., Spencer, J., and Tusn\'ady, G., The degree
  sequence of a scale-free random graph process, \textit{Random Structures and
  Algorithms} \textbf{18}, 279--290 (2001).

\bibitem{Bonacich72}
Bonacich, P.~F., A technique for analyzing overlapping memberships, in
  H.~Costner (ed.), \textit{Sociological Methodology}, Jossey-Bass, San
  Francisco (1972).

\bibitem{Bonacich87}
Bonacich, P.~F., Power and centrality: A family of measures, \textit{Am. J.
  Sociol.} \textbf{92}, 1170--1182 (1987).

\bibitem{BG00}
Bordens, M. and G\'omez, I., Collaboration networks in science, in H.~B. Atkins
  and B.~Cronin (eds.), \textit{The Web of Knowledge: A Festschrift in Honor of
  Eugene Garfield}, Information Today, Medford, NJ (2000).

\bibitem{BE01}
Bornholdt, S. and Ebel, H., {W}orld {W}ide {W}eb scaling exponent from
  {S}imon's 1955 model, \textit{Phys. Rev. E} \textbf{64}, 035104 (2001).

\bibitem{BS03}
Bornholdt, S. and Schuster, H.~G. (eds.), \textit{Handbook of Graphs and
  Networks}, Wiley-VCH, Berlin (2003).

\bibitem{BBA75}
Breiger, R.~L., Boorman, S.~A., and Arabie, P., An algorithm for clustering
  relations data with applications to social network analysis and comparison
  with multidimensional scaling, \textit{Journal of Mathematical Psychology}
  \textbf{12}, 328--383 (1975).

\bibitem{BP98}
Brin, S. and Page, L., The anatomy of a large-scale hypertextual {W}eb search
  engine, \textit{Computer Networks} \textbf{30}, 107--117 (1998).

\bibitem{BH57}
Broadbent, S.~R. and Hammersley, J.~M., Percolation processes: {I}. {C}rystals
  and mazes, \textit{Proc. Cambridge Philos. Soc.} \textbf{53}, 629--641
  (1957).

\bibitem{Broder00}
Broder, A., Kumar, R., Maghoul, F., Raghavan, P., Rajagopalan, S., Stata, R.,
  Tomkins, A., and Wiener, J., Graph structure in the web, \textit{Computer
  Networks} \textbf{33}, 309--320 (2000).

\bibitem{BC01}
Broida, A. and Claffy, K.~C., Internet topology: Connectivity of {IP} graphs,
  in S.~Fahmy and K.~Park (eds.), \textit{Scalability and Traffic Control in IP
  Networks}, no. 4526 in Proc. SPIE, pp. 172--187, International Society for
  Optical Engineering, Bellingham, WA (2001).

\bibitem{Buchanan02}
Buchanan, M., \textit{Nexus: Small Worlds and the Groundbreaking Science of
  Networks}, Norton, New York (2002).

\bibitem{BCK01}
Burda, Z., Correia, J.~D., and Krzywicki, A., Statistical ensemble of
  scale-free random graphs, \textit{Phys. Rev. E} \textbf{64}, 046118 (2001).

\bibitem{CCDM02}
Caldarelli, G., Capocci, A., De~Los~Rios, P., and Mu{\~n}oz, M.~A., Scale-free
  networks from varying vertex intrinsic fitness, \textit{Phys. Rev. Lett.}
  \textbf{89}, 258702 (2002).

\bibitem{CPV03}
Caldarelli, G., Pastor-Satorras, R., and Vespignani, A., Cycles structure and
  local ordering in complex networks, Preprint cond-mat/0212026 (2002).

\bibitem{Callaway01}
Callaway, D.~S., Hopcroft, J.~E., Kleinberg, J.~M., Newman, M. E.~J., and
  Strogatz, S.~H., Are randomly grown graphs really random?, \textit{Phys. Rev.
  E} \textbf{64}, 041902 (2001).

\bibitem{CNSW00}
Callaway, D.~S., Newman, M. E.~J., Strogatz, S.~H., and Watts, D.~J., Network
  robustness and fragility: Percolation on random graphs, \textit{Phys. Rev.
  Lett.} \textbf{85}, 5468--5471 (2000).

\bibitem{CGA02}
Camacho, J., Guimer\`a, R., and Amaral, L. A.~N., Robust patterns in food web
  structure, \textit{Phys. Rev. Lett.} \textbf{88}, 228102 (2002).

\bibitem{CCD03}
Capocci, A., Caldarelli, G., and De~Los~Rios, P., Quantitative description and
  modeling of real networks, Preprint cond-mat/0206336 (2002).

\bibitem{CVV03}
Castellano, C., Vilone, D., and Vespignani, A., Incomplete ordering of the
  voter model on small-world networks, Preprint cond-mat/0210465 (2002).

\bibitem{CCKF92}
Catania, J.~A., Coates, T.~J., Kegels, S., and Fullilove, M.~T., The
  population-based {AMEN} ({AIDS} in {M}ulti-{E}thnic {N}eighborhoods) study,
  \textit{Am. J. Public Health} \textbf{82}, 284--287 (1992).

\bibitem{Chen02}
Chen, Q., Chang, H., Govindan, R., Jamin, S., Shenker, S.~J., and Willinger,
  W., The origin of power laws in {I}nternet topologies revisited, in
  \textit{Proceedings of the 21st Annual Joint Conference of the {IEEE}
  Computer and Communications Societies}, IEEE Computer Society (2002).

\bibitem{CHE02}
Chowell, G., Hyman, J.~M., and Eubank, S., Analysis of a real world network:
  The {C}ity of {P}ortland, Technical Report BU-1604-M, Department of
  Biological Statistics and Computational Biology, Cornell University (2002).

\bibitem{CL02b}
Chung, F. and Lu, L., The average distances in random graphs with given
  expected degrees, \textit{Proc. Natl. Acad. Sci. USA} \textbf{99},
  15879--15882 (2002).

\bibitem{CL02a}
Chung, F. and Lu, L., Connected components in random graphs with given degree
  sequences, \textit{Annals of Combinatorics} \textbf{6}, 125--145 (2002).

\bibitem{CLDG03}
Chung, F., Lu, L., Dewey, T.~G., and Galas, D.~J., Duplication models for
  biological networks, \textit{Journal of Computational Biology}  (in press).

\bibitem{CBN90}
Cohen, J.~E., Briand, F., and Newman, C.~M., \textit{Community food webs: Data
  and theory}, Springer, New York (1990).

\bibitem{CBH02}
Cohen, R., {ben-Avraham}, D., and Havlin, S., Efficient immunization of
  populations and computers, Preprint cond-mat/0207387 (2002).

\bibitem{CEBH00}
Cohen, R., Erez, K., {ben-Avraham}, D., and Havlin, S., Resilience of the
  {I}nternet to random breakdowns, \textit{Phys. Rev. Lett.} \textbf{85},
  4626--4628 (2000).

\bibitem{CEBH01}
Cohen, R., Erez, K., {ben-Avraham}, D., and Havlin, S., Breakdown of the
  {I}nternet under intentional attack, \textit{Phys. Rev. Lett.} \textbf{86},
  3682--3685 (2001).

\bibitem{CH03}
Cohen, R. and Havlin, S., Scale-free networks are ultrasmall, \textit{Phys.
  Rev. Lett.} \textbf{90}, 058701 (2003).

\bibitem{CHB99}
Connor, R.~C., Heithaus, M.~R., and Barre, L.~M., Superalliance of bottlenose
  dolphins, \textit{Nature} \textbf{397}, 571--572 (1999).

\bibitem{CKZ01a}
Coppersmith, S.~N., Kadanoff, L.~P., and Zhang, Z., Reversible {B}oolean
  networks: {I}. {D}istribution of cycle lengths, \textit{Physica D}
  \textbf{149}, 11--29 (2000).

\bibitem{CKZ01b}
Coppersmith, S.~N., Kadanoff, L.~P., and Zhang, Z., Reversible {B}oolean
  networks: {II}. {P}hase transition, oscillation, and local structures,
  \textit{Physica D} \textbf{157}, 54--74 (2001).

\bibitem{CKMD02}
Corman, S.~R., Kuhn, T., Mcphee, R.~D., and Dooley, K.~J., Studying complex
  discursive systems: Centering resonance analysis of organizational
  communication, \textit{Human Communication Research} \textbf{28}, 157--206
  (2002).

\bibitem{CB02}
Coulumb, S. and Bauer, M., Asymmetric evolving random networks, Preprint
  cond-mat/0212371 (2002).

\bibitem{Crane72}
Crane, D., \textit{Invisible colleges: Diffusion of knowledge in scientific
  communities}, University of Chicago Press, Chicago (1972).

\bibitem{DEB02}
Davidsen, J., Ebel, H., , and Bornholdt, S., Emergence of a small world from
  local interactions: Modeling acquaintance networks, \textit{Phys. Rev. Lett.}
  \textbf{88}, 128701 (2002).

\bibitem{DGG41}
Davis, A., Gardner, B.~B., and Gardner, M.~R., \textit{Deep South}, University
  of Chicago Press, Chicago (1941).

\bibitem{DG97}
Davis, G.~F. and Greve, H.~R., Corporate elite networks and governance changes
  in the 1980s, \textit{Am. J. Sociol.} \textbf{103}, 1--37 (1997).

\bibitem{DYB01}
Davis, G.~F., Yoo, M., and Baker, W.~E., The small world of the corporate
  elite, Preprint, University of Michigan Business School (2001).

\bibitem{DH02}
de~Arcangelis, L. and Herrmann, H.~J., Self-organized criticality on small
  world networks, \textit{Physica A} \textbf{308}, 545--549 (2002).

\bibitem{DG99}
de~Castro, R. and Grossman, J.~W., Famous trails to {P}aul {E}rd{\H{o}}s,
  \textit{Mathematical Intelligencer} \textbf{21}, 51--63 (1999).

\bibitem{deGroot74}
de~Groot, M.~H., Reaching a consensus, \textit{J. Amer. Statist. Assoc.}
  \textbf{69}, 118--121 (1974).

\bibitem{MMP00}
de~Menezes, M.~A., Moukarzel, C., and Penna, T. J.~P., First-order transition
  in small-world networks, \textit{Europhys. Lett.} \textbf{50}, 574--579
  (2000).

\bibitem{DMW03}
Dodds, P.~S., Muhamad, R., and Watts, D.~J., An experiment study of social
  search and the small world problem, Preprint, Department of Sociology,
  Columbia University (2002).

\bibitem{DR01abc}
Dodds, P.~S. and Rothman, D.~H., Geometry of river networks, \textit{Phys. Rev.
  E} \textbf{63}, 016115, 016116, \& 016117 (2001).

\bibitem{DGM02b}
Dorogovtsev, S.~N., Goltsev, A.~V., and Mendes, J. F.~F., Ising model on
  networks with an arbitrary distribution of connections, \textit{Phys. Rev. E}
  \textbf{66}, 016104 (2002).

\bibitem{DGM02a}
Dorogovtsev, S.~N., Goltsev, A.~V., and Mendes, J. F.~F., Pseudofractal
  scale-free web, \textit{Phys. Rev. E} \textbf{65}, 066122 (2002).

\bibitem{DM00c}
Dorogovtsev, S.~N. and Mendes, J. F.~F., Evolution of networks with aging of
  sites, \textit{Phys. Rev. E} \textbf{62}, 1842--1845 (2000).

\bibitem{DM00a}
Dorogovtsev, S.~N. and Mendes, J. F.~F., Exactly solvable small-world network,
  \textit{Europhys. Lett.} \textbf{50}, 1--7 (2000).

\bibitem{DM00b}
Dorogovtsev, S.~N. and Mendes, J. F.~F., Scaling behaviour of developing and
  decaying networks, \textit{Europhys. Lett.} \textbf{52}, 33--39 (2000).

\bibitem{DM01d}
Dorogovtsev, S.~N. and Mendes, J. F.~F., Comment on ``{B}reakdown of the
  {I}nternet under intentional attack'', \textit{Phys. Rev. Lett.} \textbf{87},
  219801 (2001).

\bibitem{DM01a}
Dorogovtsev, S.~N. and Mendes, J. F.~F., Effect of the accelerating growth of
  communications networks on their structure, \textit{Phys. Rev. E}
  \textbf{63}, 025101 (2001).

\bibitem{DM01b}
Dorogovtsev, S.~N. and Mendes, J. F.~F., Language as an evolving word web,
  \textit{Proc. R. Soc. London B} \textbf{268}, 2603--2606 (2001).

\bibitem{DM02}
Dorogovtsev, S.~N. and Mendes, J. F.~F., Evolution of networks,
  \textit{Advances in Physics} \textbf{51}, 1079--1187 (2002).

\bibitem{DM03a}
Dorogovtsev, S.~N. and Mendes, J. F.~F., Accelerated growth of networks, in
  S.~Bornholdt and H.~G. Schuster (eds.), \textit{Handbook of Graphs and
  Networks}, pp. 318--341, Wiley-VCH, Berlin (2003).

\bibitem{DM03b}
Dorogovtsev, S.~N. and Mendes, J. F.~F., \textit{Evolution of Networks: From
  Biological Nets to the Internet and WWW}, Oxford University Press, Oxford
  (2003).

\bibitem{DMS00}
Dorogovtsev, S.~N., Mendes, J. F.~F., and Samukhin, A.~N., Structure of growing
  networks with preferential linking, \textit{Phys. Rev. Lett.} \textbf{85},
  4633--4636 (2000).

\bibitem{DMS01b}
Dorogovtsev, S.~N., Mendes, J. F.~F., and Samukhin, A.~N., Anomalous
  percolation properties of growing networks, \textit{Phys. Rev. E}
  \textbf{64}, 066110 (2001).

\bibitem{DMS01a}
Dorogovtsev, S.~N., Mendes, J. F.~F., and Samukhin, A.~N., Giant strongly
  connected component of directed networks, \textit{Phys. Rev. E} \textbf{64},
  025101 (2001).

\bibitem{DMS01c}
Dorogovtsev, S.~N., Mendes, J. F.~F., and Samukhin, A.~N., Size-dependent
  degree distribution of a scale-free growing network, \textit{Phys. Rev. E}
  \textbf{63}, 062101 (2001).

\bibitem{DMS03b}
Dorogovtsev, S.~N., Mendes, J. F.~F., and Samukhin, A.~N., Metric structure of
  random networks, Preprint cond-mat/0210085 (2002).

\bibitem{DMS03a}
Dorogovtsev, S.~N., Mendes, J. F.~F., and Samukhin, A.~N., Principles of
  statistical mechanics of random networks, Preprint cond-mat/0204111 (2002).

\bibitem{DS03}
Dorogovtsev, S.~N. and Samukhin, A.~N., Mesoscopics and fluctuations in
  networks, Preprint cond-mat/0211518 (2002).

\bibitem{Doye02}
Doye, J. P.~K., Network topology of a potential energy landscape: A static
  scale-free network, \textit{Phys. Rev. Lett.} \textbf{88}, 238701 (2002).

\bibitem{DGP97}
Du, D., Gu, J., and Pardalos, P.~M. (eds.), \textit{Satisfiability Problem:
  Theory and Applications}, no.~35 in DIMACS Series in Discrete Mathematics and
  Theoretical Computer Science, American Mathematical Society, Providence, RI
  (1997).

\bibitem{DWM02a}
Dunne, J.~A., Williams, R.~J., and Martinez, N.~D., Food-web structure and
  network theory: The role of connectance and size, \textit{Proc. Natl. Acad.
  Sci. USA} \textbf{99}, 12917--12922 (2002).

\bibitem{DWM02b}
Dunne, J.~A., Williams, R.~J., and Martinez, N.~D., Network structure and
  biodiversity loss in food webs: Robustness increases with connectance,
  \textit{Ecology Letters} \textbf{5}, 558--567 (2002).

\bibitem{Durrett03}
Durrett, R.~T., Rigorous results for the
  {C}allaway--{H}opcroft--{K}leinberg--{N}ewman--{S}trogatz model, Preprint,
  Department of Mathematics, Cornell University (2003).

\bibitem{EB02}
Ebel, H. and Bornholdt, S., Co-evolutionary games on networks, \textit{Phys.
  Rev. E} \textbf{66}, 056118 (2002).

\bibitem{EMB02}
Ebel, H., Mielsch, L.-I., and Bornholdt, S., Scale-free topology of e-mail
  networks, \textit{Phys. Rev. E} \textbf{66}, 035103 (2002).

\bibitem{EM02}
Eckmann, J.-P. and Moses, E., Curvature of co-links uncovers hidden thematic
  layers in the world wide web, \textit{Proc. Natl. Acad. Sci. USA}
  \textbf{99}, 5825--5829 (2002).

\bibitem{ER90}
Egghe, L. and Rousseau, R., \textit{Introduction to Informetrics}, Elsevier,
  Amsterdam (1990).

\bibitem{EK02}
Eguiluz, V.~M. and Klemm, K., Epidemic threshold in structured scale-free
  networks, \textit{Phys. Rev. Lett.} \textbf{89}, 108701 (2002).

\bibitem{ES79}
Eigen, M. and Schuster, P., \textit{The Hypercycle: A Principle of Natural
  Self-Organization}, Springer, New York (1979).

\bibitem{ER59}
Erd\H{o}s, P. and R\'enyi, A., On random graphs, \textit{Publicationes
  Mathematicae} \textbf{6}, 290--297 (1959).

\bibitem{ER60}
Erd\H{o}s, P. and R\'enyi, A., On the evolution of random graphs,
  \textit{Publications of the Mathematical Institute of the Hungarian Academy
  of Sciences} \textbf{5}, 17--61 (1960).

\bibitem{ER61}
Erd\H{o}s, P. and R\'enyi, A., On the strength of connectedness of a random
  graph, \textit{Acta Mathematica Scientia Hungary} \textbf{12}, 261--267
  (1961).

\bibitem{Ergun02}
Erg{\"u}n, G., Human sexual contact network as a bipartite graph,
  \textit{Physica A} \textbf{308}, 483--488 (2002).

\bibitem{ER02}
Erg{\"u}n, G. and Rodgers, G.~J., Growing random networks with fitness,
  \textit{Physica A} \textbf{303}, 261--272 (2002).

\bibitem{ESMS03}
Eriksen, K.~A., Simonsen, I., Maslov, S., and Sneppen, K., Modularity and
  extreme edges of the {I}nternet, Preprint cond-mat/0212001 (2002).

\bibitem{Everitt74}
Everitt, B., \textit{Cluster Analysis}, John Wiley, New York (1974).

\bibitem{FFF99}
Faloutsos, M., Faloutsos, P., and Faloutsos, C., On power-law relationships of
  the internet topology, \textit{Computer Communications Review} \textbf{29},
  251--262 (1999).

\bibitem{FS64}
Fararo, T.~J. and Sunshine, M., \textit{A Study of a Biased Friendship
  Network}, Syracuse University Press, Syracuse (1964).

\bibitem{FDBV01}
Farkas, I.~J., Der\'enyi, I., Barab\'asi, A.-L., and Vicsek, T., Spectra of
  ``real-world'' graphs: Beyond the semicircle law, \textit{Phys. Rev. E}
  \textbf{64}, 026704 (2001).

\bibitem{Farkas02}
Farkas, I.~J., Der\'enyi, I., Jeong, H., Neda, Z., Oltvai, Z.~N., Ravasz, E.,
  Schurbert, A., Barab\'asi, A.-L., and Vicsek, T., Networks in life: Scaling
  properties and eigenvalue spectra, \textit{Physica A} \textbf{314}, 25--34
  (2002).

\bibitem{Farkas03}
Farkas, I.~J., Jeong, H., Vicsek, T., Barab\'asi, A.-L., and Oltvai, Z.~N., The
  topology of the transcription regulatory network in the yeast,
  {S}accharomyces cerevisiae, \textit{Physica A} \textbf{381}, 601--612 (2003).

\bibitem{FW00}
Fell, D.~A. and Wagner, A., The small world of metabolism, \textit{Nature
  Biotechnology} \textbf{18}, 1121--1122 (2000).

\bibitem{FG00}
Ferguson, N.~M. and Garnett, G.~P., More realistic models of sexually
  transmitted disease transmission dynamics: Sexual partnership networks, pair
  models, and moment closure, \textit{Sex. Transm. Dis.} \textbf{27}, 600--609
  (2000).

\bibitem{FJS01}
Ferrer~i Cancho, R., Janssen, C., and Sol\'e, R.~V., Topology of technology
  graphs: Small world patterns in electronic circuits, \textit{Phys. Rev. E}
  \textbf{64}, 046119 (2001).

\bibitem{FS01b}
Ferrer~i Cancho, R. and Sol\'e, R.~V., Optimization in complex networks,
  Preprint cond-mat/0111222 (2001).

\bibitem{FS01a}
Ferrer~i Cancho, R. and Sol\'e, R.~V., The small world of human language,
  \textit{Proc. R. Soc. London B} \textbf{268}, 2261--2265 (2001).

\bibitem{FLGC02}
Flake, G.~W., Lawrence, S.~R., Giles, C.~L., and Coetzee, F.~M.,
  Self-organization and identification of {W}eb communities, \textit{IEEE
  Computer} \textbf{35}, 66--71 (2002).

\bibitem{FH01}
Fox, J.~J. and Hill, C.~C., From topology to dynamics in biochemical networks,
  \textit{Chaos} \textbf{11}, 809--815 (2001).

\bibitem{FS86}
Frank, O. and Strauss, D., Markov graphs, \textit{J. Amer. Stat. Assoc.}
  \textbf{81}, 832--842 (1986).

\bibitem{Freeman77}
Freeman, L., A set of measures of centrality based upon betweenness,
  \textit{Sociometry} \textbf{40}, 35--41 (1977).

\bibitem{Freeman96}
Freeman, L.~C., Some antecedents of social network analysis,
  \textit{Connections} \textbf{19}, 39--42 (1996).

\bibitem{French56}
French, J. R.~P., A formal theory of social power, \textit{Psychological
  Review} \textbf{63}, 181--194 (1956).

\bibitem{FFH03}
Fronczak, A., Fronczak, P., and Holyst, J.~A., Exact solution for average path
  length in random graphs, Preprint cond-mat/0212230 (2002).

\bibitem{FHJS02}
Fronczak, A., Holyst, J.~A., Jedynak, M., and Sienkiewicz, J., Higher order
  clustering coefficients in {B}arabasi-{A}lbert networks, \textit{Physica A}
  \textbf{316}, 688--694 (2002).

\bibitem{GLS00}
Gafiychuk, V., Lubashevsky, I., and Stosyk, A., Remarks on scaling properties
  inherent to the systems with hierarchically organized supplying network,
  Preprint nlin/0004033 (2000).

\bibitem{Galaskiewicz85}
Galaskiewicz, J., \textit{Social Organization of an Urban Grants Economy},
  Academic Press, New York (1985).

\bibitem{GM78}
Galaskiewicz, J. and Marsden, P.~V., Interorganizational resource networks:
  Formal patterns of overlap, \textit{Social Science Research} \textbf{7},
  89--107 (1978).

\bibitem{Garfield79}
Garfield, E., It's a small world after all, \textit{Current Contents}
  \textbf{43}, 5--10 (1979).

\bibitem{GGM02}
Garfinkel, I., Glei, D.~A., and McLanahan, S.~S., Assortative mating among
  unmarried parents, \textit{Journal of Population Economics} \textbf{15},
  417--432 (2002).

\bibitem{GN02}
Girvan, M. and Newman, M. E.~J., Community structure in social and biological
  networks, \textit{Proc. Natl. Acad. Sci. USA} \textbf{99}, 8271--8276 (2002).

\bibitem{GSWF01}
Gleiss, P.~M., Stadler, P.~F., Wagner, A., and Fell, D.~A., Relevant cycles in
  chemical reaction networks, \textit{Advances in Complex Systems} \textbf{4},
  207--226 (2001).

\bibitem{GKK01a}
Goh, K.-I., Kahng, B., and Kim, D., Spectra and eigenvectors of scale-free
  networks, \textit{Phys. Rev. E} \textbf{64}, 051903 (2001).

\bibitem{GKK01b}
Goh, K.-I., Kahng, B., and Kim, D., Universal behavior of load distribution in
  scale-free networks, \textit{Phys. Rev. Lett.} \textbf{87}, 278701 (2001).

\bibitem{Goh02}
Goh, K.-I., Oh, E., Jeong, H., Kahng, B., and Kim, D., Classification of
  scale-free networks, \textit{Proc. Natl. Acad. Sci. USA} \textbf{99},
  12583--12588 (2002).

\bibitem{GNOT92}
Goldberg, D., Nichols, D., Oki, B.~M., and Terry, D., Using collaborative
  filtering to weave an information tapestry, \textit{Comm. ACM} \textbf{35},
  61--70 (1992).

\bibitem{GR93}
Goldwasser, L. and Roughgarden, J., Construction and analysis of a large
  {C}aribbean food web, \textit{Ecology} \textbf{74}, 1216--1233 (1993).

\bibitem{GDM03}
Goltsev, A.~V., Dorogovtsev, S.~N., and Mendes, J. F.~F., Critical phenomena in
  networks, \textit{Phys. Rev. E} \textbf{67}, 026123 (2003).

\bibitem{Grassberger83}
Grassberger, P., On the critical behavior of the general epidemic process and
  dynamical percolation, \textit{Math. Biosci.} \textbf{63}, 157--172 (1983).

\bibitem{Grassberger02}
Grassberger, P., Go with the winners: A general {M}onte {C}arlo strategy,
  \textit{Computer Physics Communications} \textbf{147}, 64--70 (2002).

\bibitem{Greenhalgh86}
Greenhalgh, D., Optimal control of an epidemic by ring vaccination,
  \textit{Communications in Statistics: Stochastic Models} \textbf{2}, 339--363
  (1986).

\bibitem{GI95}
Grossman, J.~W. and Ion, P. D.~F., On a portion of the well-known collaboration
  graph, \textit{Congressus Numerantium} \textbf{108}, 129--131 (1995).

\bibitem{Guare90}
Guare, J., \textit{Six Degrees of Separation: A Play}, Vintage, New York
  (1990).

\bibitem{GBBK02}
Guelzim, N., Bottani, S., Bourgine, P., and Kepes, F., Topological and causal
  structure of the yeast transcriptional regulatory network, \textit{Nature
  Genetics} \textbf{31}, 60--63 (2002).

\bibitem{Guimera03}
Guimer\`a, R., Danon, L., D\'\i{}az-Guilera, A., Giralt, F., and Arenas, A.,
  Self-similar community structure in organisations, Preprint cond-mat/0211498
  (2002).

\bibitem{GAM89}
Gupta, S., Anderson, R.~M., and May, R.~M., Networks of sexual contacts:
  Implications for the pattern of spread of {HIV}, \textit{AIDS} \textbf{3},
  807--817 (1989).

\bibitem{Hammersley57}
Hammersley, J.~M., Percolation processes: {II}. {T}he connective constant,
  \textit{Proc. Cambridge Philos. Soc.} \textbf{53}, 642--645 (1957).

\bibitem{Harary95}
Harary, F., \textit{Graph Theory}, Perseus, Cambridge, MA (1995).

\bibitem{Hayes00a}
Hayes, B., Graph theory in practice: {P}art {I}, \textit{American Scientist}
  \textbf{88}~(1), 9--13 (2000).

\bibitem{Hayes00b}
Hayes, B., Graph theory in practice: {P}art {II}, \textit{American Scientist}
  \textbf{88}~(2), 104--109 (2000).

\bibitem{Herrero02}
Herrero, C.~P., Ising model in small-world networks, \textit{Phys. Rev. E}
  \textbf{65}, 066110 (2002).

\bibitem{Hethcote00}
Hethcote, H.~W., The mathematics of infectious diseases, \textit{SIAM Review}
  \textbf{42}, 599--653 (2000).

\bibitem{Higham02}
Higham, D.~J., Greedy pathlengths and small world graphs, Mathematics Research
  Report~8, University of Strathclyde (2002).

\bibitem{HL81}
Holland, P.~W. and Leinhardt, S., An exponential family of probability
  distributions for directed graphs, \textit{J. Amer. Stat. Assoc.}
  \textbf{76}, 33--65 (1981).

\bibitem{Holme02b}
Holme, P., Edge overload breakdown in evolving networks, \textit{Phys. Rev. E}
  \textbf{66}, 036119 (2002).

\bibitem{HEL02}
Holme, P., Edling, C.~R., and Liljeros, F., Structure and time-evolution of the
  {I}nternet community pussokram.com, Preprint cond-mat/0210514 (2002).

\bibitem{HHJ02}
Holme, P., Huss, M., and Jeong, H., Subnetwork hierarchies of biochemical
  pathways, Preprint cond-mat/0206292 (2002).

\bibitem{HK02b}
Holme, P. and Kim, B.~J., Growing scale-free networks with tunable clustering,
  \textit{Phys. Rev. E} \textbf{65}, 026107 (2002).

\bibitem{HK02a}
Holme, P. and Kim, B.~J., Vertex overload breakdown in evolving networks,
  \textit{Phys. Rev. E} \textbf{65}, 066109 (2002).

\bibitem{Holme02a}
Holme, P., Kim, B.~J., Yoon, C.~N., and Han, S.~K., Attack vulnerability of
  complex networks, \textit{Phys. Rev. E} \textbf{65}, 056109 (2002).

\bibitem{HCK02}
Hong, H., Choi, M.~Y., and Kim, B.~J., Synchronization on small-world networks,
  \textit{Phys. Rev. E} \textbf{65}, 026139 (2002).

\bibitem{HKC02}
Hong, H., Kim, B.~J., and Choi, M.~Y., Comment on ``{I}sing model on a small
  world network,'' \textit{Phys. Rev. E} \textbf{66}, 018101 (2002).

\bibitem{Huberman01}
Huberman, B.~A., \textit{The Laws of the Web}, MIT Press, Cambridge, MA (2001).

\bibitem{HBR96}
Huxham, M., Beaney, S., and Raffaelli, D., Do parasites reduce the chances of
  triangulation in a real food web?, \textit{Oikos} \textbf{76}, 284--300
  (1996).

\bibitem{IRF02}
Iamnitchi, A., Ripeanu, M., and Foster, I., Locating data in (small-world?)\
  peer-to-peer scientific collaborations, in P.~Druschel, F.~Kaashoek, and
  A.~Rowstron (eds.), \textit{Proceedings of the First International Workshop
  on Peer-to-Peer Systems}, no. 2429 in Lecture Notes in Computer Science, pp.
  232--241, Springer, Berlin (2002).

\bibitem{Ito01}
Ito, T., Chiba, T., Ozawa, R., Yoshida, M., Hattori, M., and Sakaki, Y., A
  comprehensive two-hybrid analysis to explore the yeast protein interactome,
  \textit{Proc. Natl. Acad. Sci. USA} \textbf{98}, 4569--4574 (2001).

\bibitem{JT02}
Jaffe, A. and Trajtenberg, M., \textit{Patents, Citations and Innovations: A
  Window on the Knowledge Economy}, MIT Press, Cambridge, MA (2002).

\bibitem{JMF99}
Jain, A.~K., Murty, M.~N., and Flynn, P.~J., Data clustering: A review,
  \textit{ACM Computing Surveys} \textbf{31}, 264--323 (1999).

\bibitem{JK98}
Jain, S. and Krishna, S., Autocatalytic sets and the growth of complexity in an
  evolutionary model, \textit{Phys. Rev. Lett.} \textbf{81}, 5684--5687 (1998).

\bibitem{JK01}
Jain, S. and Krishna, S., A model for the emergence of cooperation,
  interdependence, and structure in evolving networks, \textit{Proc. Natl.
  Acad. Sci. USA} \textbf{98}, 543--547 (2001).

\bibitem{JLR99}
Janson, S., \L{}uczak, T., and Rucinski, A., \textit{Random Graphs}, John
  Wiley, New York (1999).

\bibitem{Jeong01}
Jeong, H., Mason, S., Barab\'asi, A.-L., and Oltvai, Z.~N., Lethality and
  centrality in protein networks, \textit{Nature} \textbf{411}, 41--42 (2001).

\bibitem{JNB03}
Jeong, H., N\'eda, Z., and Barab\'asi, A.-L., Measuring preferential attachment
  in evolving networks, \textit{Europhys. Lett.} \textbf{61}, 567--572 (2003).

\bibitem{Jeong00}
Jeong, H., Tombor, B., Albert, R., Oltvai, Z.~N., and Barab\'asi, A.-L., The
  large-scale organization of metabolic networks, \textit{Nature} \textbf{407},
  651--654 (2000).

\bibitem{JB00}
Jespersen, S. and Blumen, A., Small-world networks: Links with long-tailed
  distributions, \textit{Phys. Rev. E} \textbf{62}, 6270--6274 (2000).

\bibitem{JSB00}
Jespersen, S., Sokolov, I.~M., and Blumen, A., Relaxation properties of
  small-world networks, \textit{Phys. Rev. E} \textbf{62}, 4405--4408 (2000).

\bibitem{JGN01}
Jin, E.~M., Girvan, M., and Newman, M. E.~J., The structure of growing social
  networks, \textit{Phys. Rev. E} \textbf{64}, 046132 (2001).

\bibitem{JH03}
Jones, J.~H. and Handcock, M.~S., An assessment of preferential attachment as a
  mechanism for human sexual network formation, Preprint, University of
  Washington (2003).

\bibitem{JBO03}
Jordano, P., Bascompte, J., and Olesen, J.~M., Invariant properties in
  coevolutionary networks of plant-animal interactions, \textit{Ecology
  Letters} \textbf{6}, 69--81 (2003).

\bibitem{JJ02}
Jost, J. and Joy, M.~P., Evolving networks with distance preferences,
  \textit{Phys. Rev. E} \textbf{66}, 036126 (2002).

\bibitem{KSM03}
Kalapala, V.~K., Sanwalani, V., and Moore, C., The structure of the {U}nited
  {S}tates road network, Preprint, University of New Mexico (2003).

\bibitem{Karinthy29}
Karinthy, F., Chains, in \textit{Everything is Different}, Budapest (1929).

\bibitem{Karonski82}
Karo\'nski, M., A review of random graphs, \textit{Journal of Graph Theory}
  \textbf{6}, 349--389 (1982).

\bibitem{Kauffman69}
Kauffman, S.~A., Metabolic stability and epigenesis in randomly connected nets,
  \textit{J. Theor. Bio.} \textbf{22}, 437--467 (1969).

\bibitem{Kauffman71}
Kauffman, S.~A., Gene regulation networks: A theory for their structure and
  global behavior, in A.~Moscana and A.~Monroy (eds.), \textit{Current Topics
  in Developmental Biology 6}, pp. 145--182, Academic Press, New York (1971).

\bibitem{Kauffman93}
Kauffman, S.~A., \textit{The Origins of Order}, Oxford University Press, Oxford
  (1993).

\bibitem{KSS97}
Kautz, H., Selman, B., and Shah, M., {R}eferral{W}eb: Combining social networks
  and collaborative filtering, \textit{Comm. ACM} \textbf{40}, 63--65 (1997).

\bibitem{Keeling99}
Keeling, M.~J., The effects of local spatial structure on epidemiological
  invasion, \textit{Proc. R. Soc. London B} \textbf{266}, 859--867 (1999).

\bibitem{KW91}
Kephart, J.~O. and White, S.~R., Directed-graph epidemiological models of
  computer viruses, in \textit{Proceedings of the 1991 IEEE Computer Society
  Symposium on Research in Security and Privacy}, pp. 343--359, IEEE Computer
  Society, Los Alamitos, CA (1991).

\bibitem{KB78}
Killworth, P.~D. and Bernard, H.~R., The reverse small world experiment,
  \textit{Social Networks} \textbf{1}, 159--192 (1978).

\bibitem{Kim02}
Kim, B.~J., Trusina, A., Holme, P., Minnhagen, P., Chung, J.~S., and Choi,
  M.~Y., Dynamic instabilities induced by asymmetric influence: Prisoners'
  dilemma game on small-world networks, \textit{Phys. Rev. E} \textbf{66},
  021907 (2002).

\bibitem{KYHJ02}
Kim, B.~J., Yoon, C.~N., Han, S.~K., and Jeong, H., Path finding strategies in
  scale-free networks, \textit{Phys. Rev. E} \textbf{65}, 027103 (2002).

\bibitem{KKKR02}
Kim, J., Krapivsky, P.~L., Kahng, B., and Redner, S., Infinite-order
  percolation and giant fluctuations in a protein interaction network,
  \textit{Phys. Rev. E} \textbf{66}, 055101 (2002).

\bibitem{Kinouchi02}
Kinouchi, O., Martinez, A.~S., Lima, G.~F., Louren\c{}co, G.~M., and
  Risau-Gusman, S., Deterministic walks in random networks: An application to
  thesaurus graphs, \textit{Physica A} \textbf{315}, 665--676 (2002).

\bibitem{KG99}
Kleczkowski, A. and Grenfell, B.~T., Mean-field-type equations for spread of
  epidemics: The `small world' model, \textit{Physica A} \textbf{274}, 355--360
  (1999).

\bibitem{KL01}
Kleinberg, J. and Lawrence, S., The structure of the {W}eb, \textit{Science}
  \textbf{294}, 1849--1850 (2001).

\bibitem{Kleinberg99a}
Kleinberg, J.~M., Authoritative sources in a hyperlinked environment,
  \textit{J. ACM} \textbf{46}, 604--632 (1999).

\bibitem{Kleinberg00}
Kleinberg, J.~M., Navigation in a small world, \textit{Nature} \textbf{406},
  845 (2000).

\bibitem{Kleinberg00proc}
Kleinberg, J.~M., The small-world phenomenon: An algorithmic perspective, in
  \textit{Proceedings of the 32nd Annual {ACM} Symposium on Theory of
  Computing}, pp. 163--170, Association of Computing Machinery, New York
  (2000).

\bibitem{Kleinberg02}
Kleinberg, J.~M., Small world phenomena and the dynamics of information, in
  T.~G. Dietterich, S.~Becker, and Z.~Ghahramani (eds.), \textit{Proceedings of
  the 2001 Neural Information Processing Systems Conference}, MIT Press,
  Cambridge, MA (2002).

\bibitem{Kleinberg99b}
Kleinberg, J.~M., Kumar, S.~R., Raghavan, P., Rajagopalan, S., and Tomkins, A.,
  The {W}eb as a graph: Measurements, models and methods, in
  \textit{Proceedings of the International Conference on Combinatorics and
  Computing}, no. 1627 in Lecture Notes in Computer Science, pp. 1--18,
  Springer, Berlin (1999).

\bibitem{KE02}
Klemm, K. and Eguiluz, V.~M., Highly clustered scale-free networks,
  \textit{Phys. Rev. E} \textbf{65}, 036123 (2002).

\bibitem{Klovdahl94}
Klovdahl, A.~S., Potterat, J.~J., Woodhouse, D.~E., Muth, J.~B., Muth, S.~Q.,
  and Darrow, W.~W., Social networks and infectious disease: The {C}olorado
  {S}prings study, \textit{Soc. Sci. Med.} \textbf{38}, 79--88 (1994).

\bibitem{Knuth93}
Knuth, D.~E., \textit{The Stanford GraphBase: A Platform for Combinatorial
  Computing}, Addison-Wesley, Reading, MA (1993).

\bibitem{KR01}
Krapivsky, P.~L. and Redner, S., Organization of growing random networks,
  \textit{Phys. Rev. E} \textbf{63}, 066123 (2001).

\bibitem{KR02b}
Krapivsky, P.~L. and Redner, S., Finiteness and fluctuations in growing
  networks, \textit{J. Phys. A} \textbf{35}, 9517--9534 (2002).

\bibitem{KR02a}
Krapivsky, P.~L. and Redner, S., A statistical physics perspective on {W}eb
  growth, \textit{Computer Networks} \textbf{39}, 261--276 (2002).

\bibitem{KR03}
Krapivsky, P.~L. and Redner, S., Rate equation approach for growing networks,
  in R.~Pastor-Satorras and J.~Rubi (eds.), \textit{Proceedings of the XVIII
  Sitges Conference on Statistical Mechanics}, Lecture Notes in Physics,
  Springer, Berlin (2003).

\bibitem{KRL00}
Krapivsky, P.~L., Redner, S., and Leyvraz, F., Connectivity of growing random
  networks, \textit{Phys. Rev. Lett.} \textbf{85}, 4629--4632 (2000).

\bibitem{KRR01}
Krapivsky, P.~L., Rodgers, G.~J., and Redner, S., Degree distributions of
  growing networks, \textit{Phys. Rev. Lett.} \textbf{86}, 5401--5404 (2001).

\bibitem{KVS96}
Kretzschmar, M., van Duynhoven, Y. T. H.~P., and Severijnen, A.~J., Modeling
  prevention strategies for gonorrhea and chlamydia using stochastic network
  simulations, \textit{Am. J. Epidemiol.} \textbf{114}, 306--317 (1996).

\bibitem{KAS99}
Kulkarni, R.~V., Almaas, E., and Stroud, D., Evolutionary dynamics in the
  {B}ak-{S}neppen model on small-world networks, Preprint cond-mat/9908216
  (1999).

\bibitem{KAS00}
Kulkarni, R.~V., Almaas, E., and Stroud, D., Exact results and scaling
  properties of small-world networks, \textit{Phys. Rev. E} \textbf{61},
  4268--4271 (2000).

\bibitem{Kumar00}
Kumar, R., Raghavan, P., Rajagopalan, S., Sivakumar, D., Tomkins, A.~S., and
  Upfal, E., Stochastic models for the {W}eb graph, in \textit{Proceedings of
  the 42st Annual {IEEE} Symposium on the Foundations of Computer Science}, pp.
  57--65, Institute of Electrical and Electronics Engineers, New York (2000).

\bibitem{KA01}
Kuperman, M. and Abramson, G., Small world effect in an epidemiological model,
  \textit{Phys. Rev. Lett.} \textbf{86}, 2909--2912 (2001).

\bibitem{KZ02}
Kuperman, M. and Zanette, D.~H., Stochastic resonance in a model of opinion
  formation on small world networks, \textit{Eur. Phys. J. B} \textbf{26},
  387--391 (2002).

\bibitem{LHCS00}
Lago-Fern\'andez, L.~F., Huerta, R., Corbacho, F., and Sig{\"u}enza, J.~A.,
  Fast response and temporal coherent oscillations in small-world networks,
  \textit{Phys. Rev. Lett.} \textbf{84}, 2758--2761 (2000).

\bibitem{LKK01}
Lahtinen, J., Kert\'esz, J., and Kaski, K., Scaling of random spreading in
  small world networks, \textit{Phys. Rev. E} \textbf{64}, 057105 (2001).

\bibitem{LKK02}
Lahtinen, J., Kert\'esz, J., and Kaski, K., Random spreading phenomena in
  annealed small world networks, \textit{Physica A} \textbf{311}, 571--580
  (2002).

\bibitem{LM01b}
Latora, V. and Marchiori, M., Efficient behavior of small-world networks,
  \textit{Phys. Rev. Lett.} \textbf{87}, 198701 (2001).

\bibitem{LM03}
Latora, V. and Marchiori, M., Economic small-world behavior in weighted
  networks, Preprint cond-mat/0204089 (2002).

\bibitem{LM02}
Latora, V. and Marchiori, M., Is the {B}oston subway a small-world network?,
  \textit{Physica A} \textbf{314}, 109--113 (2002).

\bibitem{LG99}
Lawrence, S. and Giles, C.~L., Accessibility of information on the web,
  \textit{Nature} \textbf{400}, 107--109 (1999).

\bibitem{LVVZ02}
Leone, M., V\'azquez, A., Vespignani, A., and Zecchina, R., Ferromagnetic
  ordering in graphs with arbitrary degree distribution, \textit{Eur. Phys. J.
  B} \textbf{28}, 191--197 (2002).

\bibitem{LEA03}
Liljeros, F., Edling, C.~R., and Amaral, L. A.~N., Sexual networks: Implication
  for the transmission of sexually transmitted infection, \textit{Microbes and
  Infections}  (in press).

\bibitem{Liljeros01}
Liljeros, F., Edling, C.~R., Amaral, L. A.~N., Stanley, H.~E., and \AA{}berg,
  Y., The web of human sexual contacts, \textit{Nature} \textbf{411}, 907--908
  (2001).

\bibitem{LM01a}
Lloyd, A.~L. and May, R.~M., How viruses spread among computers and people,
  \textit{Science} \textbf{292}, 1316--1317 (2001).

\bibitem{Luczak92}
\L{}uczak, T., Sparse random graphs with a given degree sequence, in A.~M.
  Frieze and T.~\L{}uczak (eds.), \textit{Proceedings of the Symposium on
  Random Graphs, Pozna\'n 1989}, pp. 165--182, John Wiley, New York (1992).

\bibitem{Mariolis75}
Mariolis, P., Interlocking directorates and control of corporations: The theory
  of bank control, \textit{Social Science Quarterly} \textbf{56}, 425--439
  (1975).

\bibitem{Maritan96}
Maritan, A., Rinaldo, A., Rigon, R., Giacometti, A., and Rodr\'\i{}guez-Iturbe,
  I., Scaling laws for river networks, \textit{Phys. Rev. E} \textbf{53},
  1510--1515 (1996).

\bibitem{Marsden90}
Marsden, P.~V., Network data and measurement, \textit{Annual Review of
  Sociology} \textbf{16}, 435--463 (1990).

\bibitem{Martinez91}
Martinez, N.~D., Artifacts or attributes? {E}ffects of resolution on the
  {L}ittle {R}ock {L}ake food web, \textit{Ecological Monographs} \textbf{61},
  367--392 (1991).

\bibitem{Martinez92}
Martinez, N.~D., Constant connectance in community food webs, \textit{American
  Naturalist} \textbf{139}, 1208--1218 (1992).

\bibitem{MS02a}
Maslov, S. and Sneppen, K., Specificity and stability in topology of protein
  networks, \textit{Science} \textbf{296}, 910--913 (2002).

\bibitem{MSZ03}
Maslov, S., Sneppen, K., and Zaliznyak, A., Pattern detection in complex
  networks: Correlation profile of the {I}nternet, Preprint cond-mat/0205379
  (2002).

\bibitem{MA88}
May, R.~M. and Anderson, R.~M., The transmission dynamics of human
  immunodeficiency virus ({HIV}), \textit{Philos. Trans. R. Soc. London B}
  \textbf{321}, 565--607 (1988).

\bibitem{ML01}
May, R.~M. and Lloyd, A.~L., Infection dynamics on scale-free networks,
  \textit{Phys. Rev. E} \textbf{64}, 066112 (2001).

\bibitem{MR96}
Meester, R. and Roy, R., \textit{Continuum Percolation}, Cambridge University
  Press, Cambridge (1996).

\bibitem{MP96}
Melin, G. and Persson, O., Studying research collaboration using
  co-authorships, \textit{Scientometrics} \textbf{36}, 363--377 (1996).

\bibitem{MB00}
Menczer, F. and Belew, R.~K., Adaptive retrieval agents: Internalizing local
  context and scaling up to the {W}eb, \textit{Machine Learning}
  \textbf{39}~(2-3), 203--242 (2000).

\bibitem{MPRS01}
Menczer, F., Pant, G., Ruiz, M., and Srinivasan, P., Evaluating topic-driven
  {W}eb crawlers, in D.~H. Kraft, W.~B. Croft, D.~J. Harper, and J.~Zobel
  (eds.), \textit{Proceedings of the 24th Annual International {ACM} {SIGIR}
  Conference on Research and Development in Information Retrieval}, pp.
  241--249, Association of Computing Machinery, New York (2001).

\bibitem{Merton68}
Merton, R.~K., The {M}atthew effect in science, \textit{Science} \textbf{159},
  56--63 (1968).

\bibitem{Milgram67}
Milgram, S., The small world problem, \textit{Psychology Today} \textbf{2},
  60--67 (1967).

\bibitem{Milo02}
Milo, R., Shen-Orr, S., Itzkovitz, S., Kashtan, N., Chklovskii, D., and Alon,
  U., Network motifs: Simple building blocks of complex networks,
  \textit{Science} \textbf{298}, 824--827 (2002).

\bibitem{Mitchell96}
Mitchell, M., \textit{Introduction to Genetic Algorithms}, MIT Press,
  Cambridge, MA (1996).

\bibitem{Mizruchi82}
Mizruchi, M.~S., \textit{The American Corporate Network, 1904--1974}, Sage,
  Beverley Hills (1982).

\bibitem{MR95}
Molloy, M. and Reed, B., A critical point for random graphs with a given degree
  sequence, \textit{Random Structures and Algorithms} \textbf{6}, 161--179
  (1995).

\bibitem{MR98}
Molloy, M. and Reed, B., The size of the giant component of a random graph with
  a given degree sequence, \textit{Combinatorics, Probability and Computing}
  \textbf{7}, 295--305 (1998).

\bibitem{Monasson99}
Monasson, R., Diffusion, localization and dispersion relations on `small-world'
  lattices, \textit{Eur. Phys. J. B} \textbf{12}, 555--567 (1999).

\bibitem{MS02b}
Montoya, J.~M. and Sol\'e, R.~V., Small world patterns in food webs, \textit{J.
  Theor. Bio.} \textbf{214}, 405--412 (2002).

\bibitem{Moody01}
Moody, J., Race, school integration, and friendship segregation in {A}merica,
  \textit{Am. J. Sociol.} \textbf{107}, 679--716 (2001).

\bibitem{Moody03}
Moody, J., The structure of a social science collaboration network, Preprint,
  Department of Sociology, Ohio State University (2003).

\bibitem{MN00a}
Moore, C. and Newman, M. E.~J., Epidemics and percolation in small-world
  networks, \textit{Phys. Rev. E} \textbf{61}, 5678--5682 (2000).

\bibitem{MN00b}
Moore, C. and Newman, M. E.~J., Exact solution of site and bond percolation on
  small-world networks, \textit{Phys. Rev. E} \textbf{62}, 7059--7064 (2000).

\bibitem{MAA03}
Moreira, A.~A., Andrade, Jr., J.~S., and Amaral, L. A.~N., Extremum statistics
  in scale-free network models, Preprint cond-mat/0205411 (2002).

\bibitem{Moreno34}
Moreno, J.~L., \textit{Who Shall Survive?}, Beacon House, Beacon, NY (1934).

\bibitem{MGP02}
Moreno, Y., G\'omez, J.~B., and Pacheco, A.~F., Instability of scale-free
  networks under node-breaking avalanches, \textit{Europhys. Lett.}
  \textbf{58}, 630--636 (2002).

\bibitem{MPVV03}
Moreno, Y., Pastor-Satorras, R., V\'azquez, A., and Vespignani, A., Critical
  load and congestion instabilities in scale-free networks, Preprint
  cond-mat/0209474 (2002).

\bibitem{MPV02}
Moreno, Y., Pastor-Satorras, R., and Vespignani, A., Epidemic outbreaks in
  complex heterogeneous networks, \textit{Eur. Phys. J. B} \textbf{26},
  521--529 (2002).

\bibitem{MV02}
Moreno, Y. and V\'azquez, A., The {B}ak-{S}neppen model on scale-free networks,
  \textit{Europhys. Lett.} \textbf{57}, 765--771 (2002).

\bibitem{MV03}
Moreno, Y. and V\'azquez, A., Disease spreading in structured scale-free
  networks, Preprint cond-mat/0210362 (2002).

\bibitem{Morris95}
Morris, M., Data driven network models for the spread of infectious disease, in
  D.~Mollison (ed.), \textit{Epidemic Models: Their Structure and Relation to
  Data}, pp. 302--322, Cambridge University Press, Cambridge (1995).

\bibitem{Morris97}
Morris, M., Sexual networks and {HIV}, \textit{AIDS 97: Year in Review}
  \textbf{11}, 209--216 (1997).

\bibitem{MDLD02}
Motter, A.~E., de~Moura, A.~P., Lai, Y.-C., and Dasgupta, P., Topology of the
  conceptual network of language, \textit{Phys. Rev. E} \textbf{65}, 065102
  (2002).

\bibitem{ML02}
Motter, A.~E. and Lai, Y.-C., Cascade-based attacks on complex networks,
  \textit{Phys. Rev. E} \textbf{66}, 065102 (2002).

\bibitem{Moukarzel99}
Moukarzel, C.~F., Spreading and shortest paths in systems with sparse
  long-range connections, \textit{Phys. Rev. E} \textbf{60}, 6263--6266 (1999).

\bibitem{MD02}
Moukarzel, C.~F. and de~Menezes, M.~A., Shortest paths on systems with
  power-law distributed long-range connections, \textit{Phys. Rev. E}
  \textbf{65}, 056709 (2002).

\bibitem{MSK00}
M{\"u}ller, J., Sch{\"o}nfisch, B., and Kirkilionis, M., Ring vaccination,
  \textit{J. Math. Biol.} \textbf{41}, 143--171 (2000).

\bibitem{Newman00b}
Newman, M. E.~J., Models of the small world, \textit{J. Stat. Phys.}
  \textbf{101}, 819--841 (2000).

\bibitem{Newman01d}
Newman, M. E.~J., Clustering and preferential attachment in growing networks,
  \textit{Phys. Rev. E} \textbf{64}, 025102 (2001).

\bibitem{Newman01b}
Newman, M. E.~J., Scientific collaboration networks: {I}. {N}etwork
  construction and fundamental results, \textit{Phys. Rev. E} \textbf{64},
  016131 (2001).

\bibitem{Newman01c}
Newman, M. E.~J., Scientific collaboration networks: {II}. {S}hortest paths,
  weighted networks, and centrality, \textit{Phys. Rev. E} \textbf{64}, 016132
  (2001).

\bibitem{Newman01a}
Newman, M. E.~J., The structure of scientific collaboration networks,
  \textit{Proc. Natl. Acad. Sci. USA} \textbf{98}, 404--409 (2001).

\bibitem{Newman02f}
Newman, M. E.~J., Assortative mixing in networks, \textit{Phys. Rev. Lett.}
  \textbf{89}, 208701 (2002).

\bibitem{Newman02c}
Newman, M. E.~J., Spread of epidemic disease on networks, \textit{Phys. Rev. E}
  \textbf{66}, 016128 (2002).

\bibitem{Newman02e}
Newman, M. E.~J., The structure and function of networks, \textit{Computer
  Physics Communications} \textbf{147}, 40--45 (2002).

\bibitem{Newman03a}
Newman, M. E.~J., Ego-centered networks and the ripple effect, \textit{Social
  Networks} \textbf{25}, 83--95 (2003).

\bibitem{Newman03c}
Newman, M. E.~J., Mixing patterns in networks, \textit{Phys. Rev. E}
  \textbf{67}, 026126 (2003).

\bibitem{Newman03b}
Newman, M. E.~J., Random graphs as models of networks, in S.~Bornholdt and
  H.~G. Schuster (eds.), \textit{Handbook of Graphs and Networks}, pp. 35--68,
  Wiley-VCH, Berlin (2003).

\bibitem{NBW03}
Newman, M. E.~J., Barab\'asi, A.-L., and Watts, D.~J., \textit{The Structure
  and Dynamics of Networks}, Princeton University Press, Princeton (2003).

\bibitem{NFB02}
Newman, M. E.~J., Forrest, S., and Balthrop, J., Email networks and the spread
  of computer viruses, \textit{Phys. Rev. E} \textbf{66}, 035101 (2002).

\bibitem{NMW00}
Newman, M. E.~J., Moore, C., and Watts, D.~J., Mean-field solution of the
  small-world network model, \textit{Phys. Rev. Lett.} \textbf{84}, 3201--3204
  (2000).

\bibitem{NSW01}
Newman, M. E.~J., Strogatz, S.~H., and Watts, D.~J., Random graphs with
  arbitrary degree distributions and their applications, \textit{Phys. Rev. E}
  \textbf{64}, 026118 (2001).

\bibitem{NW99a}
Newman, M. E.~J. and Watts, D.~J., Renormalization group analysis of the
  small-world network model, \textit{Phys. Lett. A} \textbf{263}, 341--346
  (1999).

\bibitem{NW99b}
Newman, M. E.~J. and Watts, D.~J., Scaling and percolation in the small-world
  network model, \textit{Phys. Rev. E} \textbf{60}, 7332--7342 (1999).

\bibitem{Ozana01}
Ozana, M., Incipient spanning cluster on small-world networks,
  \textit{Europhys. Lett.} \textbf{55}, 762--766 (2001).

\bibitem{PA93}
Padgett, J.~F. and Ansell, C.~K., Robust action and the rise of the {M}edici,
  1400--1434, \textit{Am. J. Sociol.} \textbf{98}, 1259--1319 (1993).

\bibitem{PBMW98}
Page, L., Brin, S., Motwani, R., and Winograd, T., The {P}agerank citation
  ranking: Bringing order to the web, Technical report, Stanford University
  (1998).

\bibitem{PA01}
Pandit, S.~A. and Amritkar, R.~E., Random spread on the family of small-world
  networks, \textit{Phys. Rev. E} \textbf{63}, 041104 (2001).

\bibitem{PR03}
Pastor-Satorras, R. and Rubi, J. (eds.), \textit{Proceedings of the XVIII
  Sitges Conference on Statistical Mechanics}, Lecture Notes in Physics,
  Springer, Berlin (2003).

\bibitem{PVV01}
Pastor-Satorras, R., V\'azquez, A., and Vespignani, A., Dynamical and
  correlation properties of the {I}nternet, \textit{Phys. Rev. Lett.}
  \textbf{87}, 258701 (2001).

\bibitem{PV01b}
Pastor-Satorras, R. and Vespignani, A., Epidemic dynamics and endemic states in
  complex networks, \textit{Phys. Rev. E} \textbf{63}, 066117 (2001).

\bibitem{PV01a}
Pastor-Satorras, R. and Vespignani, A., Epidemic spreading in scale-free
  networks, \textit{Phys. Rev. Lett.} \textbf{86}, 3200--3203 (2001).

\bibitem{PV02b}
Pastor-Satorras, R. and Vespignani, A., Epidemic dynamics in finite size
  scale-free networks, \textit{Phys. Rev. E} \textbf{65}, 035108 (2002).

\bibitem{PV02a}
Pastor-Satorras, R. and Vespignani, A., Immunization of complex networks,
  \textit{Phys. Rev. E} \textbf{65}, 036104 (2002).

\bibitem{PV03}
Pastor-Satorras, R. and Vespignani, A., Epidemics and immunization in
  scale-free networks, in S.~Bornholdt and H.~G. Schuster (eds.),
  \textit{Handbook of Graphs and Networks}, Wiley-VCH, Berlin (2003).

\bibitem{Pekalski01}
P\c{e}kalski, A., Ising model on a small world network, \textit{Phys. Rev. E}
  \textbf{64}, 057104 (2001).

\bibitem{Pennock02}
Pennock, D.~M., Flake, G.~W., Lawrence, S., Glover, E.~J., and Giles, C.~L.,
  Winners don't take all: Characterizing the competition for links on the web,
  \textit{Proc. Natl. Acad. Sci. USA} \textbf{99}, 5207--5211 (2002).

\bibitem{Pimm02}
Pimm, S.~L., \textit{Food Webs}, University of Chicago Press, Chicago, 2nd ed.
  (2002).

\bibitem{Podani01}
Podani, J., Oltvai, Z.~N., Jeong, H., Tombor, B., Barab\'asi, A.-L., and
  Szathmary, E., Comparable system-level organization of {A}rchaea and
  {E}ukaryotes, \textit{Nature Genetics} \textbf{29}, 54--56 (2001).

\bibitem{PK78}
Pool, I.~{\relax de S}. and Kochen, M., Contacts and influence, \textit{Social
  Networks} \textbf{1}, 1--48 (1978).

\bibitem{Potterat02}
Potterat, J.~J., Phillips-Plummer, L., Muth, S.~Q., Rothenberg, R.~B.,
  Woodhouse, D.~E., Maldonado-Long, T.~S., Zimmerman, H.~P., and Muth, J.~B.,
  Risk network structure in the early epidemic phase of {HIV} transmission in
  {C}olorado {S}prings, \textit{Sexually Transmitted Infections} \textbf{78},
  i159--i163 (2002).

\bibitem{Price65}
Price, D. J.~{\relax de S}., Networks of scientific papers, \textit{Science}
  \textbf{149}, 510--515 (1965).

\bibitem{Price76}
Price, D. J.~{\relax de S}., A general theory of bibliometric and other
  cumulative advantage processes, \textit{J. Amer. Soc. Inform. Sci.}
  \textbf{27}, 292--306 (1976).

\bibitem{RKM03}
Ramezanpour, A., Karimipour, V., and Mashaghi, A., Generating correlated
  networks from uncorrelated ones, Preprint cond-mat/0212469 (2002).

\bibitem{Rapoport57}
Rapoport, A., Contribution to the theory of random and biased nets,
  \textit{Bulletin of Mathematical Biophysics} \textbf{19}, 257--277 (1957).

\bibitem{Rapoport68}
Rapoport, A., Cycle distribution in random nets, \textit{Bulletin of
  Mathematical Biophysics} \textbf{10}, 145--157 (1968).

\bibitem{RH61}
Rapoport, A. and Horvath, W.~J., A study of a large sociogram,
  \textit{Behavioral Science} \textbf{6}, 279--291 (1961).

\bibitem{RB03}
Ravasz, E. and Barab\'asi, A.-L., Hierarchical organization in complex
  networks, \textit{Phys. Rev. E} \textbf{67}, 026112 (2003).

\bibitem{Ravasz02}
Ravasz, E., Somera, A.~L., Mongru, D.~A., Oltvai, Z., and Barab\'asi, A.-L.,
  Hierarchical organization of modularity in metabolic networks,
  \textit{Science} \textbf{297}, 1551--1555 (2002).

\bibitem{Redner98}
Redner, S., How popular is your paper? {A}n empirical study of the citation
  distribution, \textit{Eur. Phys. J. B} \textbf{4}, 131--134 (1998).

\bibitem{RV97}
Resnick, P. and Varian, H.~R., Recommender systems, \textit{Comm. ACM}
  \textbf{40}, 56--58 (1997).

\bibitem{RRR98}
Rinaldo, A., Rodr\'\i{}guez-Iturbe, I., and Rigon, R., Channel networks,
  \textit{Annual Review of Earth and Planetary Science} \textbf{26}, 289--327
  (1998).

\bibitem{RFI02}
Ripeanu, M., Foster, I., and Iamnitchi, A., Mapping the {G}nutella network:
  Properties of large-scale peer-to-peer systems and implications for system
  design, \textit{IEEE Internet Computing} \textbf{6}, 50--57 (2002).

\bibitem{RD01}
Rodgers, G.~J. and Darby-Dowman, K., Properties of a growing random directed
  network, \textit{Eur. Phys. J. B} \textbf{23}, 267--271 (2001).

\bibitem{RR97}
Rodr\'\i{}guez-Iturbe, I. and Rinaldo, A., \textit{Fractal River Basins: Chance
  and Self-Organization}, Cambridge University Press, Cambridge (1997).

\bibitem{RD39}
Roethlisberger, F.~J. and Dickson, W.~J., \textit{Management and the Worker},
  Harvard University Press, Cambridge, MA (1939).

\bibitem{RBTM01}
Rothenberg, R., Baldwin, J., Trotter, R., and Muth, S., The risk environment
  for {HIV} transmission: Results from the {A}tlanta and {F}lagstaff network
  studies, \textit{Journal of Urban Health} \textbf{78}, 419--431 (2001).

\bibitem{RCBH02}
Rozenfeld, A.~F., Cohen, R., ben Avraham, D., and Havlin, S., Scale-free
  networks on lattices, \textit{Phys. Rev. Lett.} \textbf{89}, 218701 (2002).

\bibitem{Sander02}
Sander, L.~M., Warren, C.~P., Sokolov, I., Simon, C., and Koopman, J.,
  Percolation on disordered networks as a model for epidemics, \textit{Math.
  Biosci.} \textbf{180}, 293--305 (2002).

\bibitem{SAB01}
Scala, A., Amaral, L. A.~N., and Barth\'el\'emy, M., Small-world networks and
  the conformation space of a short lattice polymer chain, \textit{Europhys.
  Lett.} \textbf{55}, 594--600 (2001).

\bibitem{Schwartz02}
Schwartz, N., Cohen, R., {ben-Avraham}, D., Barab\'asi, A.-L., and Havlin, S.,
  Percolation in directed scale-free networks, \textit{Phys. Rev. E}
  \textbf{66}, 015104 (2002).

\bibitem{Scott00}
Scott, J., \textit{Social Network Analysis: A Handbook}, Sage Publications,
  London, 2nd ed. (2000).

\bibitem{Seglen92}
Seglen, P.~O., The skewness of science, \textit{J. Amer. Soc. Inform. Sci.}
  \textbf{43}, 628--638 (1992).

\bibitem{SC01}
Sen, P. and Chakrabarti, B.~K., Small-world phenomena and the statistics of
  linear polymers, \textit{J. Phys. A} \textbf{34}, 7749--7755 (2001).

\bibitem{Sen03}
Sen, P., Dasgupta, S., Chatterjee, A., Sreeram, P.~A., Mukherjee, G., and
  Manna, S.~S., Small-world properties of the {I}ndian railway network,
  Preprint cond-mat/0208535 (2002).

\bibitem{SM95}
Shardanand, U. and Maes, P., Social information filtering: Algorithms for
  automating ``word of mouth'', in \textit{Proceedings of ACM Conference on
  Human Factors and Computing Systems}, pp. 210--217, Association of Computing
  Machinery, New York (1995).

\bibitem{SMMA02}
Shen-Orr, S., Milo, R., Mangan, S., and Alon, U., Network motifs in the
  transcriptional regulation network of {E}scherichia coli, \textit{Nature
  Genetics} \textbf{31}, 64--68 (2002).

\bibitem{SC02}
Sigman, M. and Cecchi, G.~A., Global organization of the {W}ordnet lexicon,
  \textit{Proc. Natl. Acad. Sci. USA} \textbf{99}, 1742--1747 (2002).

\bibitem{Simon55}
Simon, H.~A., On a class of skew distribution functions, \textit{Biometrika}
  \textbf{42}, 425--440 (1955).

\bibitem{Smith02}
Smith, R.~D., Instant messaging as a scale-free network, Preprint
  cond-mat/0206378 (2002).

\bibitem{Snijders02}
Snijders, T. A.~B., {M}arkov chain {M}onte {C}arlo estimation of exponential
  random graph models, \textit{Journal of Social Structure} \textbf{2}~(2)
  (2002).

\bibitem{SK03}
Socolar, J. E.~S. and Kauffman, S.~A., Scaling in ordered and critical random
  {B}oolean networks, \textit{PRL} \textbf{90}, 068702 (2003).

\bibitem{Soderberg02}
S{\"o}derberg, B., General formalism for inhomogeneous random graphs,
  \textit{Phys. Rev. E} \textbf{66}, 066121 (2002).

\bibitem{SM01}
Sol\'e, R.~V. and Montoya, J.~M., Complexity and fragility in ecological
  networks, \textit{Proc. R. Soc. London B} \textbf{268}, 2039--2045 (2001).

\bibitem{SP03}
Sol\'e, R.~V. and Pastor-Satorras, R., Complex networks in genomics and
  proteomics, in S.~Bornholdt and H.~G. Schuster (eds.), \textit{Handbook of
  Graphs and Networks}, pp. 145--167, Wiley-VCH, Berlin (2003).

\bibitem{SPSK02}
Sol\'e, R.~V., Pastor-Satorras, R., Smith, E., and Kepler, T.~B., A model of
  large-scale proteome evolution, \textit{Advances in Complex Systems}
  \textbf{5}, 43--54 (2002).

\bibitem{SR51}
Solomonoff, R. and Rapoport, A., Connectivity of random nets, \textit{Bulletin
  of Mathematical Biophysics} \textbf{13}, 107--117 (1951).

\bibitem{Sporns02}
Sporns, O., Network analysis, complexity, and brain function,
  \textit{Complexity} \textbf{8}~(1), 56--60 (2002).

\bibitem{STE00}
Sporns, O., Tononi, G., and Edelman, G.~M., Theoretical neuroanatomy: Relating
  anatomical and functional connectivity in graphs and cortical connection
  matrices, \textit{Cerebral Cortex} \textbf{10}, 127--141 (2000).

\bibitem{Stauffer02}
Stauffer, D., {M}onte {C}arlo simulations of {S}znajd models, \textit{Journal
  of Artificial Societies and Social Simulation} \textbf{5}~(1) (2002).

\bibitem{SADA03}
Stauffer, D., Aharony, A., da~Fontoura~Costa, L., and Adler, J., Efficient
  {H}opfield pattern recognition on a scale-free neural network, Preprint
  cond-mat/0212601 (2002).

\bibitem{Stelling02}
Stelling, J., Klamt, S., Bettenbrock, K., Schuster, S., and Gilles, E.~D.,
  Metabolic network structure determines key aspects of functionality and
  regulation, \textit{Nature} \textbf{420}, 190--193 (2002).

\bibitem{ST02}
Steyvers, M. and Tenenbaum, J.~B., The large-scale structure of semantic
  networks: Statistical analyses and a model for semantic growth, Preprint
  cond-mat/0110012 (2001).

\bibitem{Strauss86}
Strauss, D., On a general class of models for interaction, \textit{SIAM Review}
  \textbf{28}, 513--527 (1986).

\bibitem{Strogatz94}
Strogatz, S.~H., \textit{Nonlinear Dynamics and Chaos}, Addison-Wesley,
  Reading, MA (1994).

\bibitem{Strogatz01}
Strogatz, S.~H., Exploring complex networks, \textit{Nature} \textbf{410},
  268--276 (2001).

\bibitem{Svenson01}
Svenson, P., From {N}\'eel to {NPC}: Colouring small worlds, Preprint
  cs/0107015 (2001).

\bibitem{SAK03}
Szab\'o, G., Alava, M., and Kert\'esz, J., Structural transitions in scale-free
  networks, Preprint cond-mat/0208551 (2002).

\bibitem{SS00}
Sznajd-Weron, K. and Sznajd, J., Opinion evolution in closed community,
  \textit{Int. J. Mod. Phys. C} \textbf{11}, 1157--1165 (2000).

\bibitem{Tadic01}
Tadi\'c, B., Dynamics of directed graphs: The {W}orld-{W}ide {W}eb,
  \textit{Physica A} \textbf{293}, 273--284 (2001).

\bibitem{Tadic02}
Tadi\'c, B., Temporal fractal structures: Origin of power laws in the
  {W}orld-{W}ide {W}eb, \textit{Physica A} \textbf{314}, 278--283 (2002).

\bibitem{TM69}
Travers, J. and Milgram, S., An experimental study of the small world problem,
  \textit{Sociometry} \textbf{32}, 425--443 (1969).

\bibitem{Uetz00}
Uetz, P., Giot, L., Cagney, G., Mansfield, T.~A., Judson, R.~S., Knight, J.~R.,
  Lockshon, D., Narayan, V., Srinivasan, M., Pochart, P., Qureshi-Emili, A.,
  Li, Y., Godwin, B., Conover, D., Kalbfleisch, T., Vijayadamodar, G., Yang,
  M., Johnston, M., Fields, S., and Rothberg, J.~M., A comprehensive analysis
  of protein--protein interactions in saccharomyces cerevisiae, \textit{Nature}
  \textbf{403}, 623--627 (2000).

\bibitem{VCS02}
Valverde, S., Cancho, R.~F., and Sol\'e, R.~V., Scale-free networks from
  optimal design, \textit{Europhys. Lett.} \textbf{60}, 512--517 (2002).

\bibitem{Vazquez01}
V\'azquez, A., Statistics of citation networks, Preprint cond-mat/0105031
  (2001).

\bibitem{Vazquez03b}
V\'azquez, A., Growing networks with local rules: Preferential attachment,
  clustering hierarchy and degree correlations, Preprint cond-mat/0211528
  (2002).

\bibitem{Vazquez03a}
V\'azquez, A., Bogu{\~n}\'a, M., Moreno, Y., Pastor-Satorras, R., and
  Vespignani, A., Topology and correlations in structured scale-free networks,
  Preprint cond-mat/0209183 (2002).

\bibitem{VFMV03}
V\'azquez, A., Flammini, A., Maritan, A., and Vespignani, A., Modeling of
  protein interaction networks, \textit{Complexus} \textbf{1}, 38--44 (2003).

\bibitem{VM03}
V\'azquez, A. and Moreno, Y., Resilience to damage of graphs with degree
  correlations, \textit{Phys. Rev. E} \textbf{67}, 015101 (2003).

\bibitem{VPV02}
V\'azquez, A., Pastor-Satorras, R., and Vespignani, A., Large-scale topological
  and dynamical properties of the {I}nternet, \textit{Phys. Rev. E}
  \textbf{65}, 066130 (2002).

\bibitem{VW03}
V\'azquez, A. and Weigt, M., Computational complexity arising from degree
  correlations in networks, \textit{Phys. Rev. E} \textbf{67}, 027101 (2003).

\bibitem{VKR03}
Vazquez, F., Krapivsky, P.~L., and Redner, S., Constrained opinion dynamics:
  Freezing and slow evolution, \textit{J. Phys. A} \textbf{36}, L61--L68
  (2003).

\bibitem{Wagner01}
Wagner, A., The yeast protein interaction network evolves rapidly and contains
  few redundant duplicate genes, \textit{Mol. Biol. Evol.} \textbf{18},
  1283--1292 (2001).

\bibitem{WF01}
Wagner, A. and Fell, D., The small world inside large metabolic networks,
  \textit{Proc. R. Soc. London B} \textbf{268}, 1803--1810 (2001).

\bibitem{Walsh99}
Walsh, T., Search in a small world, in T.~Dean (ed.), \textit{Proceedings of
  the 16th International Joint Conference on Artificial Intelligence}, Morgan
  Kaufmann, San Francisco, CA (1999).

\bibitem{WZ98}
Wang, B.-Y. and Zhang, F., Exact counting of (0,1) matrices with given row and
  column sums, \textit{Discrete Mathematics} \textbf{187}, 211--220 (1998).

\bibitem{WSS02}
Warren, C.~P., Sander, L.~M., and Sokolov, I., Geography in a scale-free
  network model, \textit{Phys. Rev. E} \textbf{66}, 056105 (2002).

\bibitem{WF94}
Wasserman, S. and Faust, K., \textit{Social Network Analysis}, Cambridge
  University Press, Cambridge (1994).

\bibitem{WP96}
Wasserman, S. and Pattison, P., Logit models and logistic regressions for
  social networks: {I.} {A}n introduction to {M}arkov random graphs and p$^*$,
  \textit{Psychometrika} \textbf{61}, 401--426 (1996).

\bibitem{Watts99b}
Watts, D.~J., Networks, dynamics, and the small world phenomenon, \textit{Am.
  J. Sociol.} \textbf{105}, 493--592 (1999).

\bibitem{Watts99a}
Watts, D.~J., \textit{Small Worlds}, Princeton University Press, Princeton
  (1999).

\bibitem{Watts02}
Watts, D.~J., A simple model of global cascades on random networks,
  \textit{Proc. Natl. Acad. Sci. USA} \textbf{99}, 5766--5771 (2002).

\bibitem{Watts03}
Watts, D.~J., \textit{Six Degrees: The Science of a Connected Age}, Norton, New
  York (2003).

\bibitem{WDN02}
Watts, D.~J., Dodds, P.~S., and Newman, M. E.~J., Identity and search in social
  networks, \textit{Science} \textbf{296}, 1302--1305 (2002).

\bibitem{WS98}
Watts, D.~J. and Strogatz, S.~H., Collective dynamics of `small-world'
  networks, \textit{Nature} \textbf{393}, 440--442 (1998).

\bibitem{WBE97}
West, G.~B., Brown, J.~H., and Enquist, B.~J., A general model for the origin
  of allometric scaling laws in biology, \textit{Science} \textbf{276},
  122--126 (1997).

\bibitem{WBE99}
West, G.~B., Brown, J.~H., and Enquist, B.~J., A general model for the
  structure, and allometry of plant vascular systems, \textit{Nature}
  \textbf{400}, 664--667 (1999).

\bibitem{WBB76}
White, H.~C., Boorman, S.~A., and Breiger, R.~L., Social structure from
  multiple networks: {I.} {B}lockmodels of roles and positions, \textit{Am. J.
  Sociol.} \textbf{81}, 730--779 (1976).

\bibitem{WWN03}
White, H.~D., Wellman, B., and Nazer, N., Does citation reflect social
  structure? {L}ongitudinal evidence from the `{G}lobenet' interdisciplinary
  research group, Preprint, University of Toronto (2003).

\bibitem{WSTB86}
White, J.~G., Southgate, E., Thompson, J.~N., and Brenner, S., The structure of
  the nervous system of the nematode {C.} {E}legans, \textit{Phil. Trans. R.
  Soc. London} \textbf{314}, 1--340 (1986).

\bibitem{WH03}
Wilkinson, D. and Huberman, B.~A., A method for finding communities of related
  genes, Preprint, Stanford University (2002).

\bibitem{Williams02}
Williams, R.~J., Berlow, E.~L., Dunne, J.~A., Barab\'asi, A.-L., and Martinez,
  N.~D., Two degrees of separation in complex food webs, \textit{Proc. Natl.
  Acad. Sci. USA} \textbf{99}, 12913--12916 (2002).

\bibitem{Winfree00}
Winfree, A.~T., \textit{The Geometry of Biological Time}, Springer, New York,
  2nd ed. (2000).

\bibitem{Wormald81}
Wormald, N.~C., The asymptotic connectivity of labelled regular graphs,
  \textit{J. Comb. Theory B} \textbf{31}, 156--167 (1981).

\bibitem{Young03}
Young, H.~P., The diffusion of innovations in social networks, in L.~E. Blume
  and S.~N. Durlauf (eds.), \textit{The Economy as an Evolving Complex System},
  vol.~3, Oxford University Press, Oxford (2003).

\bibitem{Zanette01}
Zanette, D.~H., Critical behavior of propagation on small-world networks,
  \textit{Phys. Rev. E} \textbf{64}, 050901 (2001).

\bibitem{ZC01}
Zekri, N. and Clerc, J.~P., Statistical and dynamical study of disease
  propagation in a small world network, \textit{Phys. Rev. E} \textbf{64},
  056115 (2001).

\bibitem{ZZ03}
Zhu, J.-Y. and Zhu, H., Introducing small-world network effect to critical
  dynamics, Preprint cond-mat/0212542 (2002).

\end{thebibliography}
\end{document}